\documentclass[onecolumn]{aastex631}
\usepackage{xspace}
\usepackage{longtable}
\usepackage{placeins}
\usepackage{xcolor}
\usepackage{graphicx}
\usepackage{rotating}
\usepackage{url}
\usepackage{amsmath}
\usepackage{inputenc}
\usepackage[normalem]{ulem}

\begin{document}
\graphicspath{.}

\title{Structure in the Magnetic Field of the Milky Way Disk and Halo traced by Faraday Rotation} 

\author[0000-0002-6300-7459]{John M. Dickey}
\affiliation{School of Natural Sciences, Private Bag 37, University of Tasmania, Hobart, TAS, 7001, Australia}
\author[0000-0001-7722-8458]{Jennifer West}
\affiliation{Dunlap Institute for Astronomy and Astrophysics, University of Toronto, Toronto, ON M5S 3H4, Canada}
\author[0000-0001-9472-041X]{Alec J.M. Thomson}
\affiliation{CSIRO Space \& Astronomy, PO Box 1130, Bentley, WA 6102, Australia}
\author[0000-0003-1455-2546]{T.L. Landecker}
\affiliation{National Research Council Canada, Dominion Radio Astrophysical Observatory, P.O. Box 218, Penticton, British Columbia, V2A 6J9, Canada}
\author[0000-0003-0932-3140]{A. Bracco}
\affiliation{Laboratoire de Physique de l'Ecole Normale Sup\'erieure, ENS, Universit\'e PSL, CNRS, Sorbonne Universit\'e, Universit\'e de Paris, F-75005 Paris, France}
\author[0000-0002-3973-8403]{E. Carretti} 
\affiliation{INAF - Istituto di Radioastronomia, Via Gobetti 101, 40129 Bologna, Italy}
\author[0000-0002-0274-3092]{J.L. Han}
\affiliation{National Astronomical Observatories, Chinese Academy of Sciences, A20 Datun Road, Chaoyang District, Beijing 100101, China}
\affiliation{CAS Key Laboratory of FAST, NAOC, Chinese Academy of Sciences}
\affiliation{School of Astronomy and Space Sciences, University of Chinese Academy of Sciences, Beijing 100049, China}
\author[0000-0001-7301-5666]{A.S. Hill}
\affiliation{Department of Computer Science, Math, Physics, and Statistics, the University of British Columbia, Okanagan Campus, 3187 University Way, Kelowna, BC V1V1V7, Canada}
\affiliation{National Research Council Canada, Dominion Radio Astrophysical Observatory, P.O. Box 218, Penticton, British Columbia, V2A 6J9, Canada}
\author[0000-0003-0742-2006]{Y.K. Ma}
\affiliation{Research School of Astronomy \& Astrophysics, Australian National University, Canberra, ACT 2611, Australia}
\author[0000-0001-8906-7866]{S. A. Mao}
\affiliation{Max Planck Institute for Radio Astronomy, Auf dem H\"{u}gel 69, D-53121 Bonn, Germany}
\author[0000-0002-2465-8937]{A. Ordog}
\affiliation{National Research Council Canada, Dominion Radio Astrophysical Observatory, P.O. Box 218, Penticton, British Columbia, V2A 6J9, Canada}
\affiliation{Department of Computer Science, Math, Physics, and Statistics, the University of British Columbia, Okanagan Campus, 3187 University Way, Kelowna, BC V1V1V7, Canada}
\author[0000-0003-4781-5701]{Jo-Anne C. Brown}
\affiliation{Department of Physics and Astronomy, University of Calgary, Calgary, AB T2N 1N4, Canada}
\author[0000-0003-3320-2728]{K. A. Douglas}
\affiliation{Physics and Astronomy Department, Okanagan College, 1000 KLO Road, Kelowna, British Columbia, V1Y 4X8, Canada}
\affiliation{National Research Council Canada, Dominion Radio Astrophysical Observatory, P.O. Box 218, Penticton, British Columbia, V2A 6J9, Canada}
\affiliation{Department of Physics and Astronomy, University of Calgary, Calgary, AB T2N 1N4, Canada}
\author{A. Erceg}
\affiliation{Ru\-{dj}er Bo\v{s}kovi\'c Institute, Bijeni\v{c}ka cesta 54, 10 000 Zagreb, Croatia}
\author{V. Jeli\'{c}}
\affiliation{Ru\-{dj}er Bo\v{s}kovi\'c Institute, Bijeni\v{c}ka cesta 54, 10 000 Zagreb, Croatia}
\author{R. Kothes}
\affiliation{National Research Council Canada, Dominion Radio Astrophysical Observatory, P.O. Box 218, Penticton, British Columbia, V2A 6J9, Canada}
\author{M. Wolleben}
\affiliation{Skaha Remote Sensing Ltd., 3165 Juniper Drive, Naramata, BC V0H 1N0, Canada} 

\correspondingauthor{John Dickey}
\email{john.dickey@utas.edu.au}



\begin{abstract}
Magnetic fields in the ionized medium of the disk and halo of the Milky Way impose Faraday rotation on linearly polarized radio emission. We compare two surveys mapping the Galactic Faraday rotation, one showing the rotation measures of extragalactic sources seen through the Galaxy (from Hutschenreuter et al 2022), and one showing Faraday depth of the diffuse Galactic synchrotron emission from the Global Magneto-Ionic Medium Survey.
Comparing the two data sets in $5^{\circ}{\times}10^{\circ}$ bins shows good agreement at intermediate latitudes, ${10^{\circ}}<|b|<{50^{\circ}}$, and little correlation between them at lower and higher latitudes. Where they agree, both tracers show clear patterns as a function of Galactic longitude, $\ell$: in the Northern Hemisphere a strong $\sin{(2\ell)}$ pattern, and in the Southern hemisphere a $\sin{(\ell+\pi)}$ pattern. Pulsars with height above or below the plane $|z| > 300$ pc show similar $\ell$ dependence in their rotation measures. 
Nearby non-thermal structures show rotation measure shadows 
as does the Orion-Eridanus superbubble.  We describe families of dynamo models that could explain the observed patterns in the two hemispheres.  
 We suggest that a field reversal, known to cross the plane a few hundred pc inside the solar circle, could shift to positive $z$ with increasing Galactic radius to explain the $\sin{(2\ell)}$ pattern in the Northern Hemisphere.
 Correlation shows that rotation measures from extragalactic sources are one to two times the corresponding rotation measure of the diffuse emission, implying Faraday complexity along some lines of sight, especially in the Southern hemisphere.

\end{abstract}

\section{Introduction \label{sec:intro}}

The magneto-ionic medium is a mixture of ionized interstellar gas and magnetic field ($\vec{B}$)
that causes Faraday rotation of linearly polarized radiation at radio wavelengths.
The ionized gas can be either in classical H {\texttt{II}} regions or in the diffuse ionized medium,
in both the Milky Way disk and halo.  Although only the line of sight (LoS) component of
the $\vec{B}$ field contributes to Faraday rotation, surveys of rotation measure (RM)
provide such high precision and resolution that a useful picture of the interstellar
magnetic field emerges \citep{Han_2001, Brown_etal_2007, VanEck_etal_2011, Haverkorn_2015,Beck_2015, Han_2017, Jaffe_2019}.

To survey the RM requires a source of polarized emission, either compact sources or
the diffuse synchrotron emission by cosmic ray electrons in the Galactic $\vec{B}$ field.
Pulsars are excellent polarized sources, and
study of their RMs shows the structure of the ionized interstellar medium in the disk and lower
halo \citep{Han_etal_1999, Han_etal_2006, Han_etal_2018, Sobey_etal_2019}, but
it is limited by our imprecise knowledge of pulsar distance \citep{Cordes_Lazio_2002,Gaensler_etal_2008,Yao_etal_2017}. Most pulsars are close to the Galactic mid-plane, but a few are high enough above and below the plane that their RMs sample the magnetic field in the lower halo as well as in the disk. Extragalactic radio sources are often polarized, with intrinsic Faraday rotation that contributes to their RMs, but their measured RMs can be gridded, interpolated and smoothed using a Bayesian inference scheme to 
determine the contribution due to the Galactic foreground as a smooth function, i.e. the
Galactic foreground RM \citep{Han_etal_1997, Han_etal_1999, Oppermann_etal_2012, Xu_Han_2014, Oppermann_etal_2015, Ferriere_2016, Hutschenreuter_Ensslin_2020, Hutschenreuter_etal_2021}.  For brevity we refer
to the resulting values as the extragalactic RM, because it is based on surveys of polarized radio galaxies, but the gridded map is an estimate of the foreground, i.e. the Milky Way contribution to the RMs of the sources.

Another approach to measuring Galactic RMs is to study the Faraday
spectrum of the diffuse Galactic synchrotron emission.  The Faraday spectrum
\citep{Burn_1966,Brentjens_deBruyn_2005,Wolleben_etal_2010, Lenc_etal_2016, van_Eck_etal_2019,
Ferriere_etal_2021}
shows how the polarized brightness is distributed over a range of values of
Faraday depth, $\varphi$, corresponding to the RM of the intervening
magneto-ionic medium along the LoS between the telescope and each 
emission region.  Since diffuse Galactic synchrotron emission is widespread along every LoS,
the RM generalizes to the first moment of the Faraday spectrum 
\citep{Dickey_etal_2019}.  In this study, we make use of the GMIMS (Global Magneto-Ionic Medium Survey) high-band north (HBN) polarization dataset \citep{Wolleben_etal_2021}, in particular its first moments, which we will loosely refer to as the GMIMS or diffuse RM.

Comparing Galactic and extragalactic RMs at cm-wavelengths has been done in small areas, particularly at low latitudes, \citep{Ordog_etal_2017,Ordog_etal_2019,McKinven_2021} and in larger areas at low frequencies \citep{Riseley_2020,Erceg_etal_2022}.  Prior to GMIMS \citep{Wolleben_etal_2021}, large area surveys of the polarized synchrotron emission, e.g. \citet{Spoelstra_1984,Landecker_etal_2010}, did not have sufficient bandwidth, i.e. range in $\lambda^2$, to resolve the emission across the Faraday spectrum and allow accurate computation of the first moment of the RM.  This RM comparison over the whole sky north of $\delta=-28\arcdeg$ is the first step in a series of papers that will exploit the GMIMS RMs to understand the distribution of the Galactic $\vec{B}$ field with cosmic ray electrons that generate the synchrotron emission.

Section \ref{sec:RMdata} describes the RM data from the GMIMS survey and compares it to the extragalactic  RMs.  Section \ref{sec:loops} discusses the pulsar RMs and models for the nearby disk field, and presents a spherical harmonic expansion of the RM survey results, with a discussion of the imprint of nearby synchrotron and H$\alpha$ emission regions. There we compare the RMs of samples of pulsars that are at different heights, $z$, above or below the midplane with the extragalactic and diffuse emission RMs. Section \ref{sec:models} asks whether the asymmetry between the two Galactic hemispheres might be consistent with current dynamo models that solve the plasma equations for the global disk and halo field.  A combination of M0 and M1 dynamo solutions is promising and worth further study.  Section \ref{sec:discussion} discusses the significance of the ratios between the corresponding RM values in the extragalactic and GMIMS data, as evidence for different distributions of magneto-ionic (rotating) medium and synchrotron emission.  Section \ref{sec:conclusions} summarizes the results and suggests an overhead (positive $z$) field reversal as a possible paradigm for the RM pattern in the Northern Hemisphere.



\section{Rotation Measure Surveys Compared  \label{sec:RMdata}}

The extragalactic RM data used here is the map made from interpolation and gridding of RM catalogs by 
\citet{Hutschenreuter_etal_2021}, successor to similar maps by
\citet{Hutschenreuter_Ensslin_2020}, 
\citet{Oppermann_etal_2015}, and \citet{Oppermann_etal_2012}.  
RMs for the diffuse Galactic emission are derived from the 
GMIMS High-Band North survey (HBN, \citealp{Wolleben_etal_2021})
observed at wavelengths between 17 and 23 cm with the DRAO 26-m telescope.
The GMIMS RMs are the first moment of the Faraday cube
\citep[][sec 4.1]{Dickey_etal_2019, Ordog_2020}. 
For pixels whose maximum polarized intensity
in the Faraday cube is less than 0.03 K, the first moment is not computed, and the map is blanked. 
The GMIMS first moment map is further blanked for declinations less than $-25^{\circ}$,
to avoid systematic effects near the southern horizon of the survey at $\delta = -30^{\circ}$.

\begin{figure}
\centering

\hspace{.5in}\includegraphics[width=5.5in]{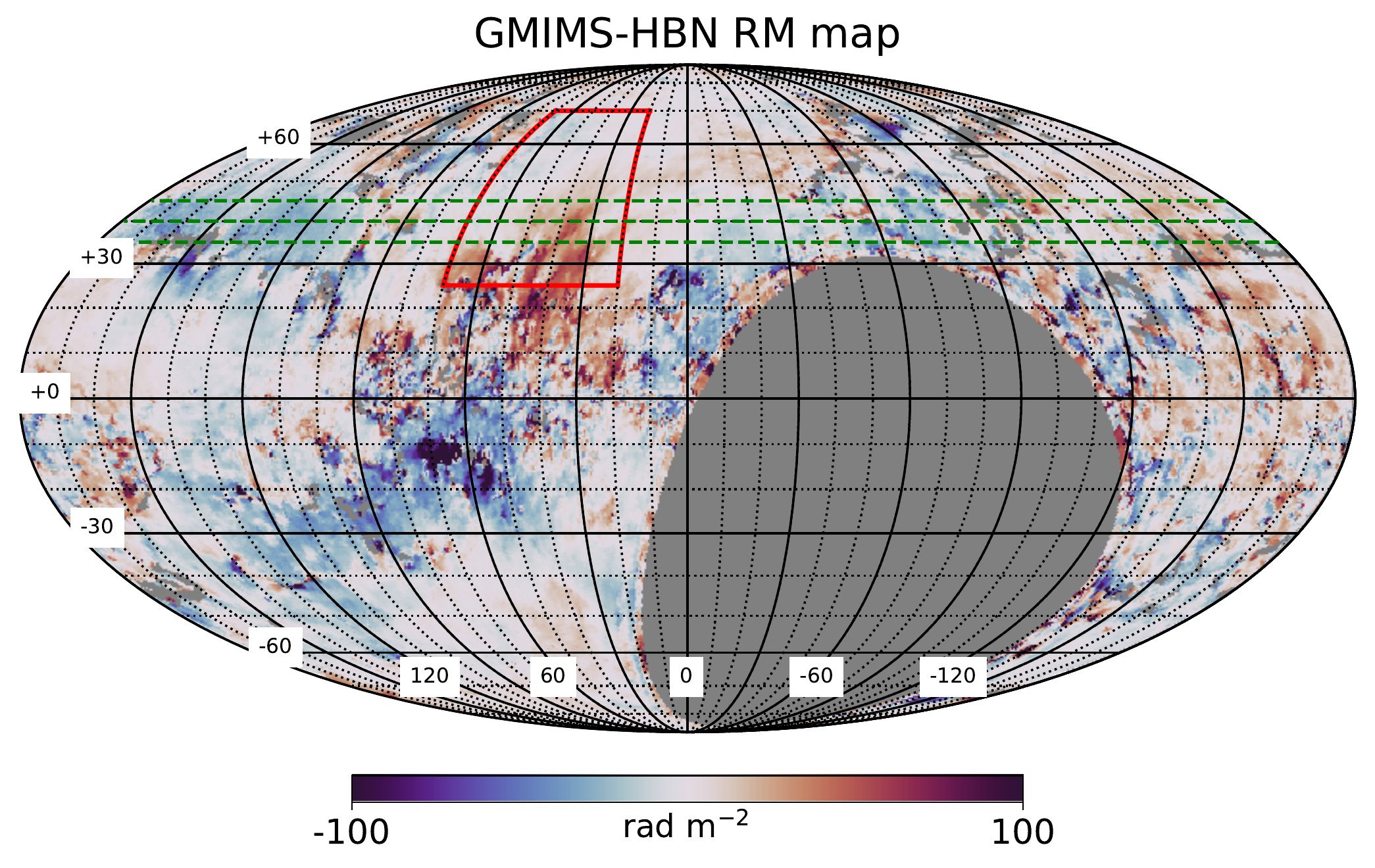}

\hspace{.5in}\includegraphics[width=5.5in]{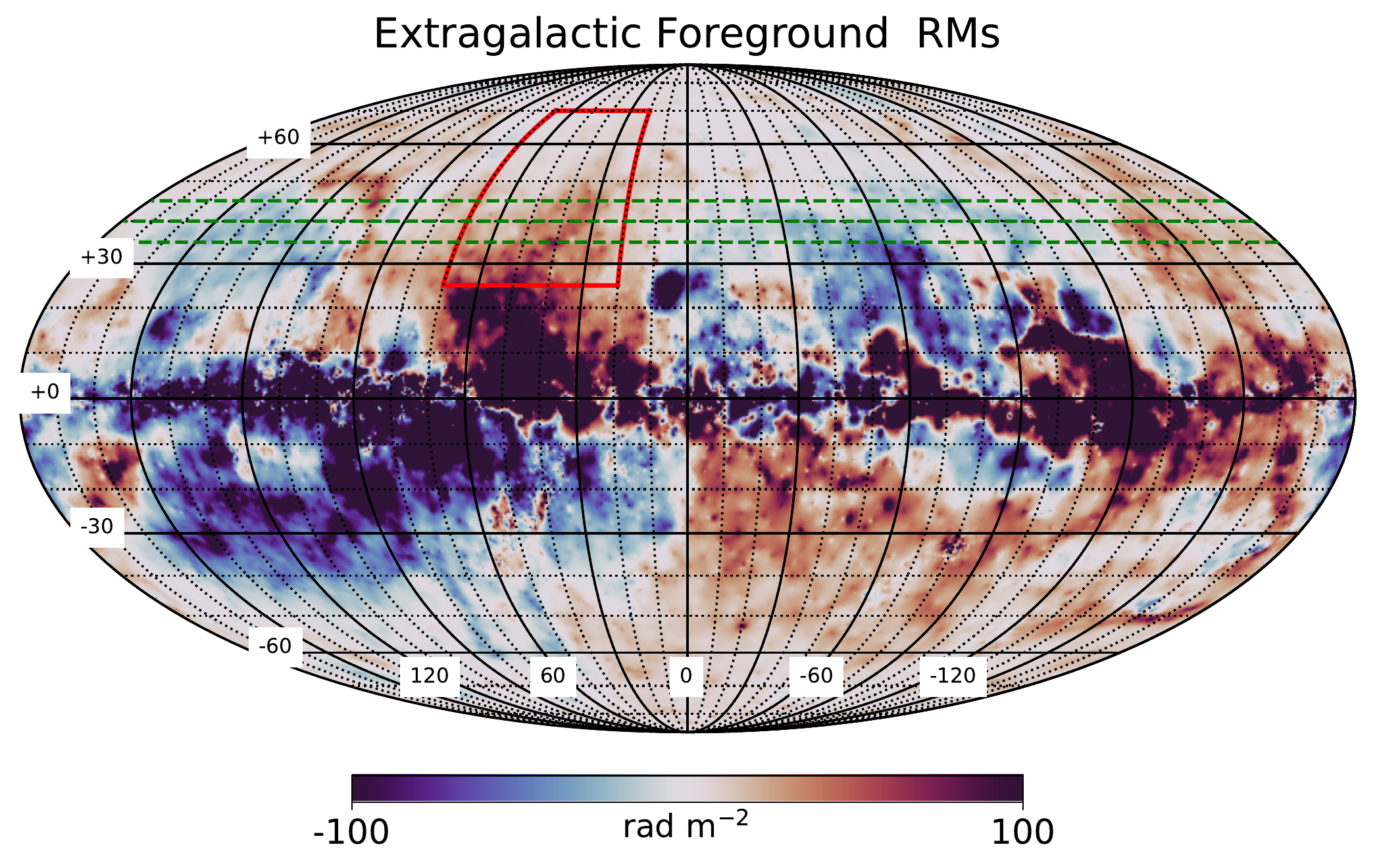} 

\caption{Two maps of the Galactic RM.  The upper panel shows the GMIMS High Band North first moment map made with data from the DRAO 26m telescope \citep{Wolleben_etal_2021}.
The lower panel shows the Galactic estimate based on extragalactic
source RMs \citep{Hutschenreuter_etal_2021}.  The red box indicates the area dominated by the North Polar Spur in Stokes I synchrotron emission, see Section \ref{sec:NPS} below.  The green dashed lines illustrate two sets of bins at constant latitude, used to make the two panels on Fig. \ref{fig:example_b40}.  
\label{fig:twomaps}   }
\end{figure}

To study the large-scale longitude variation of the RMs from the two data sets,
we sacrifice angular resolution by binning the data into cells with sizes of a few 
degrees, then compute the median value and the dispersion of the values in
each bin.  This process reduces the scatter due to small scale structure in
the RMs; the median filter attenuates the effect of spurious points with
very large positive or negative RMs that may be caused by small
regions of high electron density and/or a strong, localized, random component in the Galactic
$\vec{B}$ field.  Many different bin sizes were tried, 
all giving qualitatively similar results, described in Appendix \ref{sec:appendix_bin_sizes}.
Here we present profiles for bins with longitude width 10$^{\circ}$ and 
latitude width 5$^{\circ}$.   A reduced chi-squared measure of goodness of fit is given in Appendix \ref{sec:appendix_chisquared} with a discussion of its limitations due to the non-Gaussian distribution of RM
values in the bins.
The values from the two surveys are taken
at the same points in the maps after reprojection to 
a common Healpix\footnote{\url{http://healpix.sourceforge.net} \citep{Gorski:2005ku}} projection (Nside=512, nested).
Each bin has $\sim$50 to 140 independent values, depending on the latitude, since the beam size (FWHM) of the GMIMS observations is $\sim$40\arcmin. The density of the extragalactic sources is $\sim$1.3 per square degree on average, but lower for the south celestial pole region.

\subsection{Longitude Dependence of the RMs \label{sec:2longitude}}

\begin{figure}
\hspace{1.1in}\includegraphics[width=5in]{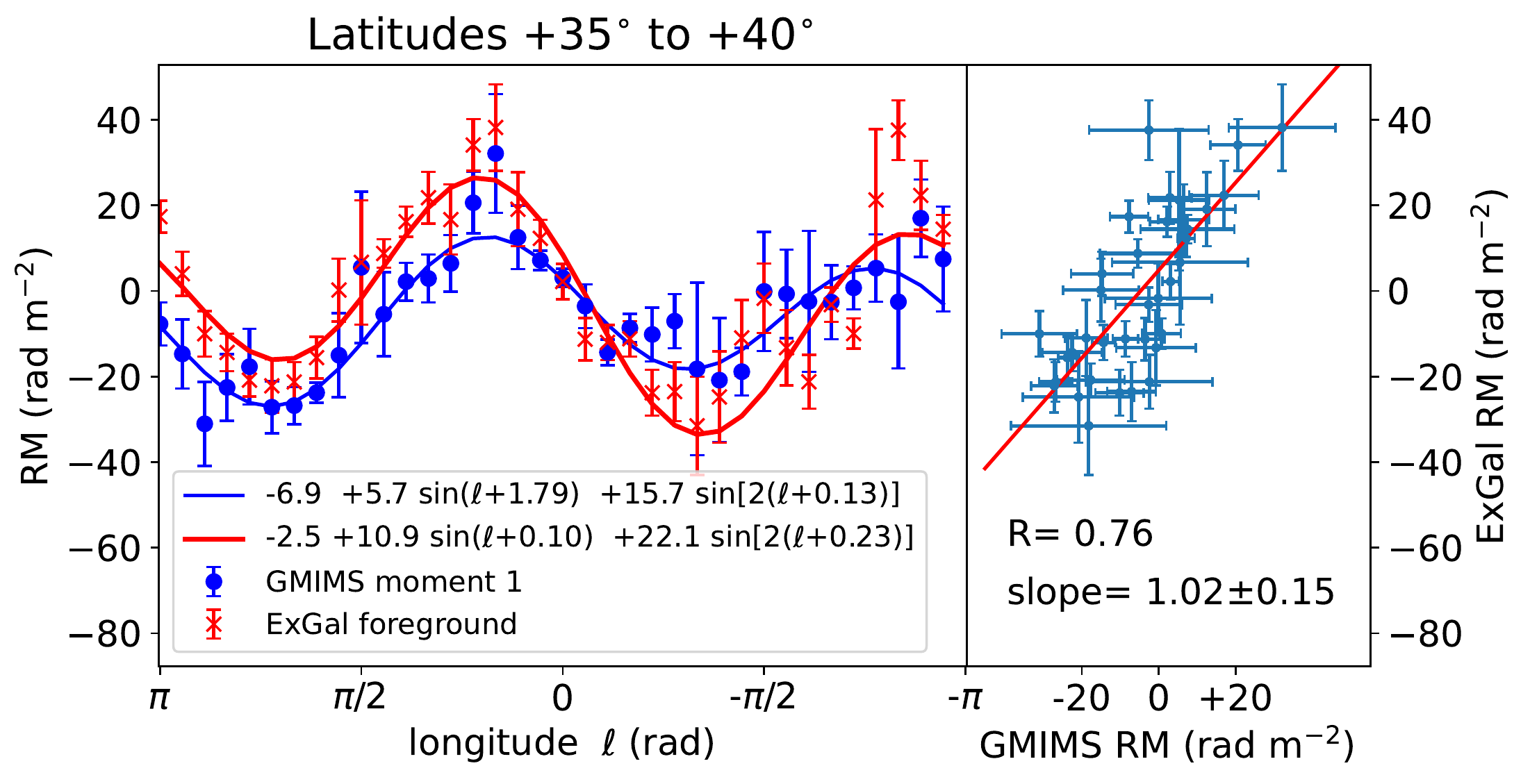} 

\hspace{1.1in} \includegraphics[width=5in]{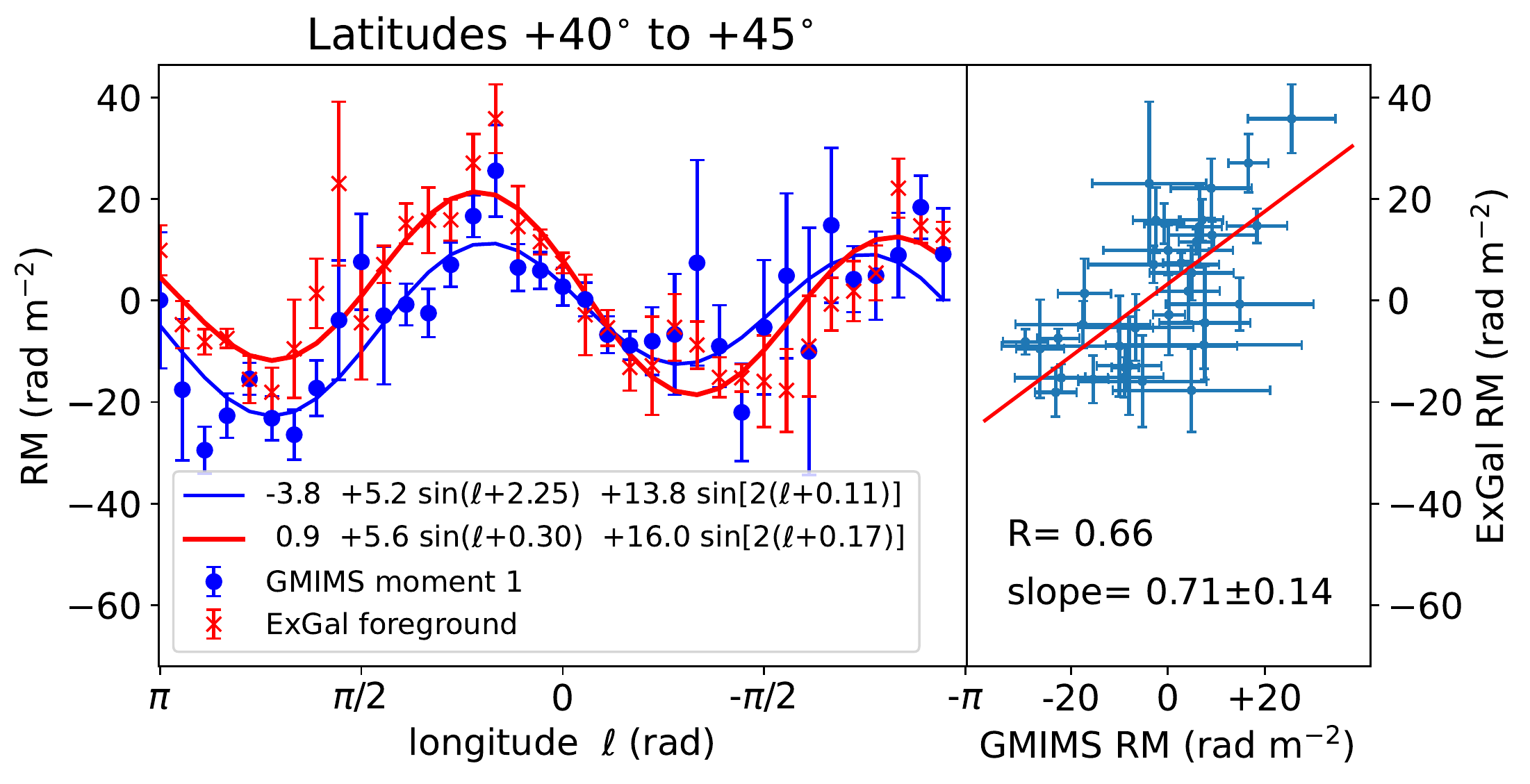}

\caption{Examples of the $\sin{(2\ell)}$ form of the RM at intermediate positive latitudes.  On the upper panel are profiles made from averaging the latitude
range $+35^{\circ}<b<+40^{\circ}$, with the GMIMS (DRAO) data shown in blue and the extragalactic
(ExGal)  RM grid \citep{Hutschenreuter_etal_2021} shown in red. Error bars on the points show $\pm 1 \sigma$ 
of the distributions of values in each bin.
The least-squares fit parameters are shown (Equation \ref{eq:5paramfit}, see Table \ref{tab:longitude_fits}).
In the lower panel is a similar pair of averages for $+40^{\circ}<b<+45^{\circ}$.
In each panel, the scatter plot on the right shows the correlation between the median 
values from GMIMS (x-axis) and the extragalactic sample.  The Pearson correlation 
coefficient ($R$) is indicated, along with the slope of the best linear fit, shown
with the red line. 
\label{fig:example_b40}   }
\end{figure}

The upper panel of Fig. \ref{fig:twomaps} shows the GMIMS High Band North (DRAO) first moment map \citep{Wolleben_etal_2021},
with a red outline showing the area influenced by the North Polar Spur (NPS, see section \ref{sec:NPS} below) at mid-latitudes.  
The lower panel of Fig. \ref{fig:twomaps} shows the \citet{Hutschenreuter_etal_2021} map based on extragalactic source RMs.
The bin edges in longitude are spaced by
10$^{\circ}$, indicated by the solid and dotted meridional lines on Fig. \ref{fig:twomaps}.
The dispersion of RM values in each bin is computed from the 16th and 84th percentiles of their distributions, corresponding to roughly plus and minus one sigma for a Gaussian distribution.  These are plotted as error bars on the data points on Fig. \ref{fig:example_b40}.

\begin{sidewaystable}[p]
\caption{Longitude Fit Parameters (Eq. \ref{eq:5paramfit}.  Bold face indicates statistically significant values.) \label{tab:longitude_fits}}
\begin{tabular}{|c|ccccc|ccccc|}
\hline
Latitude   & \multicolumn{5}{c|}{GMIMS (DRAO)}  &    \multicolumn{5}{c|}{Extragalactic} \\
range & $C_o$& $C_1$ & $\phi_1$ & $C_2$ & $\phi_2$ &
 $C_o$& $C_1$ & $\phi_1$ & $C_2$ & $\phi_2$ \\
($^{\circ}$) &
rad m$^{-2}$ & rad m$^{-2}$ & radians & rad m$^{-2}$ & radians &
rad m$^{-2}$ & rad m$^{-2}$ & radians & rad m$^{-2}$ & radians \\ \hline
-60$< b <$-55 & -15.5$\pm$ 6.4 &{20.1$\pm$10.0} &{\bf -0.18$\pm$0.15} &{12.9$\pm$ 4.5} &{0.28$\pm$0.09} &  8.1$\pm$ 0.8 &{\bf  7.8$\pm$ 1.2} &{\bf 2.99$\pm$0.12} &{ 2.7$\pm$ 1.0} &{0.65$\pm$0.22} \\ 
-55$< b <$-50 & -33.4$\pm$10.8 &{47.2$\pm$17.4} &{\bf -0.31$\pm$0.08} &{24.1$\pm$ 7.2} &{0.35$\pm$0.09} &  7.9$\pm$ 0.9 &{\bf 10.9$\pm$ 1.1} &{\bf 2.85$\pm$0.13} &{ 3.2$\pm$ 1.2} &{0.51$\pm$0.22} \\ 
-50$< b <$-45 & -29.1$\pm$10.9 &{38.0$\pm$17.1} &{\bf -0.46$\pm$0.13} &{25.5$\pm$ 7.2} &{0.24$\pm$0.10} &  5.2$\pm$ 0.8 &{\bf 13.9$\pm$ 1.2} &{\bf 2.73$\pm$0.09} &{ 2.6$\pm$ 1.2} &{0.85$\pm$0.27} \\ 
-45$< b <$-40 & -33.6$\pm$ 6.4 &{44.3$\pm$10.3} &{\bf -0.42$\pm$0.06} &{20.0$\pm$ 4.6} &{\bf 0.33$\pm$0.07} &  2.6$\pm$ 1.3 &{\bf 19.0$\pm$ 2.1} &{\bf 2.98$\pm$0.09} &{ 6.3$\pm$ 1.9} &{0.82$\pm$0.15} \\ 
-40$< b <$-35 & -20.9$\pm$ 6.8 &{20.5$\pm$10.0} &{-0.65$\pm$0.19} &{17.2$\pm$ 5.5} &{0.22$\pm$0.11} & -2.4$\pm$ 2.1 &{\bf 31.7$\pm$ 3.0} &{\bf 2.87$\pm$0.09} &{12.1$\pm$ 3.0} &{0.75$\pm$0.11} \\ 
-35$< b <$-30 & -12.5$\pm$ 4.5 &{ 3.3$\pm$ 7.9} &{-0.61$\pm$0.74} &{13.1$\pm$ 4.9} &{0.63$\pm$0.11} & -4.6$\pm$ 2.5 &{\bf 40.3$\pm$ 3.9} &{\bf 2.86$\pm$0.08} &{ 5.0$\pm$ 3.7} &{0.86$\pm$0.32} \\ 
-30$< b <$-25 & -16.3$\pm$ 5.1 &{ 8.8$\pm$ 7.6} &{-0.91$\pm$0.56} &{19.6$\pm$ 4.9} &{0.70$\pm$0.09} & -7.2$\pm$ 3.0 &{\bf 53.1$\pm$ 4.6} &{\bf 3.01$\pm$0.07} &{13.5$\pm$ 4.2} &{1.09$\pm$0.14} \\ 
-25$< b <$-20 & -14.4$\pm$ 3.9 &{ 3.8$\pm$ 3.2} &{-2.39$\pm$1.69} &{15.5$\pm$ 3.3} &{0.91$\pm$0.15} & -16.8$\pm$ 3.6 &{\bf 57.0$\pm$ 5.3} &{\bf -3.13$\pm$0.09} &{25.2$\pm$ 5.4} &{1.01$\pm$0.10} \\ 
-20$< b <$-15 & -3.2$\pm$ 6.3 &{11.6$\pm$10.2} &{-2.93$\pm$0.51} &{ 8.8$\pm$ 4.2} &{1.09$\pm$0.40} & -13.4$\pm$ 4.6 &{\bf 50.1$\pm$ 6.8} &{\bf -2.95$\pm$0.12} &{29.6$\pm$ 6.8} &{1.07$\pm$0.10} \\ 
-15$< b <$-10 & -4.0$\pm$ 6.1 &{ 7.9$\pm$ 9.4} &{-2.72$\pm$1.06} &{ 1.4$\pm$ 5.3} &{-0.11$\pm$2.29} & -9.2$\pm$ 6.2 &{44.8$\pm$ 9.3} &{-2.73$\pm$0.18} &{32.0$\pm$ 8.5} &{0.93$\pm$0.14} \\ 
-10$< b <$ -5 &  4.6$\pm$ 4.7 &{24.7$\pm$ 8.4} &{3.01$\pm$0.16} &{ 9.8$\pm$ 5.5} &{-1.30$\pm$0.17} & -2.3$\pm$ 7.1 &{\bf 68.2$\pm$10.7} &{\bf -2.98$\pm$0.14} &{12.4$\pm$ 9.2} &{0.64$\pm$0.36} \\ 
 -5$< b <$ +0 & 11.9$\pm$ 4.8 &{29.9$\pm$ 7.8} &{3.10$\pm$0.20} &{ 9.3$\pm$ 4.6} &{-1.09$\pm$0.26} & 11.0$\pm$11.4 &{\bf 100.9$\pm$17.3} &{-3.09$\pm$0.15} &{32.9$\pm$14.4} &{0.06$\pm$0.19} \\ 
 +0$< b <$ +5 & -7.6$\pm$ 4.3 &{21.1$\pm$ 7.6} &{-0.81$\pm$0.22} &{14.9$\pm$ 5.1} &{\bf 0.30$\pm$0.07} &  6.0$\pm$13.3 &{44.1$\pm$19.3} &{2.84$\pm$0.43} &{\bf 78.3$\pm$15.2} &{0.19$\pm$0.10} \\ 
 +5$< b <$+10 & -11.6$\pm$ 3.3 &{30.4$\pm$ 6.2} &{\bf -0.47$\pm$0.10} &{\bf 23.7$\pm$ 3.7} &{\bf 0.28$\pm$0.04} & 19.4$\pm$ 7.7 &{17.0$\pm$ 9.2} &{1.75$\pm$0.71} &{38.4$\pm$ 8.7} &{0.25$\pm$0.12} \\ 
+10$< b <$+15 &  1.5$\pm$ 2.5 &{ 9.2$\pm$ 3.7} &{-1.59$\pm$0.49} &{13.2$\pm$ 4.3} &{0.01$\pm$0.08} & 12.3$\pm$ 4.9 &{26.8$\pm$ 7.9} &{0.19$\pm$0.19} &{\bf 36.4$\pm$ 6.0} &{0.14$\pm$0.08} \\ 
+15$< b <$+20 & -7.8$\pm$ 3.7 &{19.6$\pm$ 6.5} &{\bf -0.17$\pm$0.14} &{\bf 21.8$\pm$ 3.6} &{\bf 0.32$\pm$0.05} &  3.5$\pm$ 3.1 &{\bf 30.7$\pm$ 4.9} &{\bf 0.18$\pm$0.12} &{\bf 27.2$\pm$ 4.1} &{0.28$\pm$0.08} \\ 
+20$< b <$+25 & -10.2$\pm$ 3.2 &{20.9$\pm$ 5.1} &{\bf -0.37$\pm$0.13} &{\bf 22.9$\pm$ 3.3} &{\bf 0.31$\pm$0.05} &  3.6$\pm$ 2.7 &{16.7$\pm$ 3.8} &{0.23$\pm$0.23} &{\bf 26.2$\pm$ 3.4} &{\bf 0.25$\pm$0.06} \\ 
+25$< b <$+30 & -2.6$\pm$ 2.6 &{ 8.3$\pm$ 4.3} &{-0.13$\pm$0.29} &{\bf 19.7$\pm$ 3.1} &{0.16$\pm$0.08} &  2.5$\pm$ 1.9 &{\bf 14.8$\pm$ 2.9} &{0.32$\pm$0.16} &{\bf 27.9$\pm$ 2.3} &{\bf 0.22$\pm$0.04} \\ 
+30$< b <$+35 & -5.3$\pm$ 1.4 &{ 3.6$\pm$ 2.2} &{-0.08$\pm$0.60} &{\bf 16.5$\pm$ 1.6} &{\bf 0.19$\pm$0.07} &  2.0$\pm$ 1.7 &{10.5$\pm$ 2.7} &{-0.10$\pm$0.19} &{\bf 23.2$\pm$ 2.5} &{\bf 0.08$\pm$0.05} \\ 
+35$< b <$+40 & -6.9$\pm$ 0.9 &{ 5.7$\pm$ 1.3} &{1.79$\pm$0.25} &{\bf 15.7$\pm$ 1.2} &{\bf 0.13$\pm$0.04} & -2.5$\pm$ 1.5 &{10.9$\pm$ 2.3} &{0.10$\pm$0.18} &{\bf 22.1$\pm$ 2.4} &{\bf 0.23$\pm$0.04} \\ 
+40$< b <$+45 & -3.8$\pm$ 1.3 &{ 5.2$\pm$ 1.5} &{2.25$\pm$0.39} &{\bf 13.8$\pm$ 1.6} &{\bf 0.11$\pm$0.07} &  0.9$\pm$ 0.9 &{ 5.6$\pm$ 1.4} &{0.30$\pm$0.19} &{\bf 16.0$\pm$ 1.3} &{\bf 0.17$\pm$0.04} \\ 
+45$< b <$+50 & -0.5$\pm$ 1.3 &{ 6.2$\pm$ 1.9} &{2.70$\pm$0.19} &{\bf 10.6$\pm$ 1.4} &{0.31$\pm$0.08} &  4.1$\pm$ 0.9 &{ 3.3$\pm$ 1.4} &{0.06$\pm$0.34} &{\bf  8.7$\pm$ 1.5} &{\bf 0.30$\pm$0.06} \\ 
+50$< b <$+55 &  0.5$\pm$ 1.0 &{ 6.6$\pm$ 1.5} &{\bf 2.75$\pm$0.14} &{\bf  7.3$\pm$ 1.4} &{0.66$\pm$0.08} &  3.9$\pm$ 0.7 &{ 1.1$\pm$ 1.2} &{0.10$\pm$0.96} &{ 6.1$\pm$ 1.3} &{0.47$\pm$0.08} \\ 
+55$< b <$+60 &  2.2$\pm$ 1.0 &{ 4.7$\pm$ 1.5} &{2.45$\pm$0.25} &{ 4.6$\pm$ 1.3} &{1.52$\pm$0.15} &  6.4$\pm$ 0.7 &{\bf  5.6$\pm$ 0.9} &{-1.02$\pm$0.18} &{ 0.6$\pm$ 0.9} &{-0.26$\pm$0.75} \\ 
\hline
\end{tabular}
\end{sidewaystable}

Two examples of the median filtered RM versus longitude data using
latitudes $+35^{\circ}< b <+40^{\circ}$ 
and latitudes $+40^{\circ}< b <+45^{\circ}$ (marked by the green dashed lines on Fig. \ref{fig:twomaps})
are shown in Fig. 
\ref{fig:example_b40}.  The GMIMS (DRAO) values are shown in blue, the extragalactic
 values in red.  The formulae indicated on the figure are least squares
fits to the points using a five-parameter function to determine the first three terms
of a Fourier series in longitude, $\ell$, i.e.
\begin{equation}\label{eq:5paramfit}
RM(\ell) \ = \ C_0 \ + \ C_1 \sin{(\ell + \phi_1)} \ + \ C_2 \sin{2 (\ell + \phi_2)}.
\end{equation}
Values of the constants $C_0, C_1, C_2, \phi_1,$ and $\phi_2$, with errors, are given on Table \ref{tab:longitude_fits} for the range of latitudes $-60^{\circ}<b<+60^{\circ}$.
Errors on the parameters are the square roots of the diagonal elements of the covariance
matrix, from \textsc{SCIPY} routine {\it optimize.curve\_fit} \citep{Jones_etal_2001}.
Amplitudes and phases shown in {\bf bold face} on Table \ref{tab:longitude_fits}
are statistically significant, either because the amplitude is more than five
times the error, or because the phase error is less than 0.15 radians (8$^{\circ}$) for $\phi_1$
or 0.075 radians (4$^{\circ}$) for $\phi_2$. 
The fitted phases have offsets of $\pm \pi$ so that all phases are in the
ranges $-\pi < \phi_1 < +\pi$ and $-\frac{\pi}{2} < \phi_2 < +\frac{\pi}{2}$ and
the amplitudes, $C_1$ and $C_2$, are positive.  

\begin{figure}
\hspace{.85in}\includegraphics[width=5.2in]{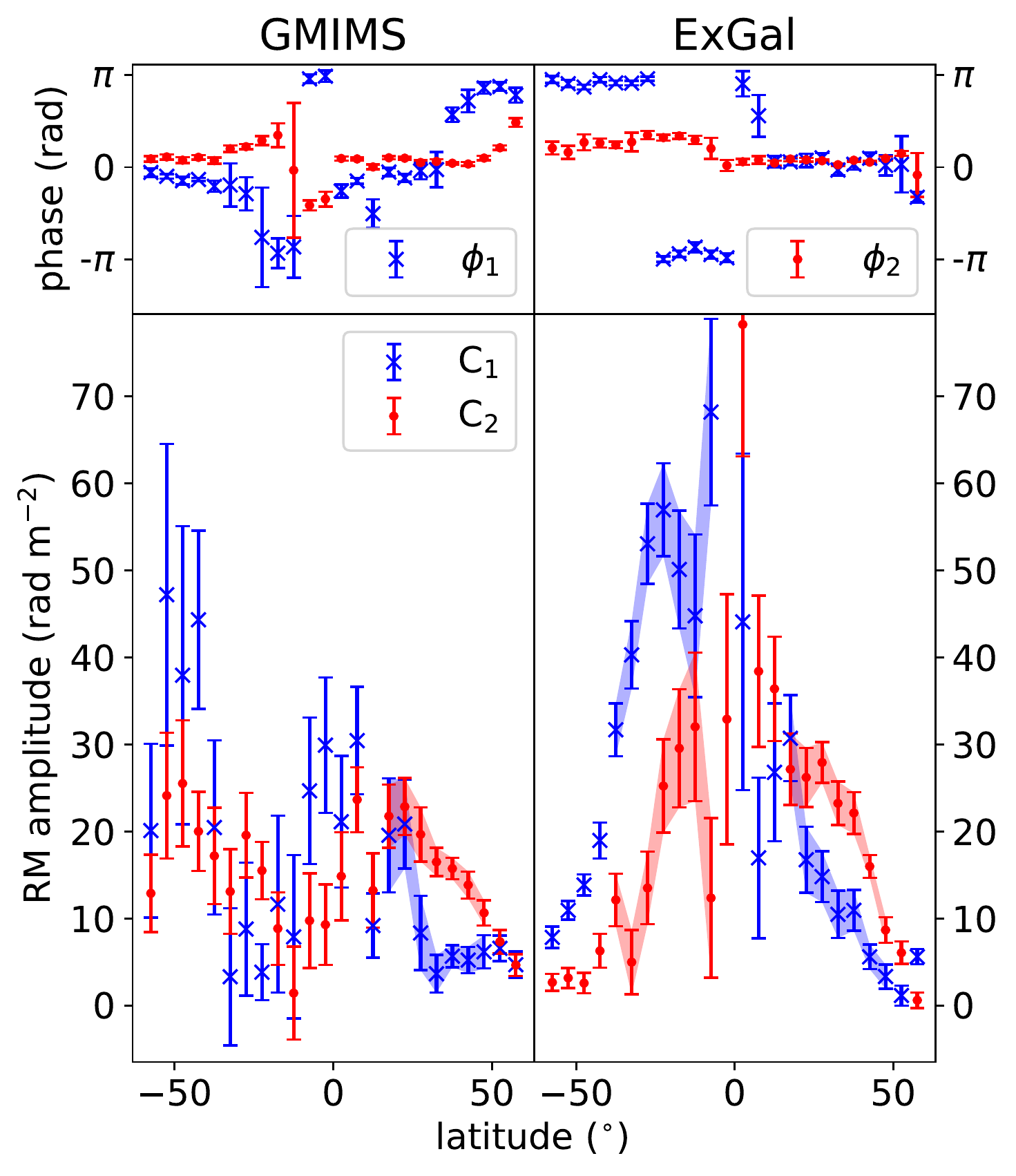} 
\caption{Amplitudes ($C_1$ and $C_2$) and phases ($\phi_1$ and $\phi_2$) of the $\sin{(\ell)}$ and $\sin{(2 \ell)}$
terms, from Table \ref{tab:longitude_fits}.  The GMIMS results are on the left,
the extragalactic on the right.  The values of the fitted parameters
(Equation \ref{eq:5paramfit}) are shown in blue for the $\sin{(\ell)}$ and red for the
$\sin{(2 \ell)}$ terms.  The curves are shaded for latitudes where the amplitudes, $C_1$ and/or $C_2$, are greater than 5$\sigma$.
At negative latitudes much of the third and fourth quadrants are not observable in
the GMIMS survey.  Because of this, points on the left side of the left panel have large errors, and they are not shaded.
\label{fig:constants_vs_lat} }
\end{figure}

The values of the amplitudes and phases from Table \ref{tab:longitude_fits} are 
displayed on Fig. \ref{fig:constants_vs_lat}.  The GMIMS phases and amplitudes
are on the left panel, the extragalactic values are on the right panel, with
phases at the top and amplitudes below.  The amplitudes and phases of the $\sin{(\ell)}$
terms are shown in blue, of the $\sin{(2 \ell)}$ terms in red.
Some low latitude points, $(|b|\leq 5\arcdeg$), are off scale on the lower panels of Fig. \ref{fig:constants_vs_lat}: to include them would collapse the scale so that the intermediate latitude points would be compressed at the bottom.  At such low latitudes the
path lengths through the disk are very long, several kpc, so RMs can be very high, and they vary dramatically on
angles smaller than the DRAO telescope beam.
For the extragalactic RMs (right panel),
the $\sin{(\ell)}$ term is much stronger than the $\sin{(2 \ell)}$ term for negative
latitudes, with the blue curve above the red for $b < 0^{\circ}$.  For the GMIMS
RMs, the negative latitudes are not fully sampled in longitude, so the amplitudes of the 
two terms are not well determined; all the GMIMS $C_1$ and $C_2$ values at $b < 0^{\circ}$ are
less than 5$\sigma$ on Table \ref{tab:longitude_fits}.  The latitude range where $\sin{(2 \ell)}$ dominates is $+20^{\circ} < b < +50^{\circ}$,
where on the two lower panels the red curves are well above the blue on Fig. \ref{fig:constants_vs_lat}.

For the latitude range $+20^{\circ} < b < +50^{\circ}$, all of the longitude slices
of the extragalactic survey show fitted amplitudes $C_2$
on Table \ref{tab:longitude_fits} that are greater than five sigma (4.9\,$\sigma$ in one case)
and also greater than $C_1$, mostly by a factor of two or more.
For all these latitudes the fits show small errors in $\phi_2$, $\sigma_{\phi 2} \leq 0.08$ radians.
These latitudes show similar domination by the $\sin{(2 \ell)}$ term in the GMIMS data.
All have values of $C_2$ from fits to the GMIMS data that are also above five sigma
with the exception of $+25^{\circ} < b < +30^{\circ}$, where the $C_2$ value is at the 
4.5\,$\sigma$ level. The phases are well determined, the noise in the phase, 
$\sigma_{\phi 2} \le 0.08$ radians. 

\subsection{Correlation Results \label{sec:correlation}}

At high latitudes ($|b|>50\arcdeg$), the RM values from the two surveys show little correlation. In both surveys the RMs are close to zero at both poles, with means +3.9 and +0.5 rad m$^{-2}$ for 50$\arcdeg < b < 70\arcdeg$ for the extragalactic and GMIMS surveys, respectively.  The standard deviations of the binned median RMs in this range are 6.5 rad m$^{-2}$ for the GMIMS data and 2.4 rad m$^{-2}$ for the extragalactic data.  In the South, the GMIMS survey covers only about half of the high latitude region.  The GMIMS survey has a broad RM spread function (RMSF), $\delta \varphi = 140$ rad m$^{-2}$ \citep{Wolleben_etal_2021}, as well as a large beam size ($\delta \theta = 40\arcmin$).  The lack of correlation between the two surveys at high Galactic latitudes may be due in part to poor Faraday spectral resolution of the GMIMS data in an area of very small values of RM, to the low surface brightness of the diffuse polarized emission, and to the dominance of the random field component, as the projection of the ordered field on the line of sight is small in this direction.

\begin{figure}

\includegraphics[width=3.5in]{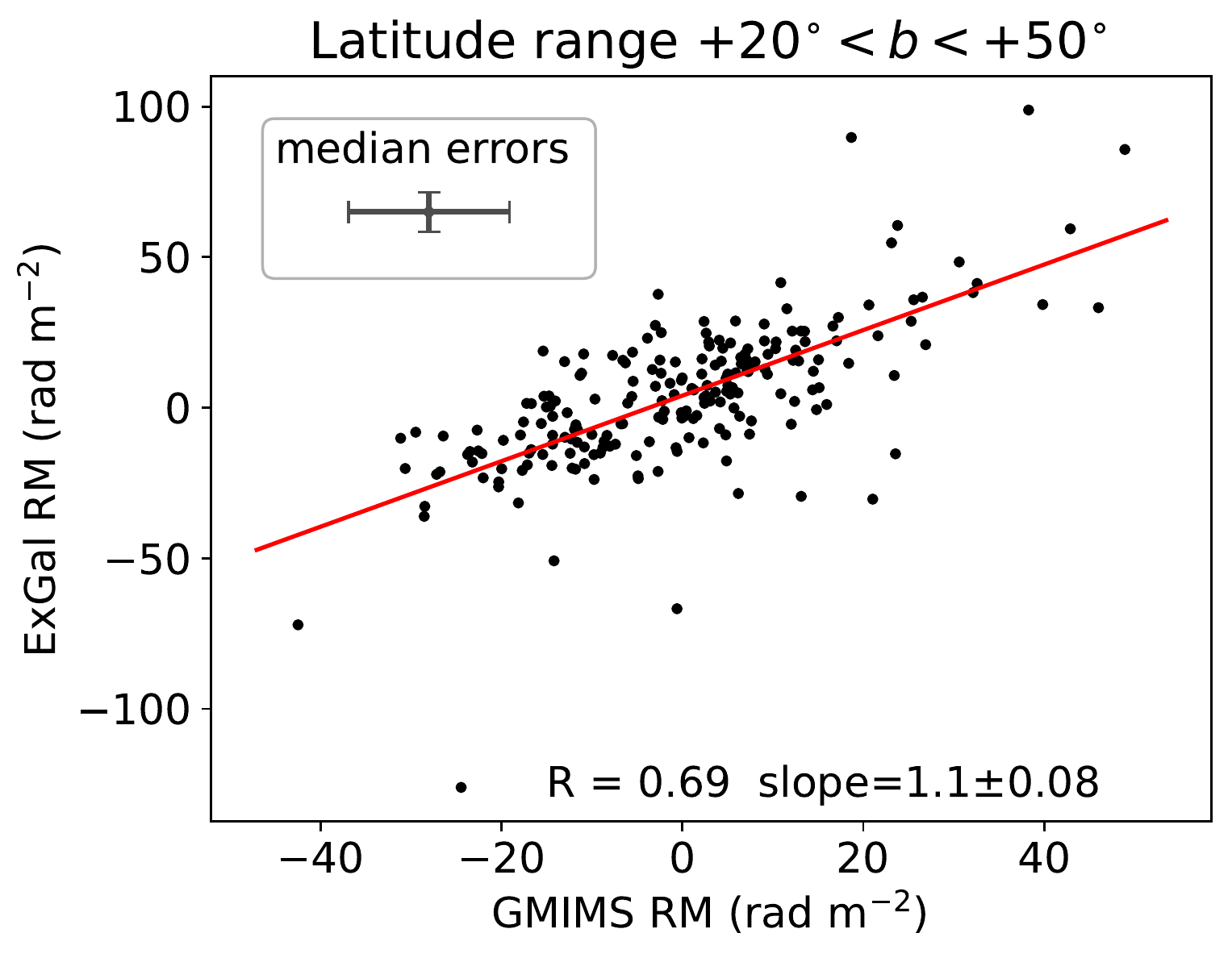} 
\includegraphics[width=3.5in]{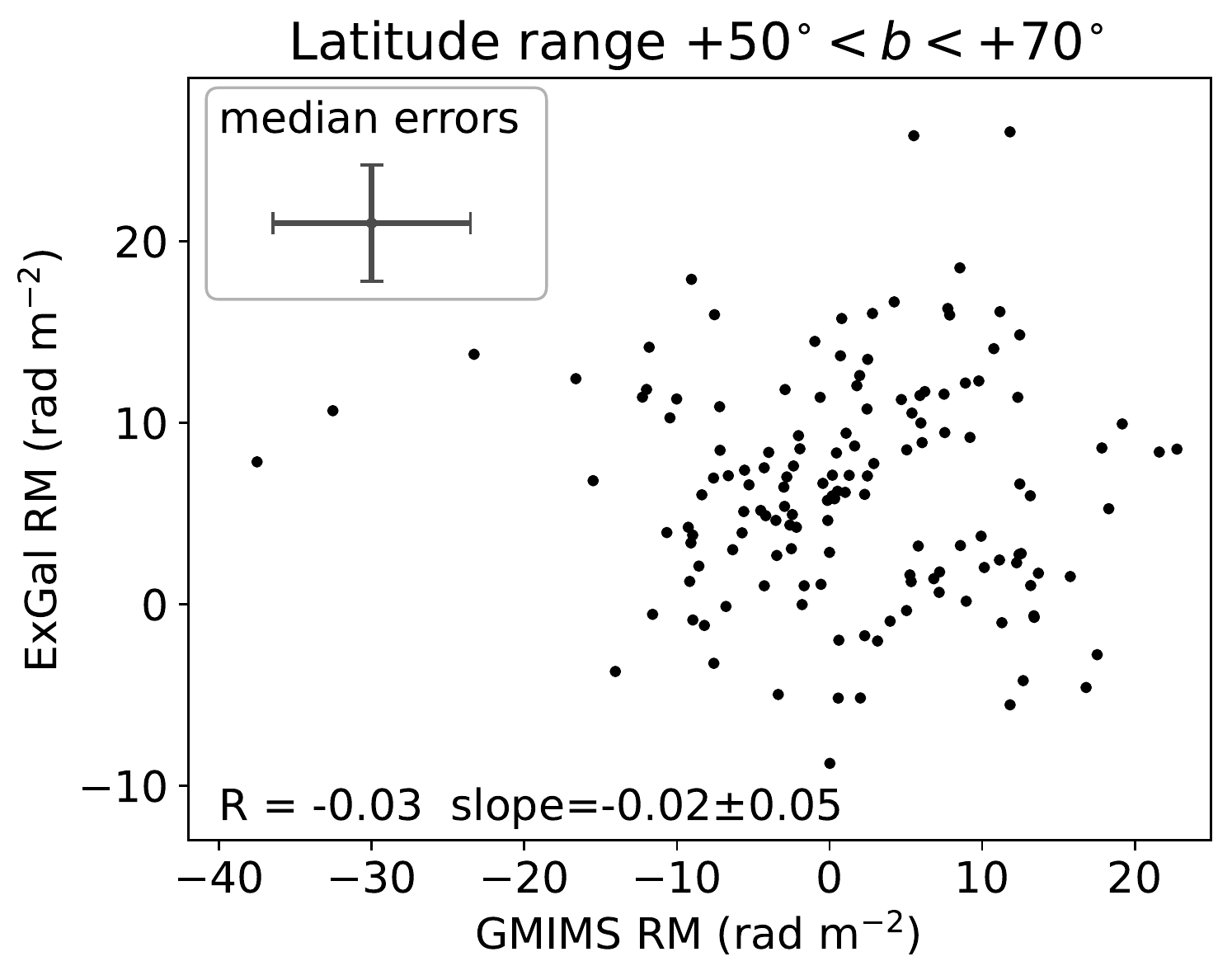}

\caption{Scatter plots of median values from the extragalactic RM map
of \citet{Hutschenreuter_etal_2021} versus corresponding median values of RM from the GMIMS survey for the latitude range $+20^{\circ}<b<+50^{\circ}$ (left panel)
and $+50^{\circ}<b<+70^{\circ}$ (right panel).  
On the left, the correlation coefficient is R=0.69, and the slope of the best fit line is 1.1$\pm$0.08,  
whereas on the right there is no correlation ($R=-0.03$).  The median errors on the points are shown in the insets.  On the right (high latitudes) the standard deviation of the GMIMS values is 9.8 (x axis) and for the extragalactic values the standard deviation is 6.1 (y axis). 
\label{fig:scatter} }
\end{figure}

\begin{figure}
\hspace{2.1in}\includegraphics[width=4.2in]{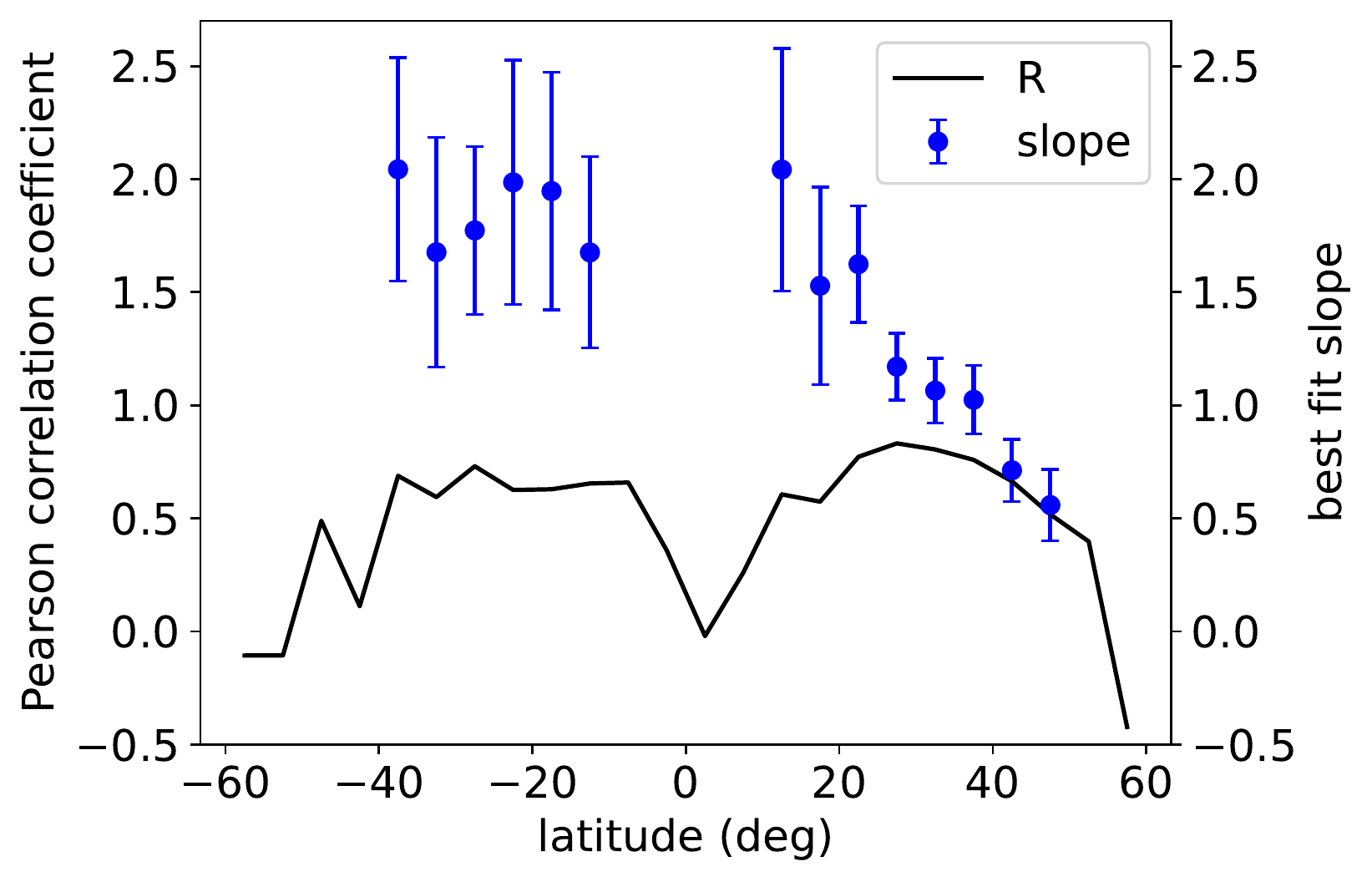} \hspace{.1in}
\caption{The Pearson correlation coefficient, $R$, versus Galactic latitude, from 
Table \ref{tab:correl}.  The blue dots show the slope, $\frac{\Delta RM_{ExGal}}{\Delta RM_{GMIMS}}$,
of the best fit line (right hand axis). Dots are plotted only for latitudes $10\arcdeg<|b|<50\arcdeg$
where the correlation is strong ($R>0.5$).  \label{fig:R_vs_b} }
\end{figure}

For latitudes between 20$\arcdeg$ and 50$\arcdeg$, we plot a scatter diagram of the extragalactic versus GMIMS median RMs, calculated in the bins described above
in the left panel of Fig. \ref{fig:scatter}.
The correlation coefficient is R=0.69, and the slope of the best fit line is 1.1,
using \textsc{SCIPY} regression analysis routine {\texttt{stats.linregress}}. 
In contrast, for latitudes above $b=+50^{\circ}$ there is no correlation, $R=-0.03$, as shown
on the right panel of Fig. \ref{fig:scatter}.
Values of $R$ for each 5$^{\circ}$ of latitude  are given on Table \ref{tab:correl} and illustrated on Fig. \ref{fig:R_vs_b}.


\begin{table}
\caption{Correlation Results \label{tab:correl}}
\begin{center}
\begin{tabular}{|lcc|}
\hline
Latitude & $R$ & slope \\
\hline
-60$< b <$-55 & -0.11 & -0.16$\pm$0.35 \\
-55$< b <$-50 & -0.11 & -0.20$\pm$0.43 \\
-50$< b <$-45 & 0.49 & 0.48$\pm$0.19 \\
-45$< b <$-40 & 0.11 & 0.19$\pm$0.39 \\
-40$< b <$-35 & 0.69 & 2.04$\pm$0.49 \\
-35$< b <$-30 & 0.59 & 1.68$\pm$0.51 \\
-30$< b <$-25 & 0.73 & 1.77$\pm$0.37 \\
-25$< b <$-20 & 0.63 & 1.99$\pm$0.54 \\
-20$< b <$-15 & 0.63 & 1.95$\pm$0.53 \\
-15$< b <$-10 & 0.65 & 1.68$\pm$0.42 \\
-10$< b <$ -5 & 0.66 & 4.01$\pm$0.98 \\
 -5$< b <$ +0 & 0.36 & 2.90$\pm$1.58 \\
 +0$< b <$ +5 & -0.02 & -0.26$\pm$2.59 \\
 +5$< b <$+10 & 0.26 & 1.45$\pm$1.10 \\
+10$< b <$+15 & 0.61 & 2.04$\pm$0.54 \\
+15$< b <$+20 & 0.57 & 1.53$\pm$0.44 \\
+20$< b <$+25 & 0.77 & 1.62$\pm$0.26 \\
+25$< b <$+30 & 0.83 & 1.17$\pm$0.15 \\
+30$< b <$+35 & 0.80 & 1.06$\pm$0.14 \\
+35$< b <$+40 & 0.76 & 1.02$\pm$0.15 \\
+40$< b <$+45 & 0.66 & 0.71$\pm$0.14 \\
+45$< b <$+50 & 0.52 & 0.56$\pm$0.16 \\
+50$< b <$+55 & 0.40 & 0.37$\pm$0.15 \\
+55$< b <$+60 & -0.42 & -0.25$\pm$0.09 \\
\hline
\end{tabular}
\end{center}
\end{table}

\begin{figure}
\hspace{2.1in}\includegraphics[width=4.2in]{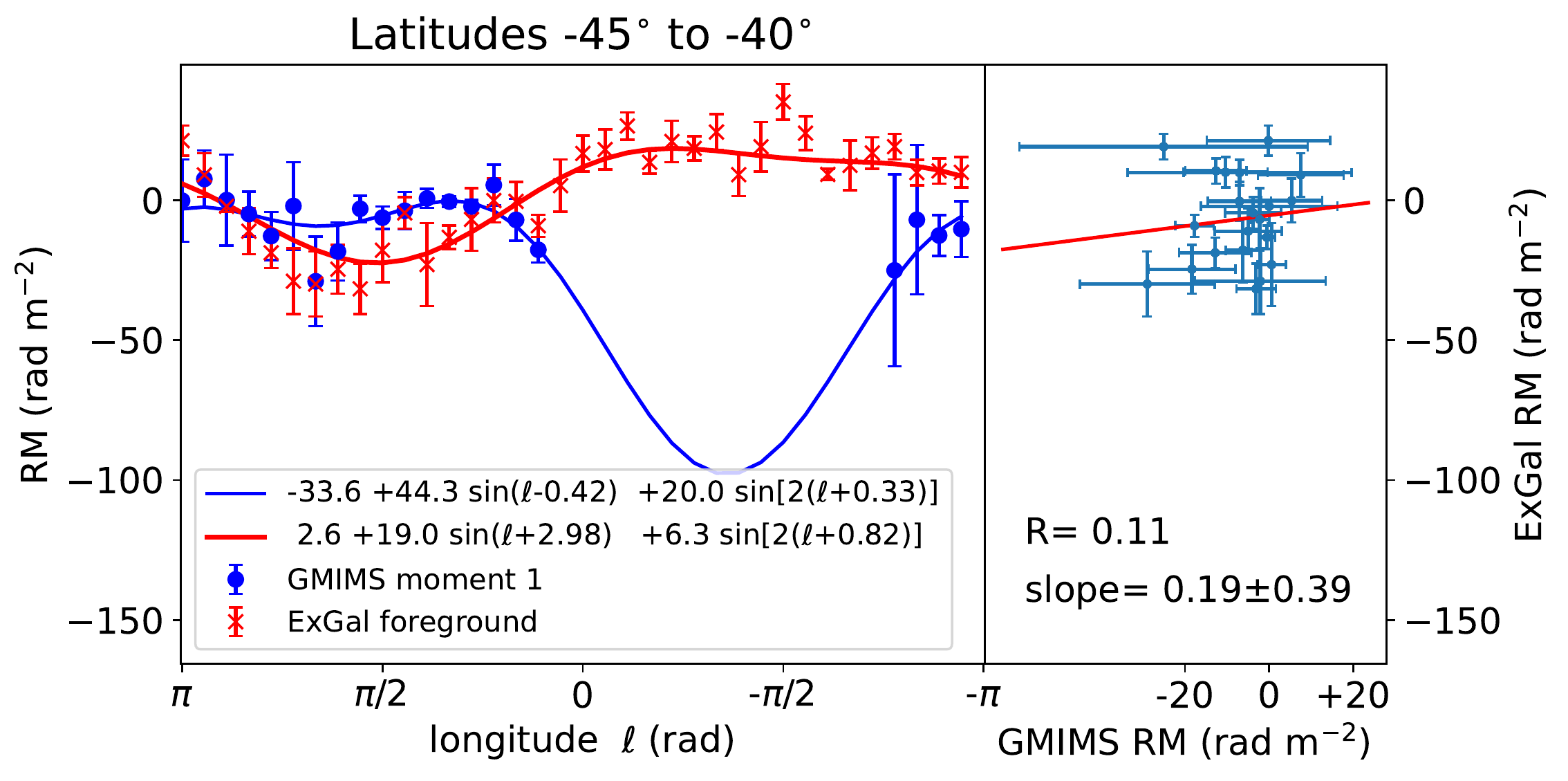} \hspace{.1in}
\caption{Comparison of GMIMS and extragalactic RMs at latitudes $-45^{\circ} < b < -40^{\circ}$.
In this case the two surveys give roughly similar results in the first and second Galactic
quadrants ($0 < \ell < \pi$), but the GMIMS survey misses most of the third and fourth 
quadrants.  The fitting is poorly constrained as a result, and the constants are ill determined.
In addition, there is very little correlation between the RMs from the two surveys (right panel).
\label{fig:ex-45} }
\end{figure}

The negative latitudes all have a longitude range that is not sampled by the GMIMS survey
(south of $\delta = -25^{\circ}$), so their scatter plots have fewer points, and the correlation tests
only a limited area.  These $R$ values are less secure.  Even so, a pattern of correlation
at mid-latitudes emerges in both hemispheres, with little or no correlation at low latitudes ($-5^{\circ} < b< +5^{\circ}$)
and at high latitudes ($|b| > 50^{\circ}$), illustrated in Fig. \ref{fig:R_vs_b}.
At mid-latitudes in the Southern Hemisphere most 5$^{\circ}$ strips show $R > 0.5$, with the exception of $-45^{\circ} < b < -40^{\circ}$
shown on Fig. \ref{fig:ex-45}.  So much of the longitude range is blanked in the GMIMS data that the fit
results are not significant, as shown by the large spurious excursion in the fit in the unobserved region.

\subsection{Effect of the North Polar Spur \label{sec:NPS}}

\begin{figure}
\hspace{1.2in}\includegraphics[width=5in]{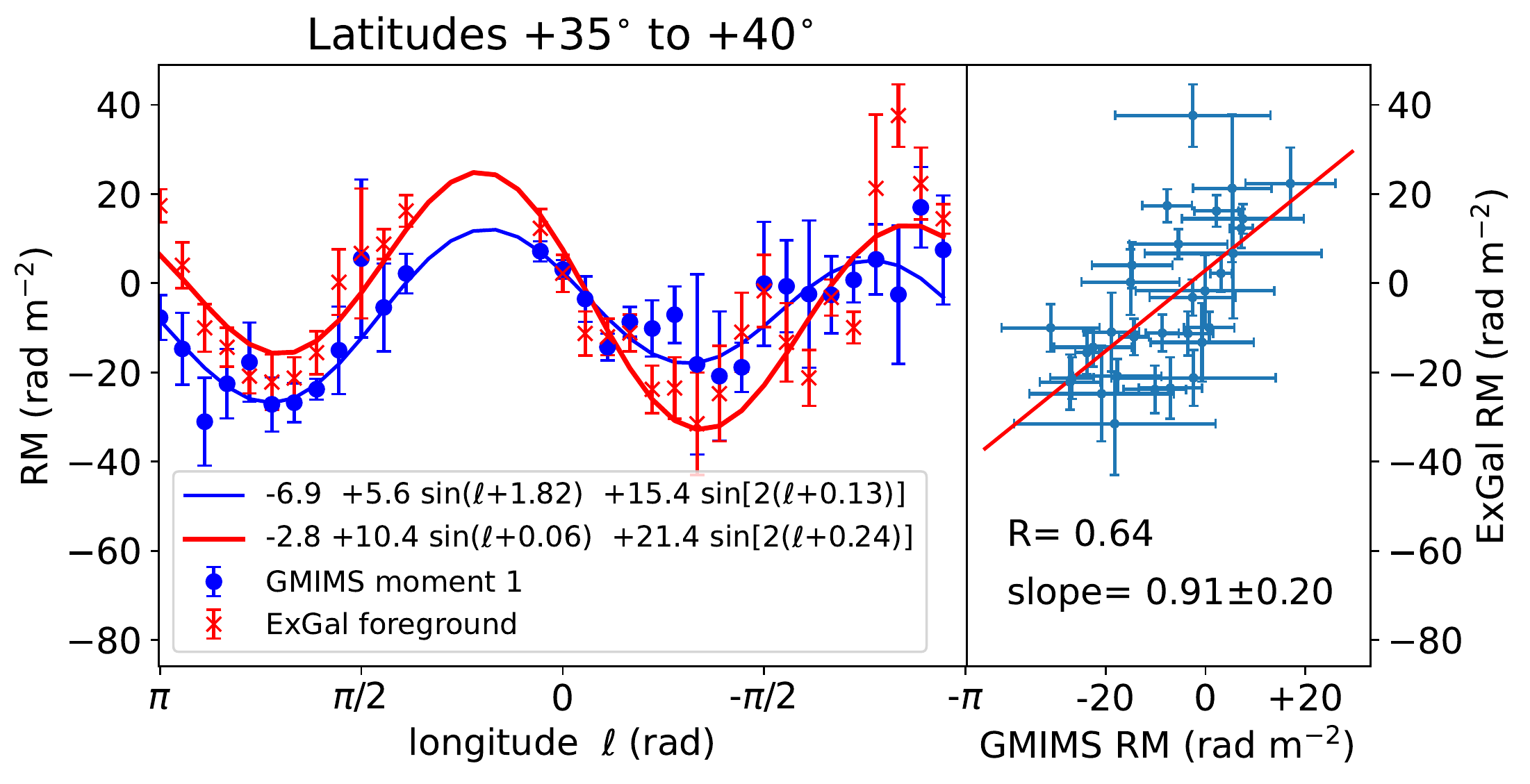} 

\hspace{1.2in}
\includegraphics[width=5in]{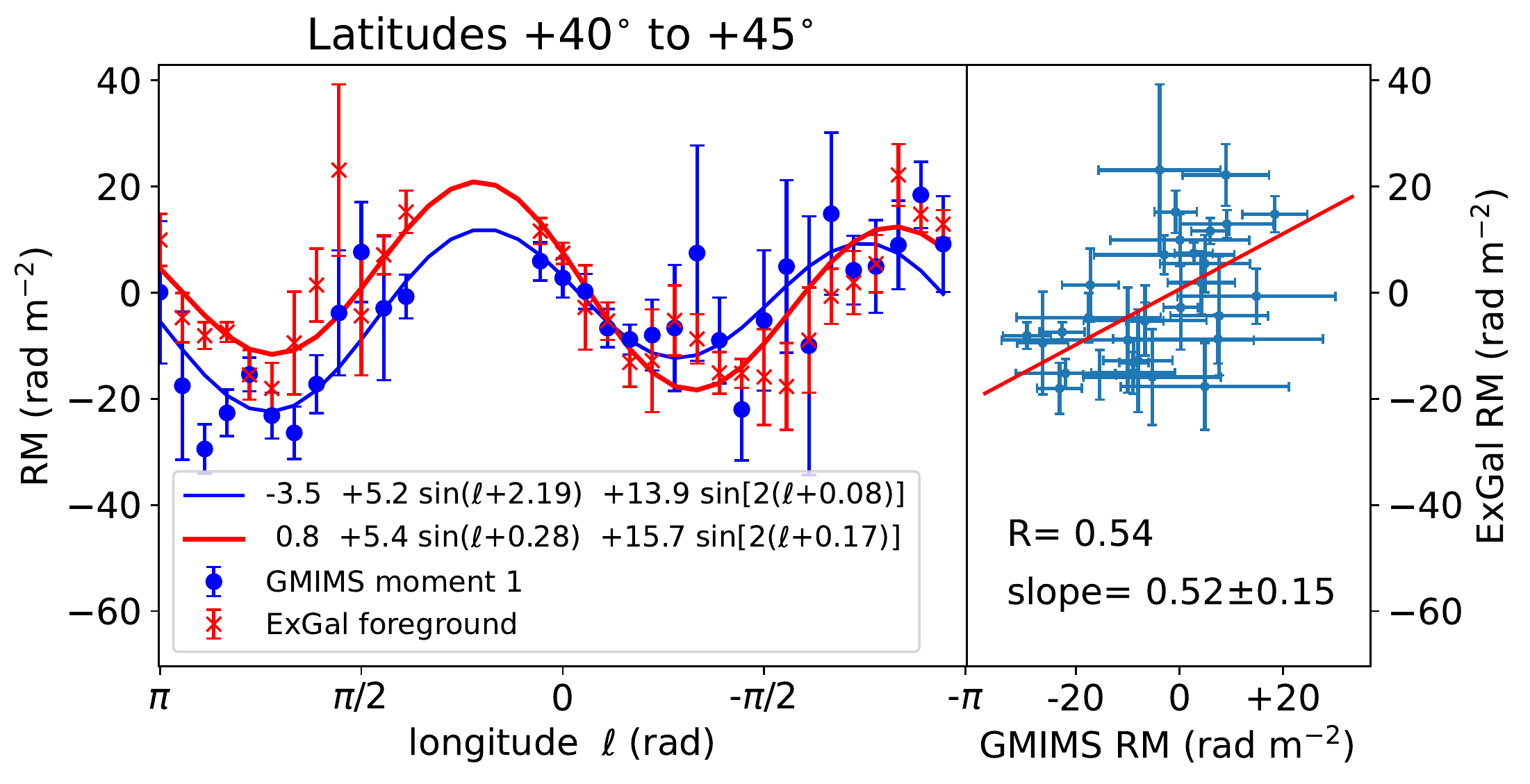}
\caption{The effect of blanking an area that covers the bright emission region corresponding to the
North Polar Spur ($25^{\circ} < b < 70^{\circ}$ and $20^{\circ} < \ell < 70^{\circ}$).  The latitude ranges
are the same as those shown on Fig. \ref{fig:example_b40}
\label{fig:b40_noNPS} }
\end{figure}

The large angular scale pattern of RMs at intermediate latitudes, that is apparent in the Galactic Northern Hemisphere
as the $\sin{(2 \ell)}$ pattern discussed here, has been ascribed to the effect of the North Polar Spur \citep[NPS][also called Loop I, in Sec. \ref{sec:loops} below]{Gardner_etal_1969,Lallement_2022}.  It may
be that the NPS is part of a larger structure that shapes the direction of the $\vec{B}$ field throughout
the hemisphere, a structure that could explain many large features in the synchrotron emission, optical and far-IR
polarization, and cosmic ray propagation \citep{West_etal_2021}.  In that case the $\sin{(2 \ell)}$ RM pattern 
may be a useful tracer of the direction of the LoS component of the field in this structure.  On
the other hand, if the effect of the NPS is restricted to the region of the first quadrant where the Stokes I emission shows
a large loop (illustrated in Sec. \ref{sec:loops}), then it is worth checking whether the values of RM in these longitudes alone can cause the $\sin{(2 \ell)}$
term to dominate over the $\sin{(\ell)}$ term, unlike in the Southern Hemisphere.  To check, we blank
the longitude range $20^{\circ} < \ell < 70^{\circ}$ for latitudes $+25^{\circ} < b < +70^{\circ}$, and repeat the
analysis above.  This area is shown by the red outlines on  Fig. \ref{fig:twomaps}.
Blanking the NPS area gives results like those shown on Fig. \ref{fig:b40_noNPS} and Table \ref{tab:fits_noNPS}.
The effect on the fitted amplitude and phase of the $\sin{(2 \ell)}$ term of blanking the NPS in the first quadrant is 
small.  All statistically significant values of
$C_2$ and $\phi_2$ (in bold face on Table \ref{tab:longitude_fits}) agree with their values for the unblanked maps within their errors,
e.g. for latitudes $+35^{\circ} < b < +40^{\circ}$, $C_2$ is decreased from 15.7$\pm$1.2 to 15.4$\pm$1.4 rad m$^{-2}$ for
the GMIMS profile, and similarly from 22.1$\pm$2.4 to 21.4$\pm$2.6 rad m$^{-2}$ for the extragalactic profile.  The correlation
between the two RM samples is reduced from $R=0.76$ to $R=0.64$.  Comparing Figs \ref{fig:example_b40} and \ref{fig:b40_noNPS}
shows that the highest peaks in both profiles are in the blanked area, but the $\sin{(2 \ell)}$ shapes are not
significantly diminished when those peaks are removed. We conclude that the North Polar Spur does not by itself generate the $\sin{(2 \ell)}$ pattern in the Northern Galactic Hemisphere.  In the following sections the
NPS area is not blanked, but similar results are found
if the blanking is applied.

\begin{sidewaystable}[p]
\caption{Fit Parameters with North Polar Spur Blanked (Eq. \ref{eq:5paramfit})\label{tab:fits_noNPS}}
\begin{tabular}{|c|ccccc|ccccc|}
\hline
Latitude   & \multicolumn{5}{c|}{GMIMS (DRAO)} & \multicolumn{5}{c|}{Extragalactic} \\
range & $C_o$& $C_1$ & $\phi_1$ & $C_2$ & $\phi_2$ &
 $C_o$& $C_1$ & $\phi_1$ & $C_2$ & $\phi_2$ \\
($^{\circ}$) &
rad m$^{-2}$ & rad m$^{-2}$ & radians & rad m$^{-2}$ & radians &
rad m$^{-2}$ & rad m$^{-2}$ & radians & rad m$^{-2}$ & radians \\ \hline
+25$< b <$+30 & -3.5$\pm$ 1.9 &{ 5.4$\pm$ 2.8} &{-0.56$\pm$0.48} &{\bf 15.1$\pm$ 2.7} &{0.16$\pm$0.08} &  3.0$\pm$ 2.0 &{15.8$\pm$ 3.2} &{0.39$\pm$0.15} &{\bf 28.7$\pm$ 2.5} &{\bf 0.23$\pm$0.04} \\ 
+30$< b <$+35 & -5.7$\pm$ 1.3 &{ 3.3$\pm$ 2.0} &{-0.42$\pm$0.58} &{\bf 15.8$\pm$ 1.5} &{\bf 0.24$\pm$0.07} &  1.2$\pm$ 1.7 &{ 9.0$\pm$ 2.7} &{-0.20$\pm$0.22} &{\bf 21.4$\pm$ 2.6} &{\bf 0.09$\pm$0.05} \\ 
+35$< b <$+40 & -6.9$\pm$ 1.0 &{ 5.6$\pm$ 1.3} &{1.82$\pm$0.27} &{\bf 15.4$\pm$ 1.4} &{\bf 0.13$\pm$0.05} & -2.8$\pm$ 1.5 &{10.4$\pm$ 2.5} &{0.06$\pm$0.20} &{\bf 21.4$\pm$ 2.6} &{\bf 0.24$\pm$0.05} \\ 
+40$< b <$+45 & -3.5$\pm$ 1.4 &{ 5.2$\pm$ 1.5} &{2.19$\pm$0.43} &{\bf 13.9$\pm$ 1.8} &{\bf 0.08$\pm$0.07} &  0.8$\pm$ 0.9 &{ 5.4$\pm$ 1.4} &{0.28$\pm$0.18} &{\bf 15.7$\pm$ 1.3} &{\bf 0.17$\pm$0.04} \\ 
+45$< b <$+50 & -0.1$\pm$ 1.3 &{ 6.0$\pm$ 2.0} &{2.74$\pm$0.26} &{\bf 10.1$\pm$ 1.7} &{0.24$\pm$0.09} &  4.7$\pm$ 1.0 &{ 4.4$\pm$ 1.5} &{0.16$\pm$0.26} &{\bf  9.0$\pm$ 1.4} &{\bf 0.26$\pm$0.06} \\ 
+50$< b <$+55 &  1.6$\pm$ 1.1 &{ 5.2$\pm$ 1.4} &{2.39$\pm$0.31} &{ 7.3$\pm$ 1.5} &{0.52$\pm$0.10} &  5.8$\pm$ 0.9 &{ 3.8$\pm$ 1.4} &{0.42$\pm$0.28} &{\bf  7.2$\pm$ 1.3} &{\bf 0.33$\pm$0.07} \\ 
+55$< b <$+60 &  1.9$\pm$ 1.4 &{ 4.8$\pm$ 1.9} &{2.56$\pm$0.36} &{ 5.1$\pm$ 1.8} &{1.53$\pm$0.16} &  6.6$\pm$ 0.8 &{\bf  5.6$\pm$ 0.9} &{-0.97$\pm$0.21} &{ 1.1$\pm$ 1.1} &{-0.26$\pm$0.46} \\ 
\hline
\end{tabular}
\end{sidewaystable}

\subsection{Slopes, Amplitude Ratios, and Phases \label{sec:correl}}

There are many reasons why surveys of RMs with different telescopes may give different, even
uncorrelated, results.  Differences in the {\it u,v} plane coverage for different instruments leads to different angular resolution and spatial filtering of the polarized brightness distribution on the sky.
In particular, single-dish surveys of diffuse polarization like GMIMS suffer from depolarization
due to several physical effects that do not apply to observations of compact, extragalactic
sources.  Two very significant processes are beam depolarization and depth depolarization
\citep{Burn_1966, Tribble_1991, Sokoloff_etal_1998, Dickey_etal_2019}.  The large beam of the DRAO telescope
blends together 
emission from a large enough area that polarized flux with many different position angles
averages so as to attenuate the measured polarized intensity.  This is particularly problematic
at low Galactic latitudes where the polarization angle varies rapidly with position on the sky.
The extragalactic sources used to construct the  RM grid are compact enough 
(a few arc seconds to tens of arc seconds) that
variations in the foreground Galactic RM are too small to cause much beam depolarization, except where H{\tt{II}} regions or other small scale RM structure causes polarization shadows \citep{Stil_Taylor_2007,Harvey-Smith_etal_2011,Thomson_etal_2019}.

Depth depolarization of the diffuse Galactic emission occurs when synchrotron emission and Faraday rotation coexist within the same volume. Emission arising at different depths along the LoS suffers different rotation, and vector averaging reduces the observed polarized intensity. In the simplest case, where magnetic field, synchrotron emissivity, and electron density are constant, the RM of the diffuse emission is exactly half that of an extragalactic source seen through the region (Burn 1966). If the ionized gas that causes the Faraday rotation is all in front of the diffuse polarized Galactic emission, there is no depth depolarization, and the extragalactic RM and the diffuse RM will be the same. If the synchrotron emission is in front of most of the rotating medium, there will be little or no correlation between the extragalactic and diffuse RMs.

\begin{figure}
\hspace{1.5in}\includegraphics[width=4in]{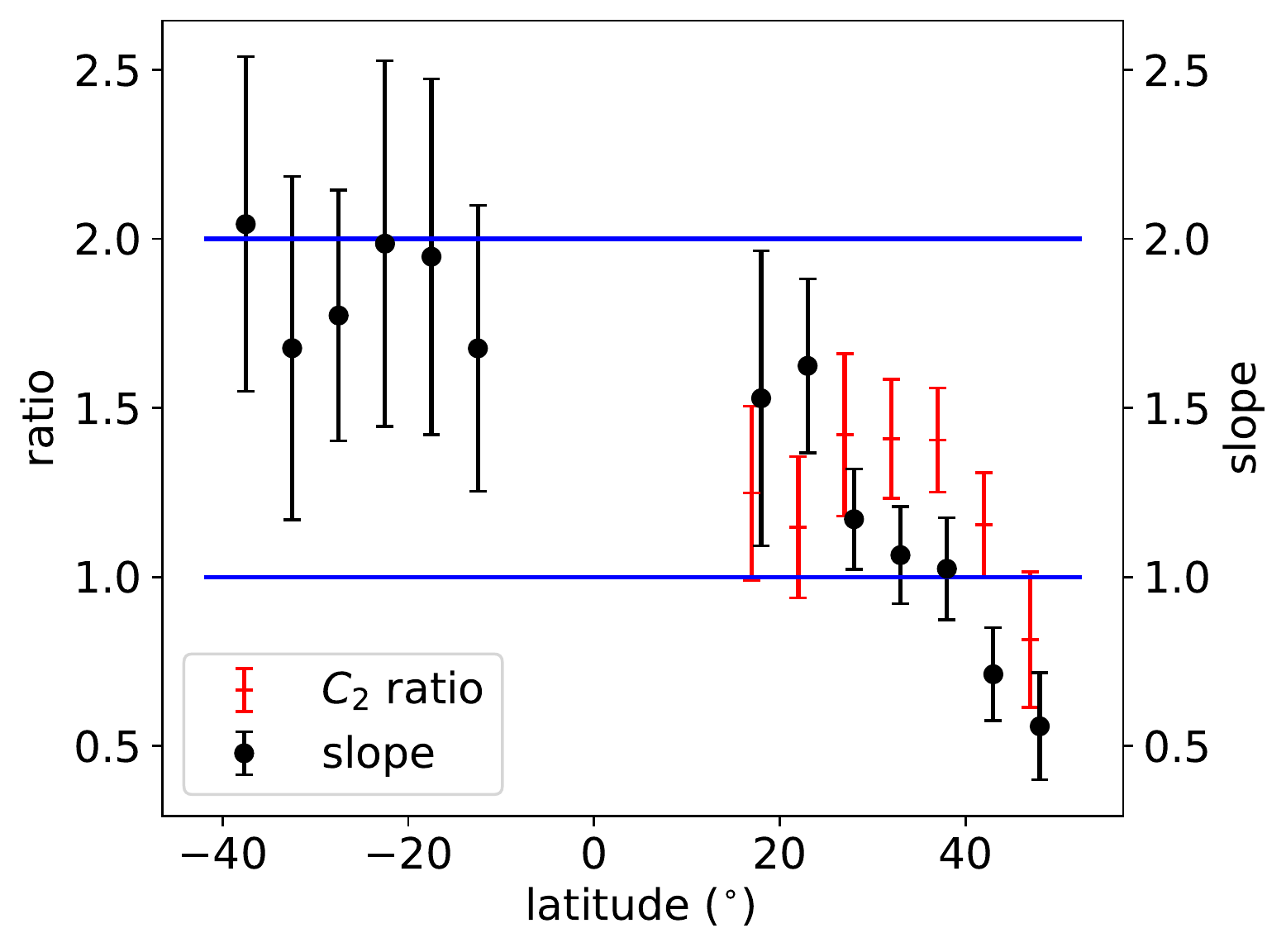} 
\caption{Correlation slopes and amplitude ratios for the $\sin{(2 \ell)}$ terms
at mid-latitudes.  Slopes are plotted for latitudes having $R>0.5$ only.
Amplitude ratios are plotted only for $+15 < b < +50$, for which all but one of the 
values of $C_2$ are greater than 5 $\sigma$ in both surveys.  (The exceptions are
latitudes 25$^{\circ}$ to 30$^{\circ}$ in the GMIMS survey that have $C_2=18.0\pm 4.0$ rad m$^{-2}$, i.e. 4.5 $\sigma$, and
latitudes 45$^{\circ}$ to 50$^{\circ}$ in the extragalactic survey that have $C_2=7.9\pm 1.6$ rad m$^{-2}$, i.e. 4.9 $\sigma$.)
The blue lines indicate the range of slopes expected for a uniform, Faraday thin slab.
Values above two can arise in various ways, e.g.
the diffuse emission is beginning to show depth depolarization.
Values close to one suggest that the diffuse emission is behind most of the 
magneto-ionic medium that causes the Faraday rotation in the extragalactic
sample.  A field reversal along the line of sight could explain values of the slope less than one.
\label{fig:slope_n_ratio}
}
\end{figure}

Fig. \ref{fig:slope_n_ratio} plots the slopes determined from the regression
analysis in Sec. \ref{sec:correlation}, for only those latitudes having correlation coefficient $R>0.5$, as on Fig. \ref{fig:R_vs_b}.
Also plotted in red is the ratio of the amplitudes of the $\sin{(2 \ell)}$ terms
of the extragalactic sample divided by the GMIMS amplitude, i.e. 
\begin{equation} \label{eq:C2ratio}
C_2 \textrm{\ ratio} \ = \ \frac{C_{2-ExGal}}{C_{2-GMIMS}}
\end{equation}
Red points are plotted only for latitudes having amplitudes for both extragalactic and GMIMS data greater than 5\,$\sigma$ (with one exception each, as noted in
the caption, see Table \ref{tab:longitude_fits}).  These criteria select only
$-40^{\circ}<b<-10^{\circ}$ and $+10^{\circ}<b<+50^{\circ}$ for the slopes, and
$+15^{\circ}<b<+45^{\circ}$ for the amplitude ratios.  
All the Southern Hemisphere
slopes are consistent with a value of two (the upper blue line on Fig. \ref{fig:slope_n_ratio}),
the maximum expected from a uniform slab of mixed emission and rotating medium.   
This suggests that the Southern mid-latitudes have polarized emission and 
Faraday rotation distributed mostly together along the LoS.  On the other
hand, the Northern Hemisphere points 
have lower values of the slopes, with values  dropping from 1.62 to 0.56 as the
latitude increases over the range $+15\arcdeg < b < +50\arcdeg$
 approaching and passing
the lower limit value of one for foreground rotation (the lower blue
line on Fig. \ref{fig:slope_n_ratio}). In this case the diffuse polarized
emission and the extragalactic sources are on average showing roughly the same
RMs, suggesting that the synchrotron emission
is further away than the medium that causes the Faraday rotation (Sec. \ref{sec:pulsar_comparison} below).
The amplitude ratios suggest an intermediate result for the component of the
RMs that is modulated by the $\sin{(2 \ell)}$ pattern. All of the red points 
are between one and two on Fig. \ref{fig:slope_n_ratio}, suggesting that the synchrotron emission and the 
Faraday rotation are coextensive over part of the LoS, but with 
some background emission that is beyond the rotating medium.

\begin{figure}
\hspace{1in}\includegraphics[width=5in]{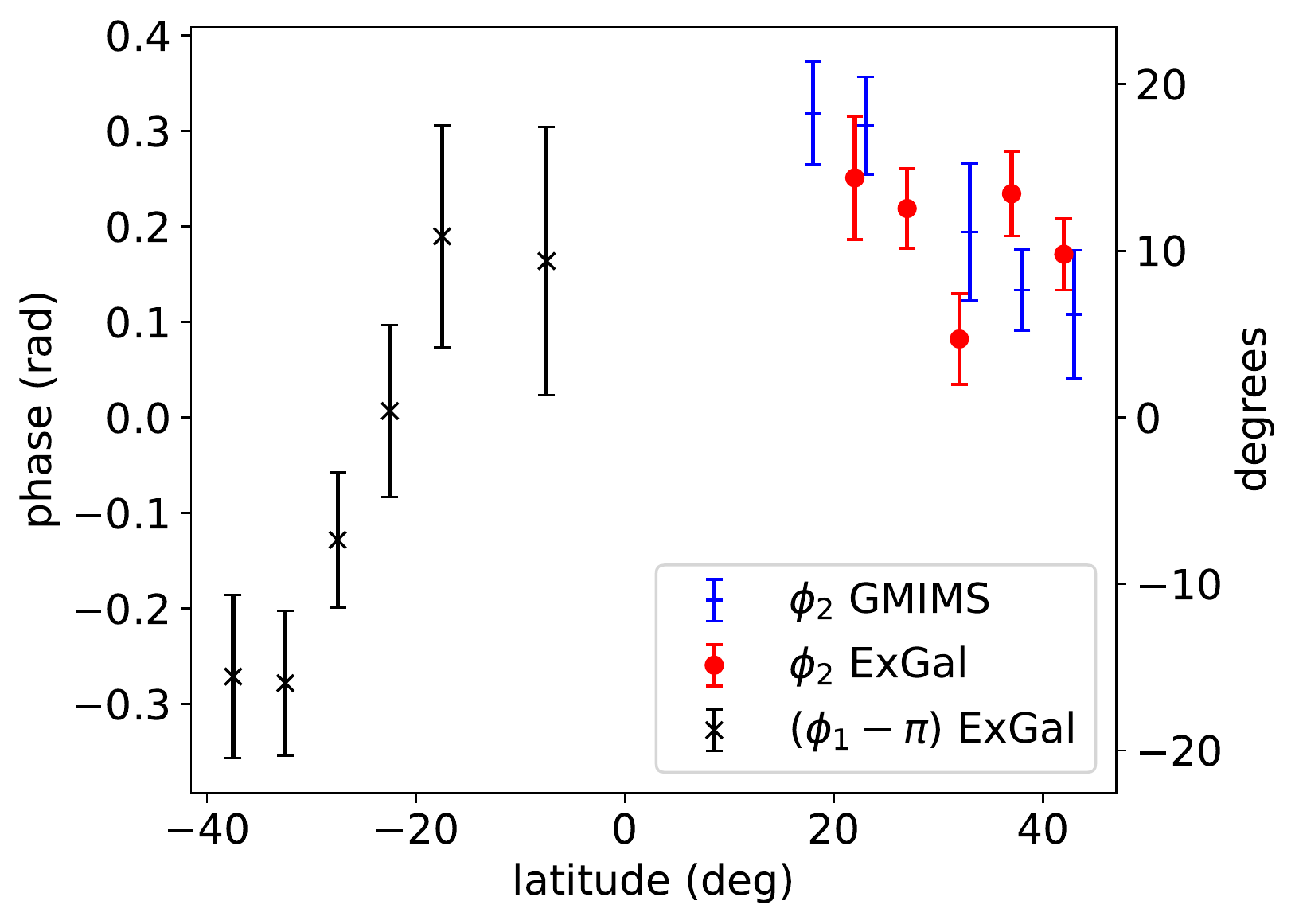}

\caption{The fitted phases of the $\sin{\ell}$ and $\sin{(2 \ell)}$
functions, i.e. $\phi_1$ and $\phi_2$ in Equation \ref{eq:5paramfit}.
In the Northern Hemisphere, $\phi_2$ from both the GMIMS and the extragalactic fits are shown, in blue and red respectively.  In the South, the extragalactic fits give $\phi_1 \simeq \pi$.  In the figure we subtract
$\pi$ to get the black points on the same scale as the $\phi_2$ values in the North.  Note that both the error bars and the range of values for $\phi_1$ naturally have about twice the range as for $\phi_2$ because of
the factor of two in the third term on the right of Eq. \ref{eq:5paramfit}. The conditions for including points on the plot are that $R>0.5$ and the error in $\phi$ is small, i.e. $\sigma_{\phi2} < 0.075$ radians or $\sigma_{\phi1} < 0.15$ radians.
\label{fig:phases}}
\end{figure}

The phases of the fitted functions, $\phi_1$ and $\phi_2$ in Eq. \ref{eq:5paramfit}, show good consistency in the intermediate latitude ranges where the fits show either a strong $\sin{(2 \ell)}$ term (the Northern Galactic
Hemisphere) or a strong $\sin{(\ell)}$ term (the Southern Galactic Hemisphere) as indicated by the shaded regions on Fig. \ref{fig:constants_vs_lat}.  But there is an offset of roughly $\pi$ between $\phi_1$ and $\phi_2$. Using the Pearson correlation coefficient $R > 0.5$ as a filter, and plotting only $\phi_2$ values with error less than 0.075 radians (= 4.3$^{\circ}$), gives the points on the right side of Fig. \ref{fig:phases}.
On the left are values of $\phi_1$ with corresponding errors less than 0.15 radians.  In the Southern Hemisphere, only the extragalactic survey has sufficient longitude coverage to give good fits in Eq. \ref{eq:5paramfit}.
The fact that the phases of the $\sin{(2 \ell)}$ terms in the North are close to zero,  $0\arcdeg < \phi_2 < 20\arcdeg$, suggests that the field sampled by these RM surveys is nearly aligned, either parallel or perpendicular, to the direction to the Galactic center.  If the large angular scale pattern in the Northern mid-latitude
RMs is due primarily to a few nearby, large structures, then this alignment would be fortuitous.  Thus Fig. \ref{fig:phases} strengthens the case for a global field configuration as the cause of the longitudinal modulation in the RMs, as discussed in Sec. \ref{sec:models} below.  The close alignment of the zero phase direction in both the Northern Hemisphere $\sin{(2 \ell)}$ and the Southern Hemisphere $\sin{(\ell+\pi)}$ functions with the Galactic Center direction ($\ell = 0^\circ$) suggests that these patterns are both aligned by a global field pattern, e.g. an azimuthal or spiral field.  The smooth decrease 
in $\phi_2$ with increasing latitude in the North is suggestive of a transition between disk-dominated and
halo-dominated fields, or perhaps the effect of flow in
or out of the disk \citep{Henriksen_Irwin_2021}.

\section{Contributions to the RM at Intermediate Latitudes from the Nearby Disk \label{sec:loops}}

The distinct, contrasting patterns in the RM at intermediate latitudes in the Northern and Southern Hemispheres, described above, trace the magnetic field and the diffuse ionized
medium along the entire line of sight through the Galaxy, for the extragalactic sources, or along the line of sight to and through the synchrotron emission, for the GMIMS survey.
Knowing the distances to the regions where most of the rotation takes place would help to interpret these patterns in terms of the magnetic field configuration.
In particular, the contributions to the RMs due to electrons and magnetic field in the disk versus the halo of the Galaxy need to be distinguished \citep{Mao_etal_2012}.  Distances to the sources of
polarized radiation are needed in order to model the
LoS distribution of the rotating medium, and to subtract
the contribution of the nearby disk from the RMs at mid-latitudes.

Pulsars are useful for tracing the three-dimensional distribution of RMs because their distances can be measured, either approximately by their dispersion measure (DM) or more precisely by parallax.  Large samples of pulsar RMs \citep{Han_etal_1999, Han_etal_2006, Han_etal_2018, Sobey_etal_2019} have been used to develop models of the field in the disk, \citep[e.g.][]{Han_Qiao_1994, Indrani_Deshpande_1999, Sun_etal_2008, Sun_Reich_2010,  VanEck_etal_2011, Jansson_Farrar_2012, Xu_Han_2019}, and to estimate the scale heights of both the magnetic field and the diffuse electron layers.  
In this section we consider the contribution to the RM by the medium in the nearby disk, based on several such empirical models of the electron density and
magnetic field configuration (section \ref{sec:pulsar_comparison}), then we model the RM due to
the nearby disk (section \ref{sec:disk_contrib}), and
finally in section \ref{sec:nearby} we match the largest RM 
features with an inventory of nearby radio continuum structures that contribute to 
both the synchrotron emission and the RM in both hemispheres.

\subsection{RMs of Pulsars with Parallax Distances \label{sec:pulsar_comparison}}

Many pulsars have approximate distances based on their dispersion measures and models of the electron density in the disk \citep[][and references therein]{Yao_etal_2017}.
Much more accurate distances come from parallax, so we start with these.
Using the ATNF Pulsar Catalog\footnote{\url{https://www.atnf.csiro.au/research/pulsar/psrcat/psrcat_help.html}} (v. 1.67, \citealp{Manchester_etal_2005}), we first consider pulsars in the range $10^{\circ}<|b|<50^{\circ}$ with 
accurate parallax distances, $\frac{\sigma_D}{D}<1$, where 
$\sigma_D$ is the error in the distance, $D$. This gives a sample of 57 pulsars.  We then separate these by height above the plane, $z = D \times \sin{b}$, and compute the correlation with the extragalactic RMs in the same directions, i.e. the healpix cell containing the pulsar position.

\begin{table} \caption{RM Correlations: Pulsars vs. Extragalactic Sources \label{tab:pulsar_correlations}}
\begin{center}
\begin{tabular}{|lccc|}
\hline
\hline
Sample & number & Pearson R & slope \\
\hline
\multicolumn{4}{|c|}{57 Pulsars with Parallax Distances, $10\arcdeg<|b|<50\arcdeg$} \\
\hline
$0<|z|<0.3$ kpc & 12 & 0.57 & 0.72\\
$0.3<|z|<0.6$ kpc & 18 & 0.95 & 0.91\\
$0.6<|z|<1$ kpc & 15 & 0.93 & 1.12\\
$1<|z|<6$ kpc & 12 & 0.97 & 0.95\\
\hline
\multicolumn{4}{|c|}{296 Pulsars with DM Distances, $10\arcdeg<|b|<50\arcdeg$} \\
\hline
$0<|z|<0.3$ kpc & 65 & 0.80 & 0.72 \\
$0.3<|z|<0.6$ kpc & 80 & 0.81 & 0.91\\
$0.6<|z|<1$ kpc & 62 & 0.84 & 0.92 \\
$1<|z|<5$ kpc & 89 & 0.88 & 0.81\\
\hline
\end{tabular}\end{center}
\end{table}

\begin{figure}

\hspace{-.1in} \includegraphics[width=3.5in]{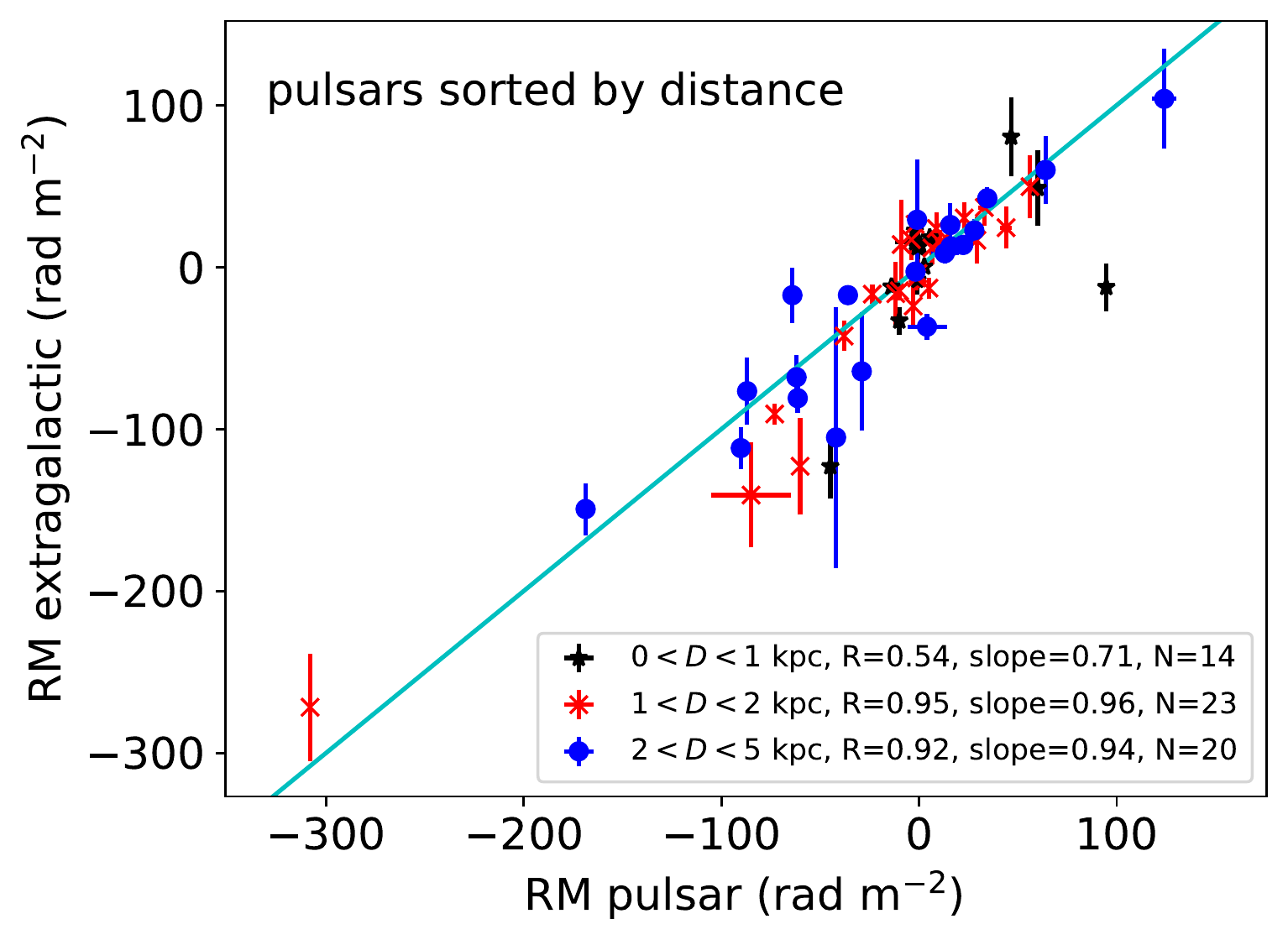} \includegraphics[width=3.5in]{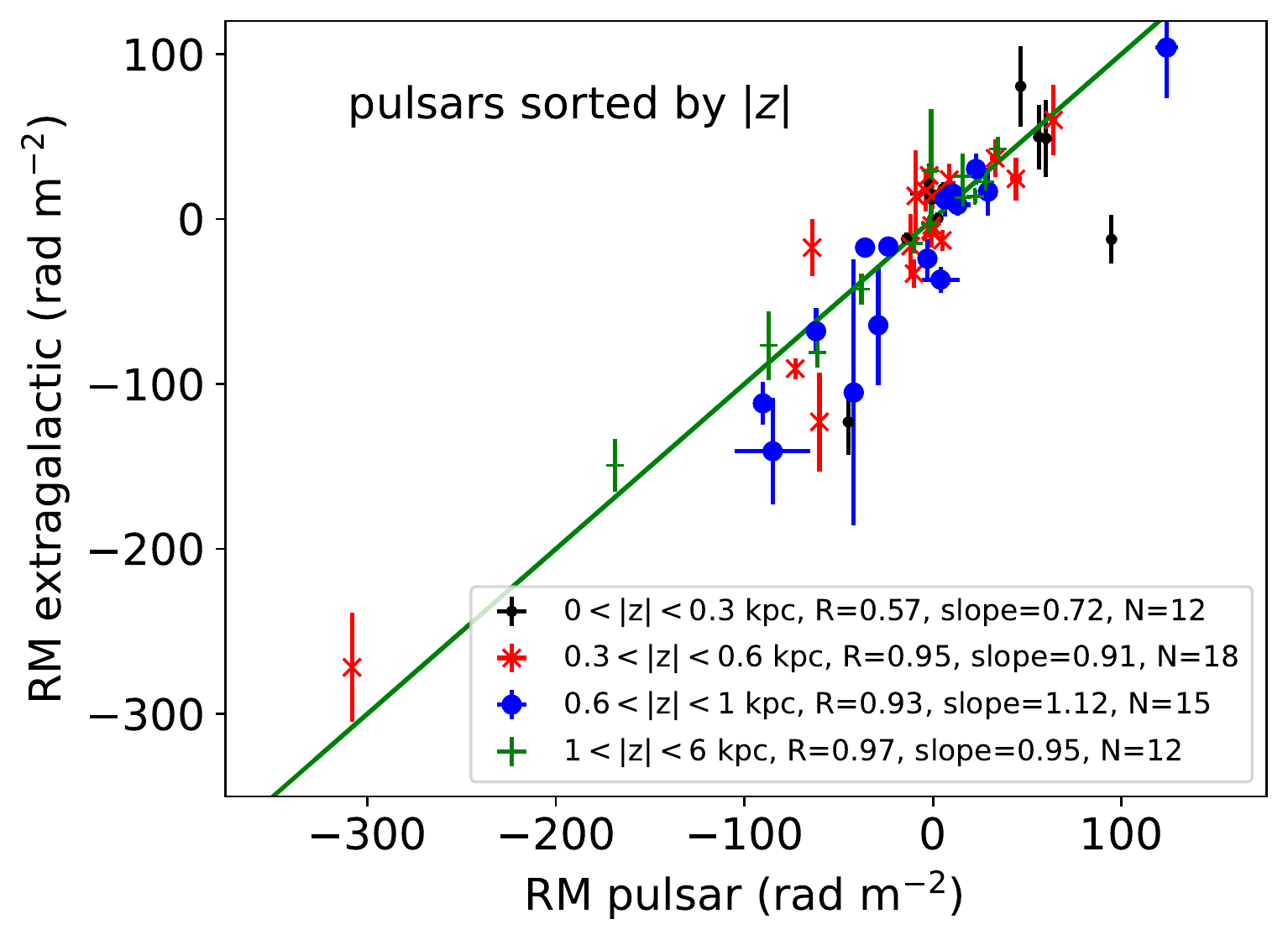}

\caption{
Scatter plots between rotation measures (RM) of Galactic pulsars with $10\arcdeg < |b| < 50\arcdeg$ taken from the ATNF catalog and RMs of extragalactic sources from \citet{Hutschenreuter_etal_2021}. The scatter plots are organized by distance ($D$, left panel) and height above and below the Galactic plane ($|z|$, right panel). 
On the right panel there is a tight correlation for pulsars with $|z|>0.6$ kpc, as indicated by the Pearson coefficients, R, shown along with the number of pulsars in each sample, N, and
the slope of the best fit line, which approaches one as $z$ increases.  These numbers are summarized on Table \ref{tab:pulsar_correlations}. The pulsar RMs are very precisely measured, in most cases the error bars are smaller than the symbols.  The errors in the extragalactic RMs are the standard deviations in the \citet{Hutschenreuter_etal_2021} map at the positions of the pulsars.   \label{fig:pulsar_scatter1}}
\end{figure}

Considering sub-samples at different distances, $D$, and height above or below the plane, $|z|$, shown on Fig. \ref{fig:pulsar_scatter1} and Table \ref{tab:pulsar_correlations}, the correlation between the extragalactic RMs and the pulsar RMs gets stronger rapidly with $|z|$ above about 0.3 kpc.  For the 12 pulsars in the sample with $|z|>1$ kpc the Pearson correlation coefficient is a remarkable 0.97.  Using a much larger sample of 296 pulsars with distances estimated from their dispersion measures and the electron density model of \citet{Yao_etal_2017} gives weaker correlation
coefficients, but still suggests that most of the RM toward the extragalactic sources is generated below $|z|<\ \sim$1 kpc (Table \ref{tab:pulsar_correlations}).  For this larger sample the correlation coefficients vary from  0.81 to  0.88 between $0.6 < |z| < 1$ kpc.  The correlation is still strong, but degraded somewhat for the second sample, perhaps because of the less precise DM distances compared with the parallax distances used for the first sample. The increasing correlation between pulsar and extragalactic RMs for pulsars with $|z|$ increasing from about 0.3 to 1 kpc agrees with the finding of \citet{Mao_etal_2012} for longitude $\ell \sim 110\arcdeg$ that the symmetric disk $\vec{B}$ dominates RMs for $|z|<0.5$\,kpc.

\begin{figure}
    \centering
        \includegraphics[width=4.5in]{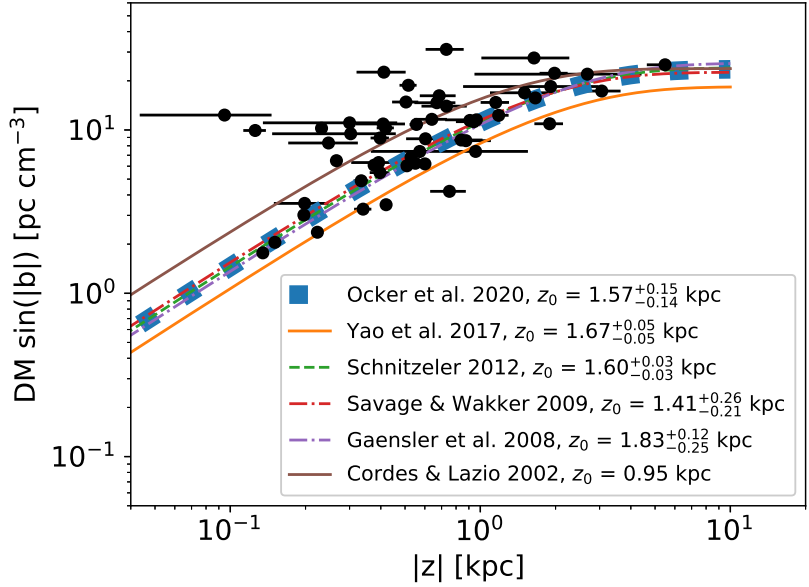}
\caption{
Dispersion measure (DM) as a function of $|z|$ for our selected 54 pulsars. Estimated trends from the literature are listed in the legend with corresponding values of the scale height of the ionized gas in the Milky Way $h_e$.  Although the points show large scatter, the data
are consistent with estimates for $h_e$ in the range 1.0 to 1.8 kpc. \label{fig:Andrea_B}}
\end{figure}

\begin{figure}
    \centering
        \includegraphics[width=4.5in]{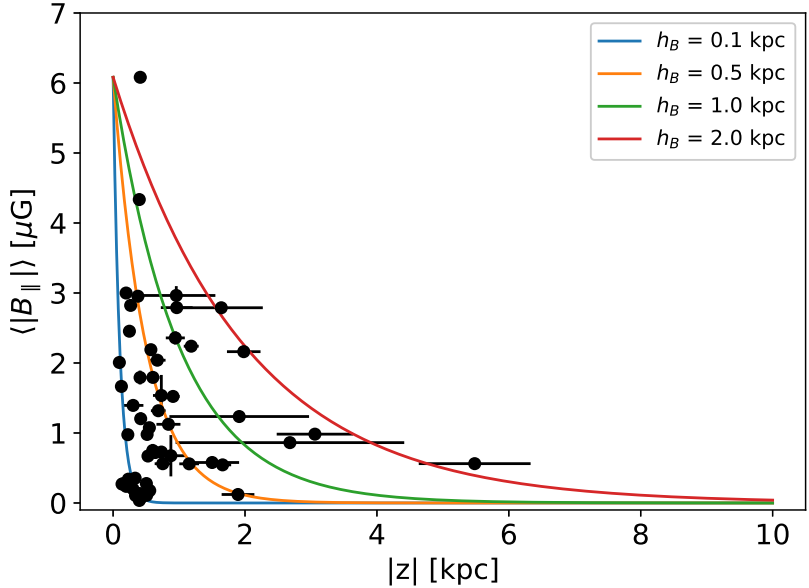}
\caption{
Correlation between the average LoS magnetic field, $\langle B_\parallel \rangle$, and $|z|$ for our selected 54 pulsars. Four analytical trends of the form $\langle B_\parallel \rangle = \langle B{_\parallel, 0} \rangle e^{(-|z|/h_B)}$ are overlaid as functions of the magnetic field scale height, $h_B$.  As on Fig. \ref{fig:Andrea_B}, the points are scattered over a wide range, but they are mostly consistent with $h_B<\sim0.5$ kpc. \label{fig:Andrea_C}}
\end{figure}

Fig. \ref{fig:Andrea_B} shows the trend of dispersion measure, DM, versus $z$ for the pulsars used in Fig. \ref{fig:pulsar_scatter1}, along with the expected DM given by 
various estimates for $h_e$, the scale height of the ionized gas layer \citep[][Table 2]{Ocker_etal_2020}, assuming that the electron density, $n_e$, depends on $z$ as
\begin{equation}\label{eq:scale_height}
n_e(z) \ = \ n_{e,0} \ e^{-|z|/h_e} 
\end{equation}
with $n_{e,0}$ the average mid-plane electron density. Recent values of $h_e$ are $\approx 1.5$ kpc.  This is roughly a factor of three greater than the corresponding scale height of the $\vec{B}$ field causing the pulsar RM.
This is supported by Fig. \ref{fig:Andrea_C}, that shows the average line-of-sight magnetic field strength, $\langle B_\parallel \rangle = 1.232 \frac{RM}{DM}$, as a function of $|z|$ for this sample of pulsars.  The curves on Fig. \ref{fig:Andrea_C} show the predictions assuming an exponential $z$ dependence of $B$ with different values of $h_B$, the magnetic field scale height \citep[e.g.][]{Sobey_etal_2019} and assuming a midplane value $B(0)= 6\  \mu$G for the ordered component of the field. 
Most of the pulsars in our sample are between the curves for $h_B=0.1$ kpc and $h_B=0.5$ kpc, which is indicative of a large scale, ordered magnetic field mostly confined to the Galactic thick disk, with a considerably smaller scale height than the thermal electrons, $h_e$.
\textcolor{red}{}This result is also consistent with theoretical expectations from numerical simulations by \citet{Pakmor_etal_2018}, where it is shown that, because the magnetic field strength decreases exponentially with height above the disk, the Faraday rotation for an observer at the solar circle is dominated by the local environment (distance $D\sim$ a few kpc in the simulations).  

The scale height of the $\vec{B}$ field derived above applies only to the field as measured with Faraday rotation, i.e. the field in regions where the thermal electron density is high enough to cause significant RMs.  The synchrotron emission may extend beyond the thermal electrons, since the
cosmic rays and magnetic fields are not so strongly confined to the disk in regions where the mass density of the interstellar gas is low, e.g. in bubbles or chimneys of hot gas  \citep{ McClure-Griffiths_etal_2000}.

\begin{figure}
    \centering
        \includegraphics[width=5.in]{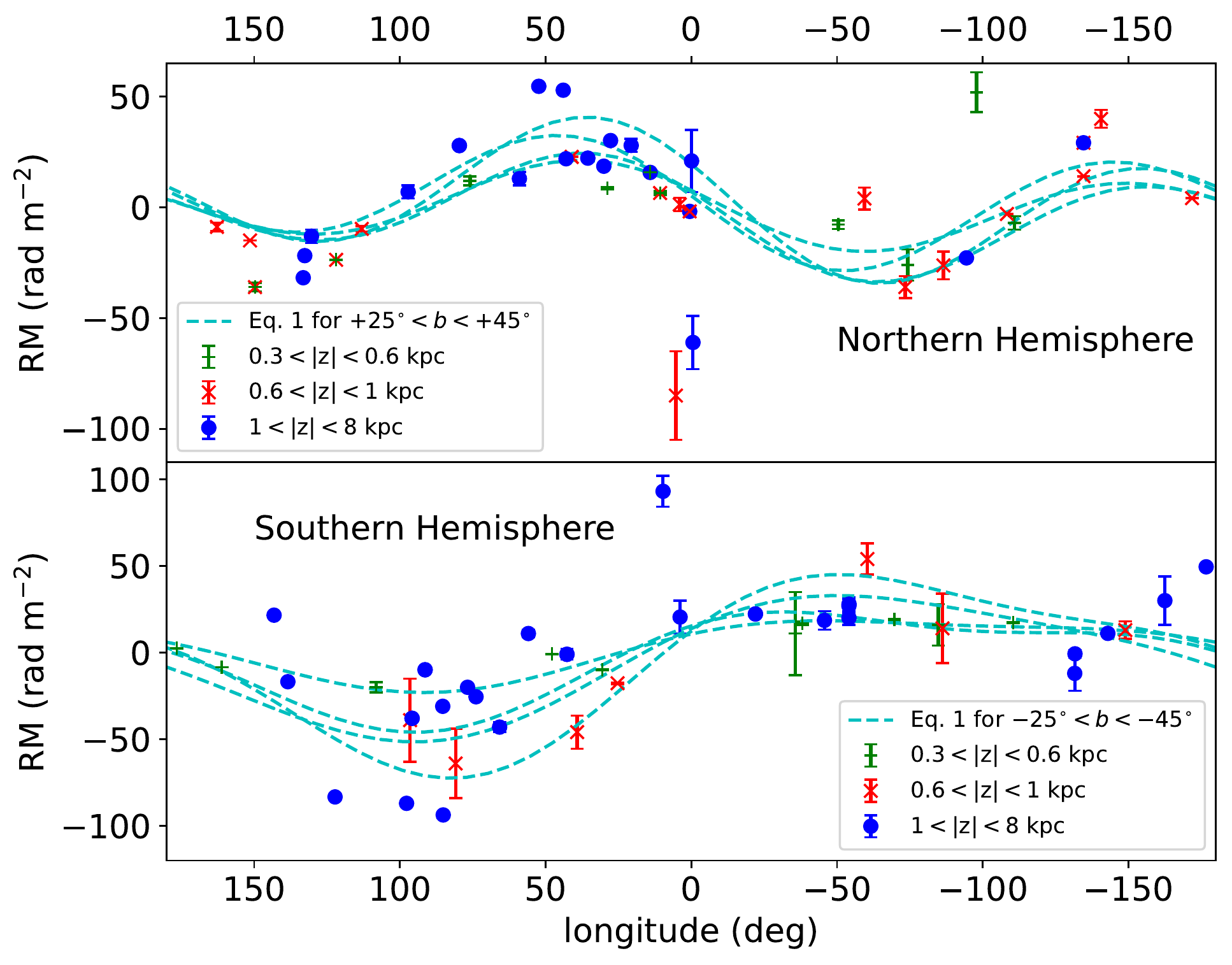}
\caption{Rotation measure versus longitude for pulsars with latitudes $+25\arcdeg \leq b \leq +45\arcdeg$ and $z > 0.3$ kpc (upper panel) or 
$-25\arcdeg \geq b \geq -45\arcdeg$ and $z < -0.3$ kpc (lower panel).  For comparison, the Eq. \ref{eq:5paramfit} predictions using the extragalactic fit parameters (Table \ref{tab:longitude_fits}) are indicated by the dashed curves. Many of the pulsars have such precisely measured RMs that the error bars are smaller than the symbols.  In both hemispheres, the agreement between the extragalactic RMs and the pulsar RMs is very good, particularly for pulsars with $|z|>0.6$ kpc.
\label{fig:pulsar_RM_vs_long}}
\end{figure}

The RMs of pulsars with $|z| > 0.6$ kpc correlate well with extragalactic RMs.  Considering the longitude dependence of the pulsar RMs, and restricting the sample to pulsars with $25\arcdeg < |b| < 45\arcdeg$ and $|z|> 0.3$\,kpc, gives the points shown on Fig. \ref{fig:pulsar_RM_vs_long}.  For comparison, results of fits of Equation \ref{eq:5paramfit} to the extragalactic RMs  (Table \ref{tab:longitude_fits}) in these latitude ranges are shown as dashed curves.  The extragalactic fits show good agreement with the pulsar points in both hemispheres.  

The conclusion from this comparison with RMs of pulsars at intermediate latitudes is that they are quite consistent with the extragalactic RMs if the pulsar is more than
$\sim$0.6 kpc above the plane. At mid-latitudes (30\arcdeg\xspace to 45\arcdeg) this gives distance $D > \sim1$ kpc.  There are many large structures more nearby that cast shadows on the RM sky, and we discuss them below in Sec. \ref{sec:nearby}.  

\subsection{Comparison with Empirical Models of the Disk Field} \label{sec:disk_contrib}

\begin{figure}
\hspace{.03in}\includegraphics[width=3.5in]{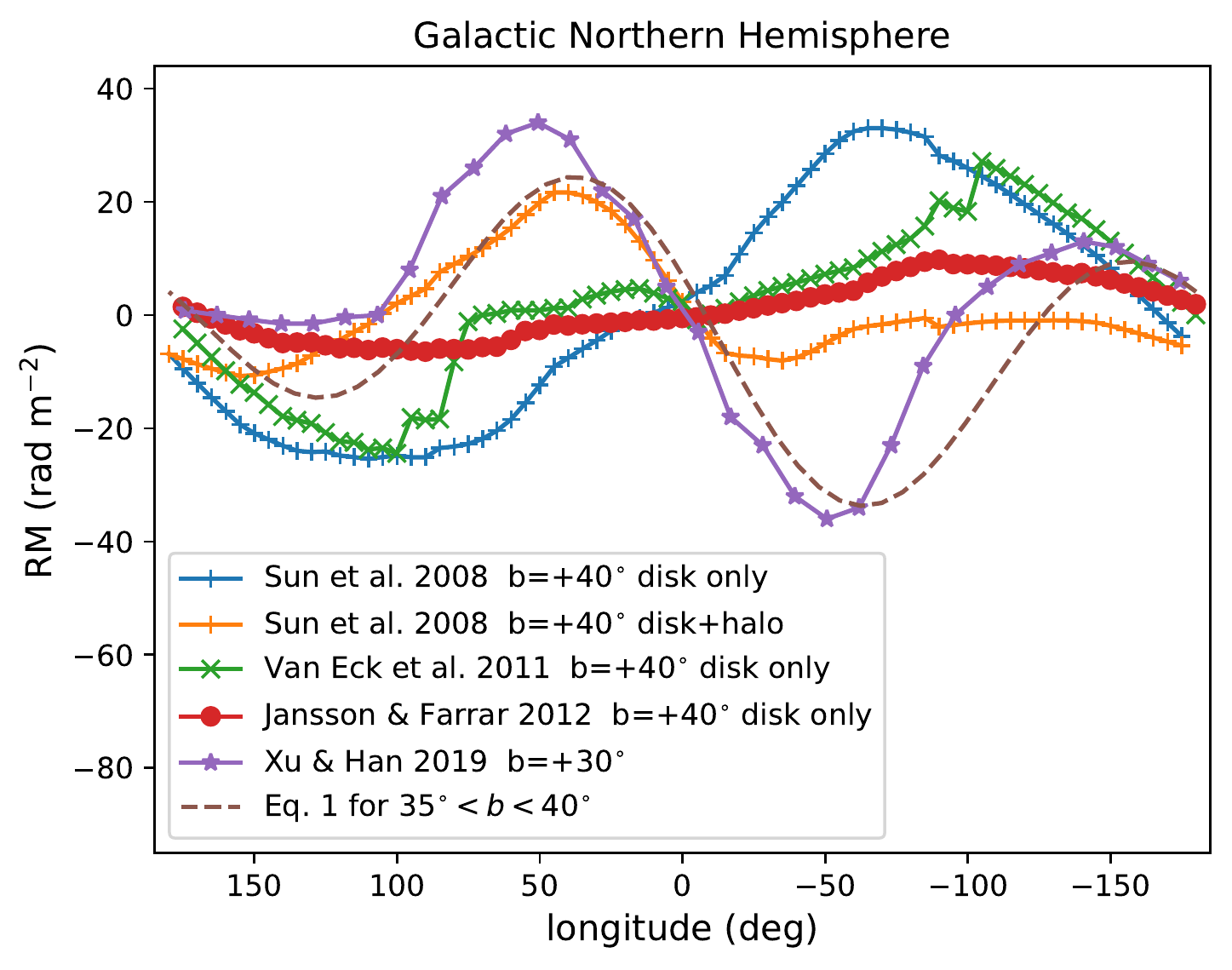}
\hspace{.1in}\includegraphics[width=3.5in]{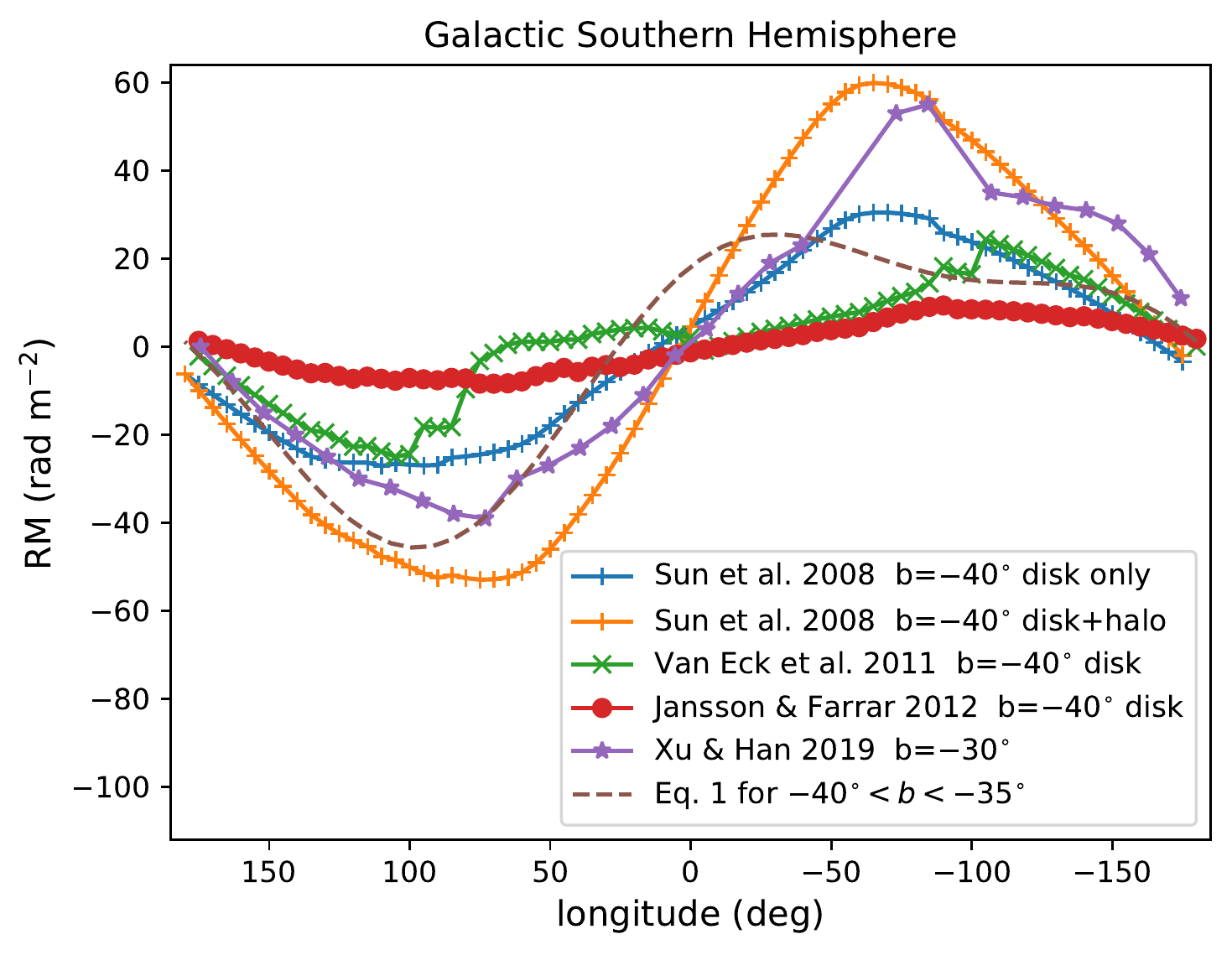}

\caption{Models for the contribution of the disk to the RMs observed
at intermediate latitudes.  The left panel shows North latitudes ($b=+40^{\circ}$) and the right panel shows the corresponding south latitudes ($b=-40^{\circ}$). These are computed following the method described
in \citet[][sec. 2.2]{Ma_etal_2020}.  The blue and orange curves result from the model of \citet{Sun_etal_2008} for the disk component and disk plus halo, respectively.  The green curve is the disk component of the \citet{VanEck_etal_2011} model, and the red is the disk component of the model by \citet{Jansson_Farrar_2012}.  The recent model of the halo contribution to RMs of pulsars
\citet{Xu_Han_2019} is shown in purple for $b=\pm30^{\circ}$ (copied from their Fig. 17).  For comparison the Equation \ref{eq:5paramfit} fits to the extragalactic RMs for similar latitude ranges are shown by the dashed curves. The \citet{Xu_Han_2019} halo model matches
the models for the GMIMS and extragalactic RMs quite well in both hemispheres, as does the \citet{Sun_etal_2008} disk plus halo model, but the disk only models do not fit the RM pattern in the Northern
Hemisphere. \label{fig:disk_contrib} }
\end{figure}

The correlation with pulsar RMs discussed above suggests that a significant contribution to the extragalactic and GMIMS RMs may be coming from the disk field, in addition to the field in the lower halo (roughly $|z|>0.6$ kpc).
Using models for the disk field 
we can predict the strength of the RMs expected at mid-latitudes 
from the line of sight path length through the disk.
Figure \ref{fig:disk_contrib} shows three models for the disk contribution, corresponding to $\vec{B}_{disk}$ field models by \citet{Sun_etal_2008, VanEck_etal_2011}, and \citet{Jansson_Farrar_2012}, combined with models for the thermal electron density in the disk, following the method for projection described in \citet{Ma_etal_2020}.  Since the disk field is primarily
azimuthal, these necessarily give roughly $\sin{(\ell + \pi)}$
dependence on longitude, but they  
include a field reversal inside the solar circle, which adds a weak $\sin{(2 \ell)}$ component, along with higher terms in a Fourier expansion.  Overall, these disk field models are in fair agreement with the
extragalactic RMs in the southern hemisphere (right hand panel, Fig. \ref{fig:disk_contrib}), but they are completely inconsistent with the RMs at positive latitudes (left hand panel).  Comparison of these disk field models with the extragalactic and diffuse RM data at mid-latitudes suggests that the disk field cannot explain the $\sin{(2 \ell)}$ behavior of the RMs at positive latitudes, but it may be sufficient to explain the $\sin{(\ell + \pi)}$ functions seen at negative latitudes.

An empirical approach to modelling the mid-latitude RM pattern, also based on pulsar data, is that of \citet{Xu_Han_2019}.  Combining pulsar RMs and corresponding dispersion measures, they find approximate $\sin{(2\ell)}$ and $\sin{(\ell + \pi)}$ functions for the intermediate latitude behavior of RM on longitude, reproduced on Fig. \ref{fig:disk_contrib} as the red curves (see their Fig. 17).  To explain the asymmetry between the two hemispheres, they invoke antisymmetric toroidal fields in the halo \citep{Han_etal_1997, Han_etal_1999}, plus a disk field with spiral shape and two field reversals inside the solar circle \citep{Han_etal_2006, Han_etal_2018},
the nearest is at a distance of just 0.14 kpc.  On Fig. \ref{fig:disk_contrib}, the \citet{Xu_Han_2019} model, which includes disk and halo fields, shows the same $\sin{(\ell+\pi)}$ dependence as the disk models in the southern hemisphere, but it gives roughly a $\sin{(2\ell)}$ behavior in the North, that is much more consistent with the extragalactic and GMIMS RM data.

\subsection{Nearby RM Structures \label{sec:nearby}}

The differences between the RM patterns in the two hemispheres have been ascribed to nearby features such as the NPS \citep{Gardner_etal_1969}. For the NPS, we show in Section 2.2 that this feature alone does not generate the observed pattern of RMs in the Northern hemisphere. Here we evaluate the contribution of other discrete nearby structures whose RM variations match the morphology of Stokes I synchrotron emission or H-alpha emission. To illustrate this comparison, in this section we decompose the RM surveys in spherical harmonics and display the results in orthographic projection using HEALPix tools \citep{Gorski:2005ku}.

\begin{figure}

\centering
    \includegraphics[width=5.5in]{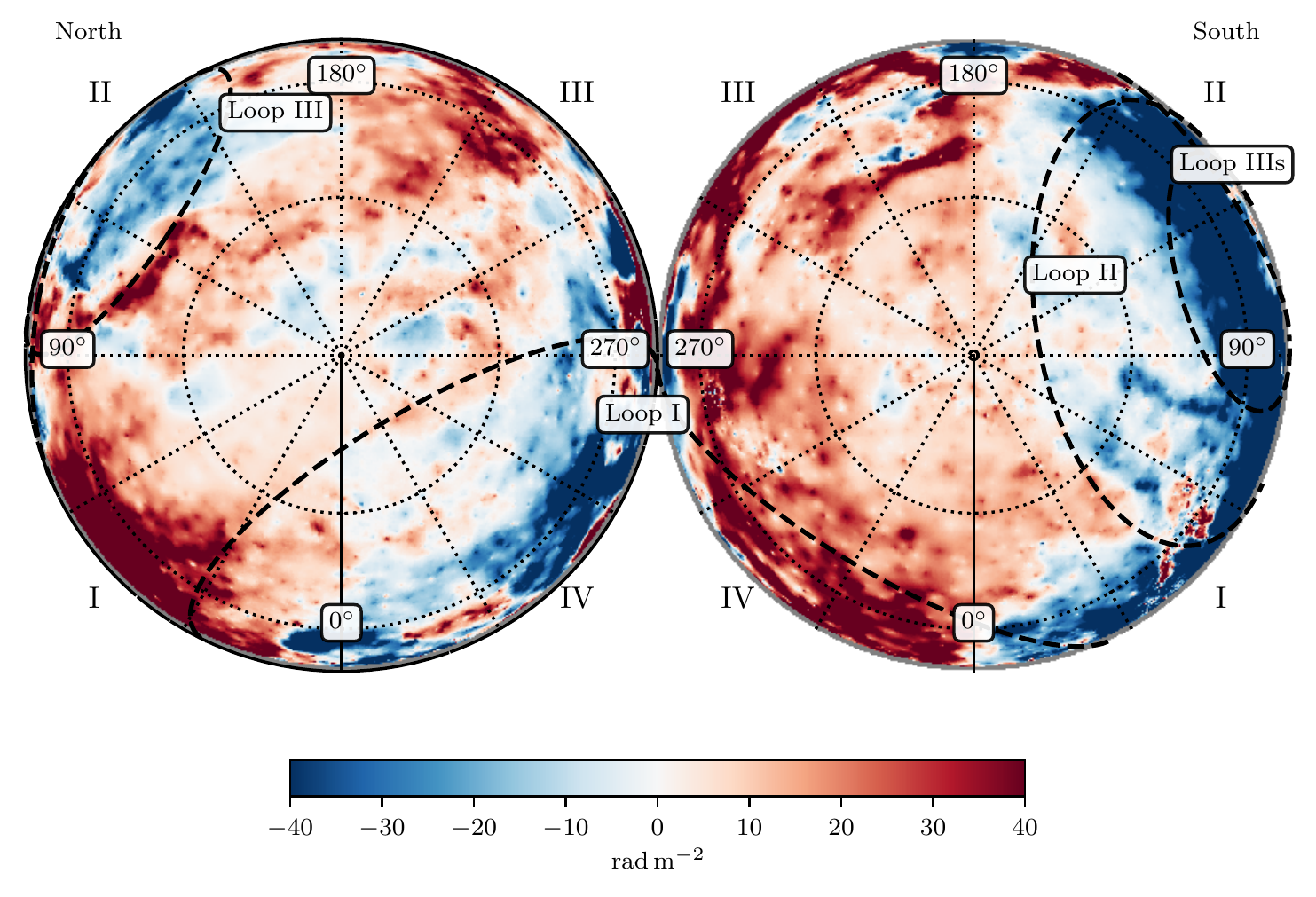}
    \includegraphics[width=5.5in]{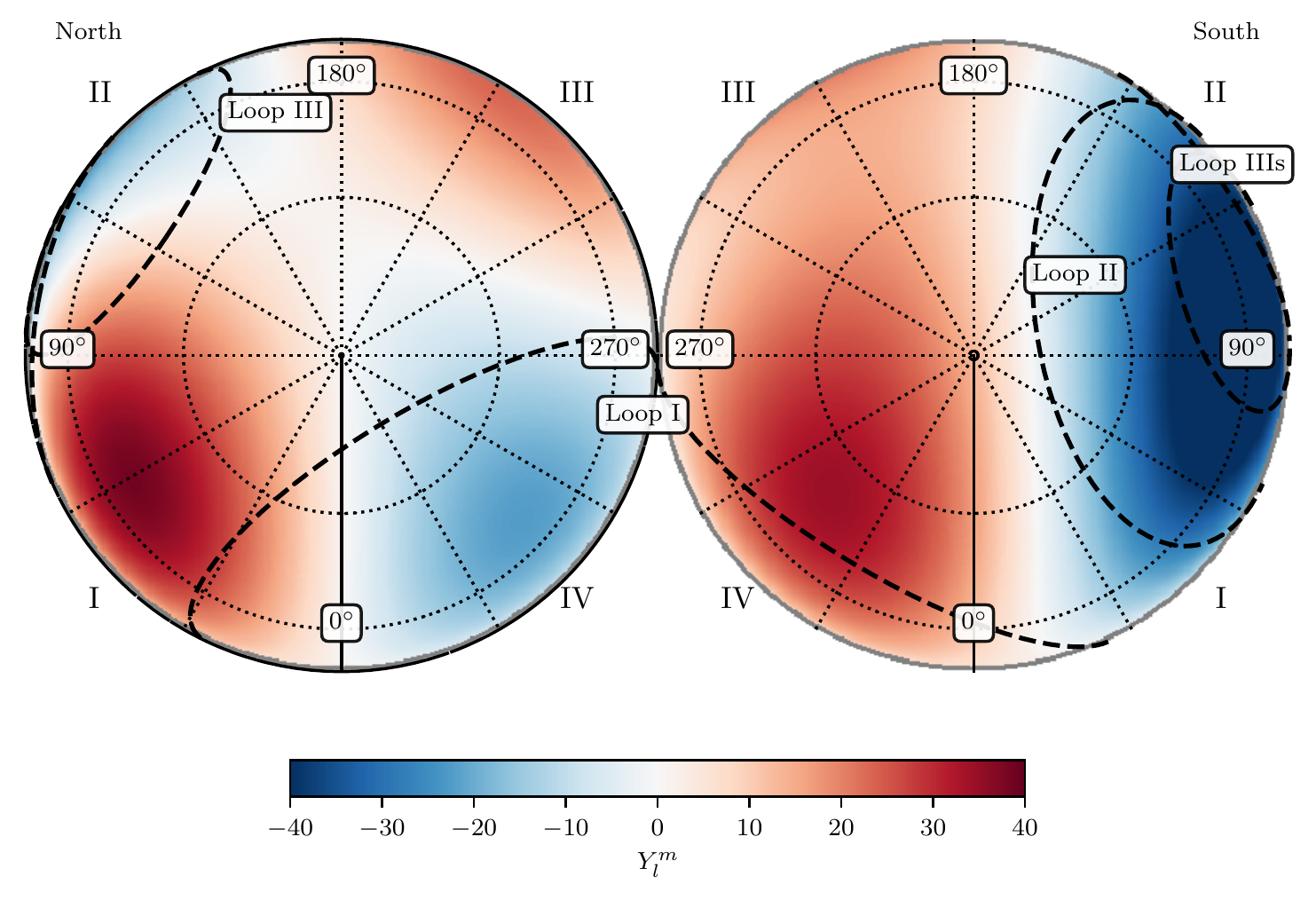}
        \caption{Upper: The extragalactic RM map of \citet{Hutschenreuter_etal_2021} in orthographic projection, centred on the Galactic poles.  The left panel shows the Northern Hemisphere, the right panel shows the south, with circles of constant latitude at $b=\pm 30\arcdeg$ and $\pm 60\arcdeg$. Lower: The spherical harmonic expansion of the \citet{Hutschenreuter_etal_2021} RM distribution with $l_\text{max}=3$. We overlay the positions of nearby radio continuum loops \citep{Vidal_2015} and label the Galactic quadrants by Roman numerals on both panels.  As these are images of the sky, parity is reversed compared with the ordinary face-on view of the Galactic plane seen from above the Northern Hemisphere.}
    \label{fig:exgal_ylm}
\end{figure}

\begin{figure}
    \centering
    \includegraphics[width=5.5in]{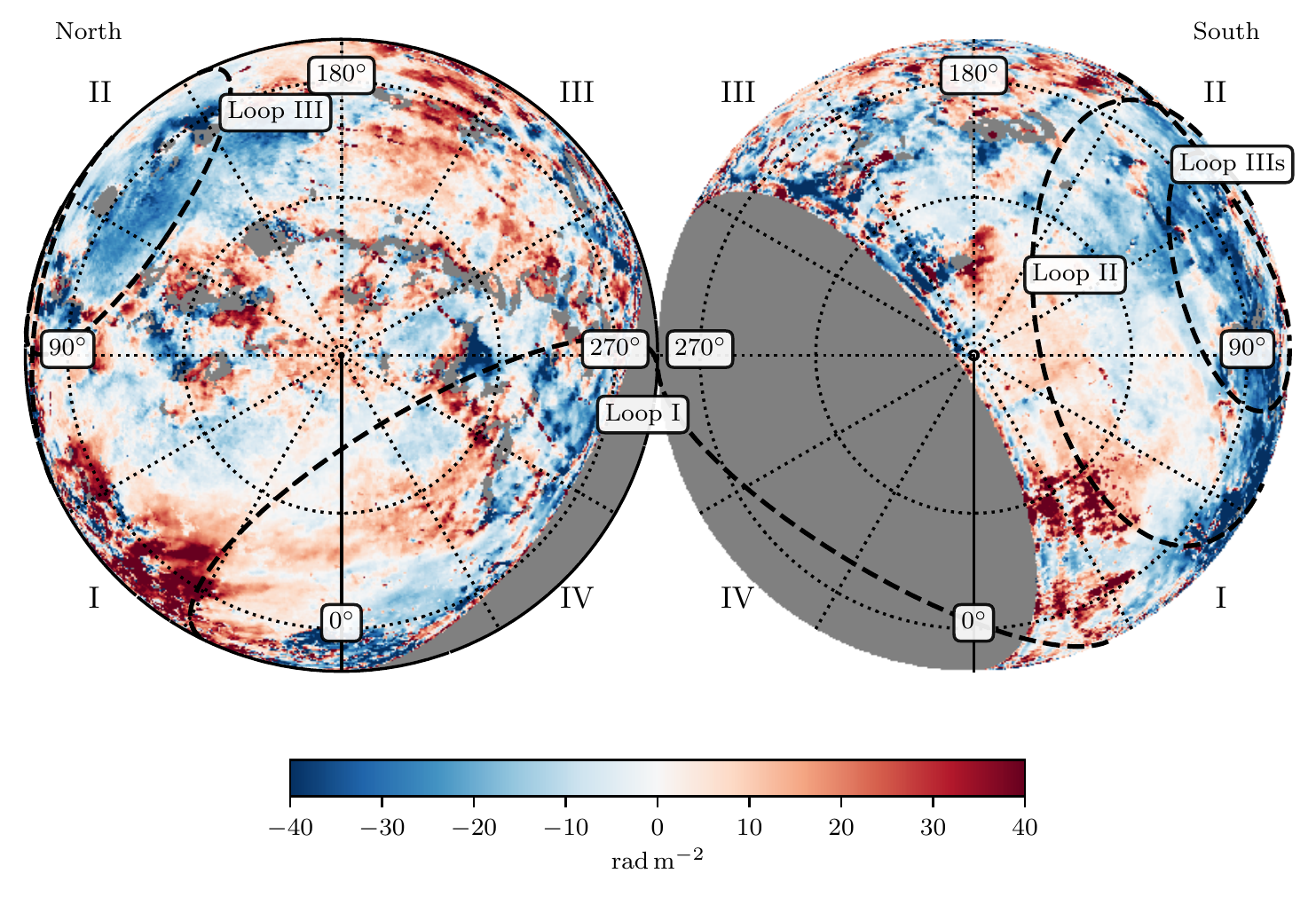}
        \includegraphics[width=5.5in]{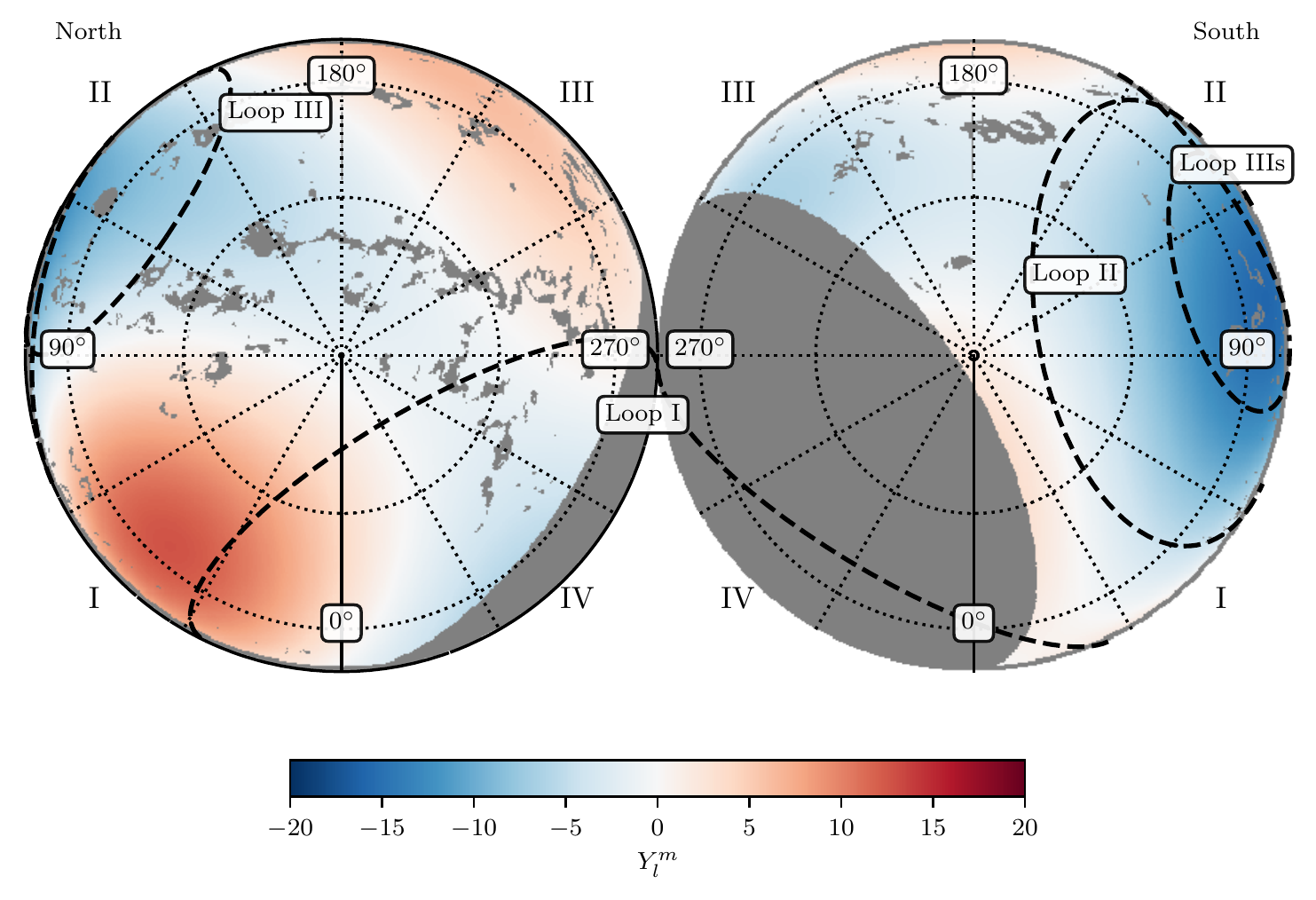}
        \caption{ The upper panels show the \citet{Wolleben_etal_2021} map of the GMIMS RMs in orthographic projection. The lower panels show the spherical harmonic expansion, as on Fig. \ref{fig:exgal_ylm}. The radio continuum loops discussed in the text are indicated as on Fig. \ref{fig:exgal_ylm}.  The grey patches are areas where the signal-to-noise is too low to allow calculation of the first moment of the Faraday spectrum. }
        \label{fig:gmims_ylm}
\end{figure}

The analysis of the functional dependence of RM on Galactic longitude, $\ell$, in Sec. \ref{sec:RMdata} by decomposing as the first few terms of a Fourier Series,
Eq. \ref{eq:5paramfit}, generalizes mathematically to a decomposition in spherical
harmonics, $Y_l^m$ \citep[e.g.][]{Dennis_Land_2008,Drake_2020}, the well-known family of orthogonal functions on a sphere.
The spherical harmonic degree, $l$, roughly corresponds to angular scale, $\theta = 180^{\circ}/l$. We compute the spherical harmonics up to degree $l=3$, which can capture dipole ($m=1$), quadrupole ($m=2$), and even octopole ($m=3$) modes (i.e. $e^{3i\ell}$) on the sphere. For both the \citet{Hutschenreuter_etal_2021} map and the GMIMS map, we mask the Galactic plane for $|b| < 10\arcdeg$ before computing the spherical harmonics. We show the sum of the first three spherical harmonics for both  maps in orthographic projection, centered on the Galactic poles, in the lower panels of Fig.s~\ref{fig:exgal_ylm} and \ref{fig:gmims_ylm}.  The $\sin{(2\ell)}$ (North) and $\sin{(\ell+\pi)}$ (south) functions stand out in this representation.

\begin{figure}
    \centering
        \includegraphics[width=5.5in]{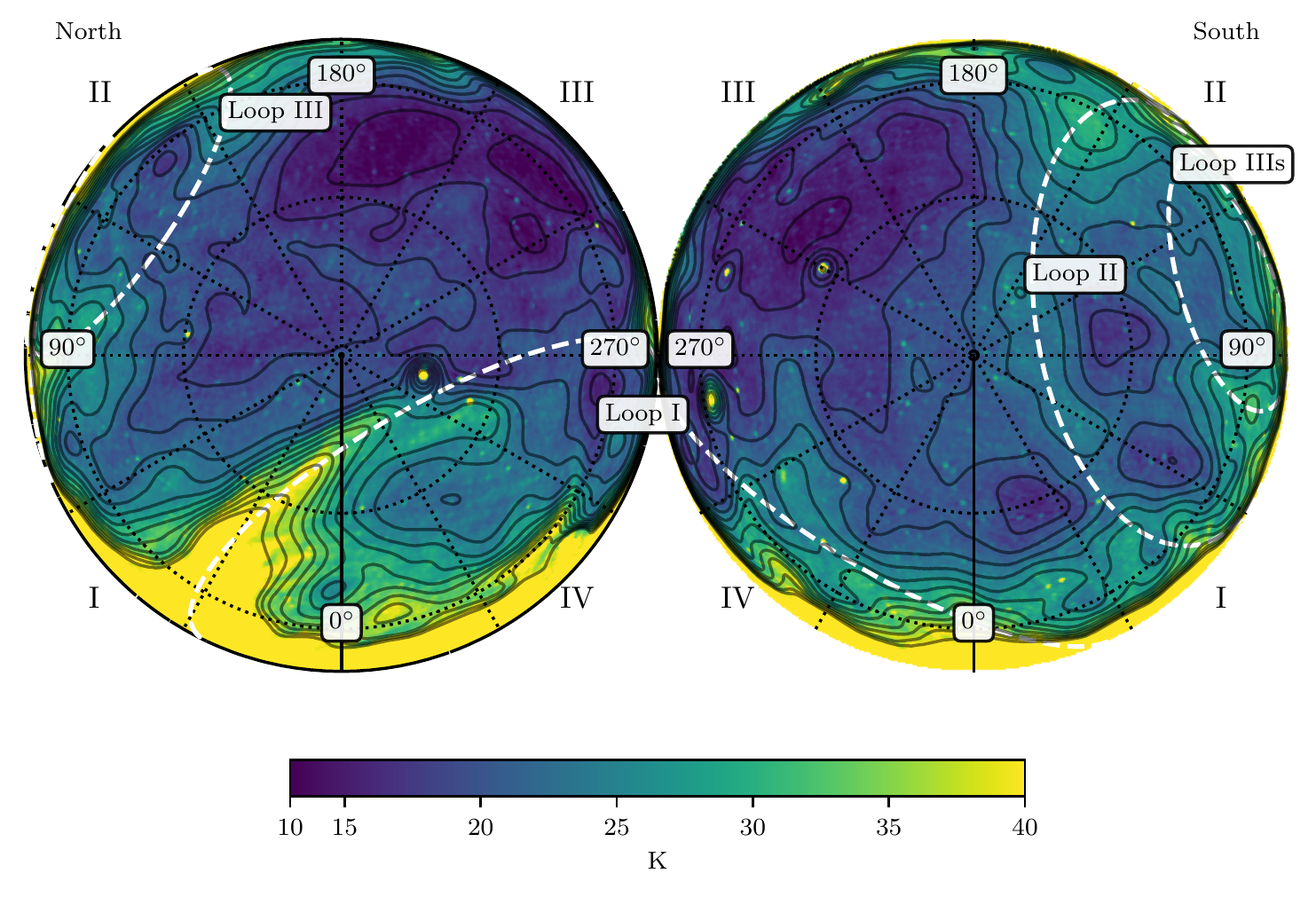}
        \includegraphics[width=5.5in]{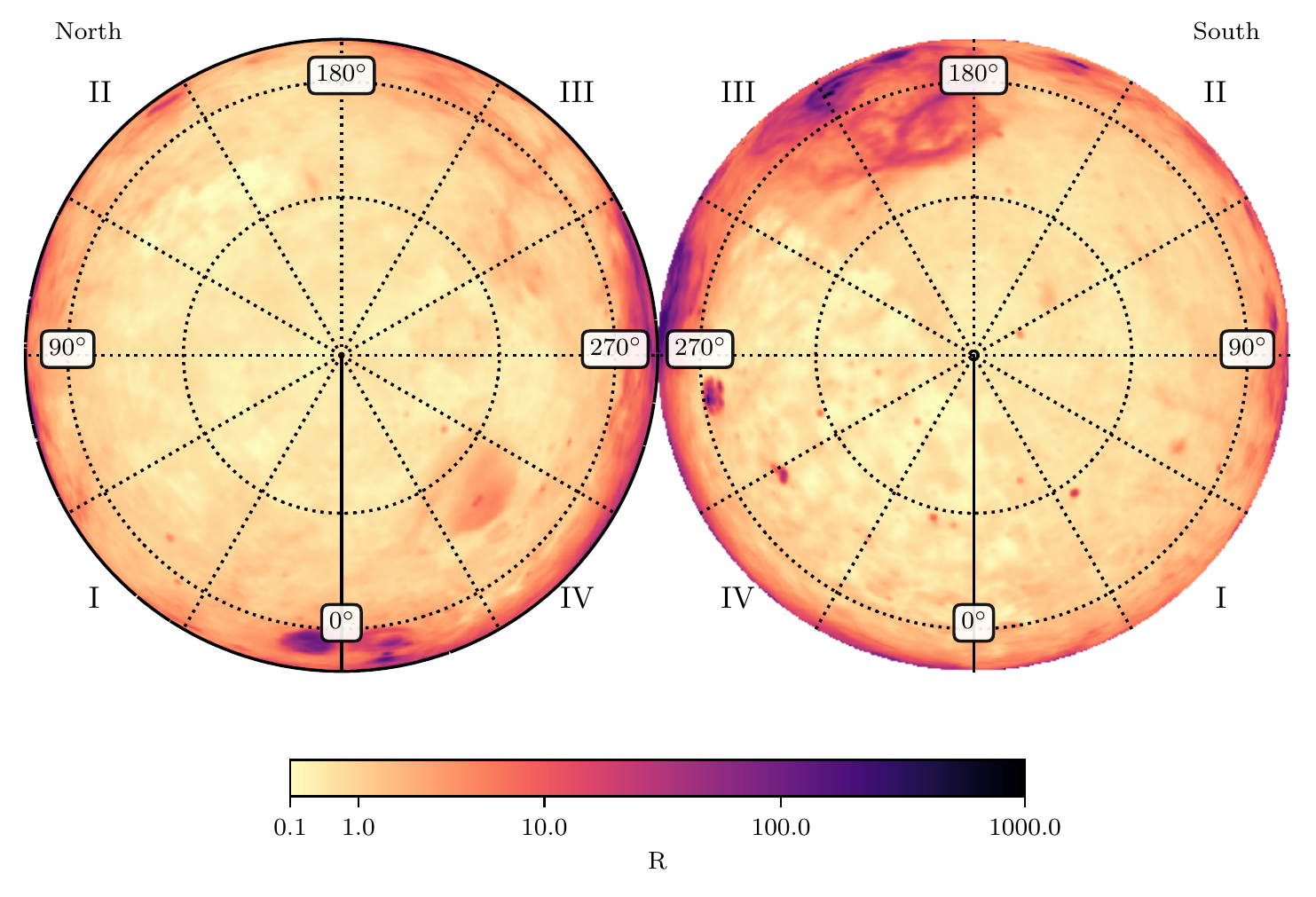}
        \caption{The upper panels show Stokes I emission at 408\,MHz (de-striped) \citep{Haslam_1982, Remazeilles_2014}. The lower panels show the H$\alpha$ emission from SHASSA, VTSS, and WHAM \citep{Finkbeiner_2003}. Alignment and overlays are the same as Fig. \ref{fig:exgal_ylm}.}
        \label{fig:ortho}
\end{figure}

The largest structures in the synchrotron sky are the Galactic loops and spurs, reviewed by \citet{Vidal_2015}, which stand out in both total synchrotron intensity (Stokes I) and polarised emission (PI). 
Using the relatively sparse RM grid of the time, \citet{Simard-Normandin_1980} described three corresponding regions, A, B, and C, which enclose the largest-scale RM features. \citet{Stil_2011} revisited this description using the RM grid of \citet{Taylor_etal_2009}. The areas covered by the three regions as defined by \citet{Simard-Normandin_1980}
are: \begin{itemize}
    \item Region A: (Loop II) a rectangle with corners $(\ell,b)=(145\arcdeg,-10\arcdeg)$ and $(170\arcdeg,-35\arcdeg)$
    \item Region B:  (the Gum Nebula) a circle centered at $(\ell,b)= (255\arcdeg,0\arcdeg)$ and radius $\sim20\arcdeg$ \citep{Vallee_Bignell_1983}
    \item Region C: (the NPS = Loop I) a rectangle with corners $(\ell,b)=(0\arcdeg,0\arcdeg)$ and $(60\arcdeg,+60\arcdeg)$
\end{itemize}

Region A is associated with radio continuum Loop II.  The outline of Loop II, as seen in diffuse, unpolarized synchrotron emission in the upper panel of Fig.~\ref{fig:ortho}, follows a boundary of RM$\sim0$ rad m$^{-2}$, the white areas on Figures \ref{fig:exgal_ylm} and \ref{fig:gmims_ylm}. Region A covers an enormous area, much of quadrants I and II in the southern hemisphere, within which the RMs are mostly negative. Nested inside Loop II is another emission feature, Loop IIIs, which roughly mirrors the Northern Loop III [center at $(\ell,b)=(124\arcdeg, +15.5\arcdeg)$ with diameter 65$\arcdeg$ \citep{Berkhuijsen_1971}]. The boundary of Loop IIIs is associated with an increase of the absolute RM value. \citet{Thomson_etal_2021} model this region as an expanding shell, following \citet{Berkhuijsen1973} and \citet{Vidal_2015}. There is an additional region of negative RM in the first Galactic quadrant of the Southern Hemisphere, 
visible in the full-resolution extragalactic map. This region is separated by a ridge of positive RM following the line of Loop II. The same positive ridge is present in the GMIMS map, but the larger area of negative of RM does not appear. However this region lies at the most southern declinations covered by the GMIMS-HBN survey.

Region B is associated with the Gum nebula, which is at lower latitude ($|b|<20\arcdeg$) and
 relatively small compared with the other two, so it is unlikely to have a strong effect on the mid-latitude RMs. In contrast, regions A and C are large enough to influence the RM patterns on sterradian scales.
 
Region C, the NPS discussed in Sec. \ref{sec:NPS} above, appears in the first quadrant as an area of relatively uniform positive RM. In the GMIMS map, the positive region only extends as high as $b\sim50$\arcdeg, whereas it extends to the North Pole in the extragalactic map.

The extragalactic data show that the RM is mostly positive at southern latitudes in the third and fourth Galactic quadrants. The same is not true for the GMIMS map. Although much of the Southern Galactic hemisphere could not be observed by the DRAO telescope, the observed portion of the third quadrant has mostly negative RMs. 

Strong correlation can be seen between the RM and the H$\alpha$ emission from nearby \ion{H}{2} regions, sometimes casting depolarization shadows that block the background diffuse polarization \citep[e.g.][]{Harvey-Smith_etal_2011, Purcell_2015,Thomson_etal_2018}.
Regions of lower density ionized gas traced by diffuse H$\alpha$ can strongly affect the RM. To study this effect, we plot the all-sky H$\alpha$ image of Finkbeiner (2003) in the lower panel of Fig. 17. Comparing this to the full-resolution maps from both surveys (upper panels of Fig.s 15 and 16) reveals a correlation with H$\alpha$ emission.
Latitudes $|b|<10\arcdeg$ are hidden in our orthographic projection, but even so the Orion-Eridanus superbubble \citep{Joubaud_etal_2019} shows up clearly at longitudes $180\arcdeg < \ell < 240\arcdeg$,
latitudes $-45\arcdeg < b < -5\arcdeg$.
There is a clear correlation with the extragalactic RMs and the Orion-Eridanus superbubble. The region itself is morphologically complex, and so is the RM distribution, but there is an enhancement in RMs along the bubble's boundary, with primarily positive RMs there.  The correlation with RM is far less clear in the GMIMS data. While there appear to be correlated RM enhancements along the H$\alpha$ filaments, the GMIMS RM structure does not match the H$\alpha$ as well as the extragalactic RMs do. The maps combining just the low order $Y_l^m$ terms (lower panels of Fig. \ref{fig:gmims_ylm}) show that the GMIMS RM is mostly negative in the Orion-Eridanus area.  The difference between the GMIMS and extragalactic RMs in this area suggests that much of the diffuse synchrotron emission is coming from the vicinity of the superbubble itself, and from the foreground.

In the Northern Galactic Hemisphere, the two RM maps show good agreement. The strongest common feature is the region of negative RM encircled by Loop III in the second quadrant. The appearance of this loop is very similar to that of Loop II in the South, with diffuse Stokes I emission following a line of RM $\sim$ 0 rad m$^{-2}$. In the full-resolution versions of both maps there is a clear ridge of positive RM which sharply changes to negative across the boundary of Loop III. 
 Loop I (the NPS) crosses the intermediate latitude range at $l\approx30$\arcdeg. The strongest postive RM features in the GMIMS map match the morphology of the spur. In both the GMIMS and the extragalatic maps, however, there is a sharp change in the strength of the RM along the ridge of the NPS \citep{Sun_etal_2015}. The distance to the high-latitude component of Loop I has been constrained to $\sim 0.1\,$kpc using starlight polarization~\citep{Panopoulou_2021}.

In summary, Loop I does not appear to contribute strongly to the $\sin(2\ell)$ pattern in the Northern Galactic Hemisphere (as shown in Sec. \ref{sec:NPS} above). 
Loops II, III, and IIIs do make significant contributions to the RMs in both surveys.  
No strong distance constraints have been placed on these loops, but their huge angular sizes and the corresponding Stokes I emission suggests that they are   local features. 
Orion-Eridanus also has a large-scale effect on the RM sky. 

The $\sin{(2\ell)}$ and $\sin{(\ell + \pi)}$ structure of the RM sky in the inner Galactic quadrants (first and fourth) does not appear to be associated with any local discrete structures. The $Y_l^m$ decompositions show that the RM pattern appears anti-symmetric about the Galactic plane at longitudes $-90\arcdeg < \ell < +90\arcdeg$. It is when we include the outer quadrants (second and third) that the asymmetry appears, along with association with local features. It is therefore possible that the RM structure from the global magnetic field is antisymmetric about the Galactic plane, but the antisymmetric pattern is obscured in the outer Galaxy by the effects of nearby objects.


\section{Self-Consistent Field Configurations based on RM Maps  \label{sec:models}}

The discussion in Sec. \ref{sec:loops} shows that the difference between the $\sin{(2\ell)}$ and $\sin{(\ell+\pi)}$ variation of RM with longitude in the Northern and Southern Hemispheres is not easily explained by the effects of nearby discrete structures such as H {\texttt{II}} regions or radio continuum loops.  The very good correlation with RMs of pulsars  with parallax distances shows that most of the Faraday rotation in the extragalactic sample occurs in the thick disk or lower halo, below $|z|\sim 1$ kpc.
Outside of the thin disk, $|z| \sim$ 0.1 kpc, the field configuration must change dramatically and differently in the two hemispheres, to explain the disagreement
between the disk $\vec{B}$ field models 
shown on Fig. \ref{fig:disk_contrib}, and the intermediate latitude extragalactic and GMIMS RM results.  How the field changes with $z$, and why it changes so differently in the North and South, is the fundamental question considered in this section.

To go beyond empirical models of the $\vec{B}$ field, like those illustrated on Fig. \ref{fig:disk_contrib}, requires a physical approach to the generation and maintenance of the magnetic field as a solution to the plasma equations \citep[e.g.][]{Ferriere_Terral_2014}.
Dynamo configurations provide the preferred model because dynamo processes in the interstellar medium can both
amplify the field (the small scale dynamo) and sustain a global mean field, i.e. the $\alpha - \Omega$ dynamo \citep{Beck_2015}. 
To try to explain the RM variation with $\ell$ described in section \ref{sec:RMdata} we consider the scale-invariant models of \citet{Henriksen:2017gp} and
\citet{Henriksen_etal_2018}, that solve the magneto-hydrodynamic equations including diffusion and a velocity field in the medium,
in a form that assumes scale invariance and solutions that are self-similar in time.  Various velocity fields in the gas are
considered by \citet{Henriksen_etal_2018}, and different amounts of diffusion.  These models do not separate disk and halo contributions to the field; they make a continuous, global solution to the plasma equations and hence to the vector potential and finally the magnetic field.

\begin{figure}
\includegraphics[width=7in]{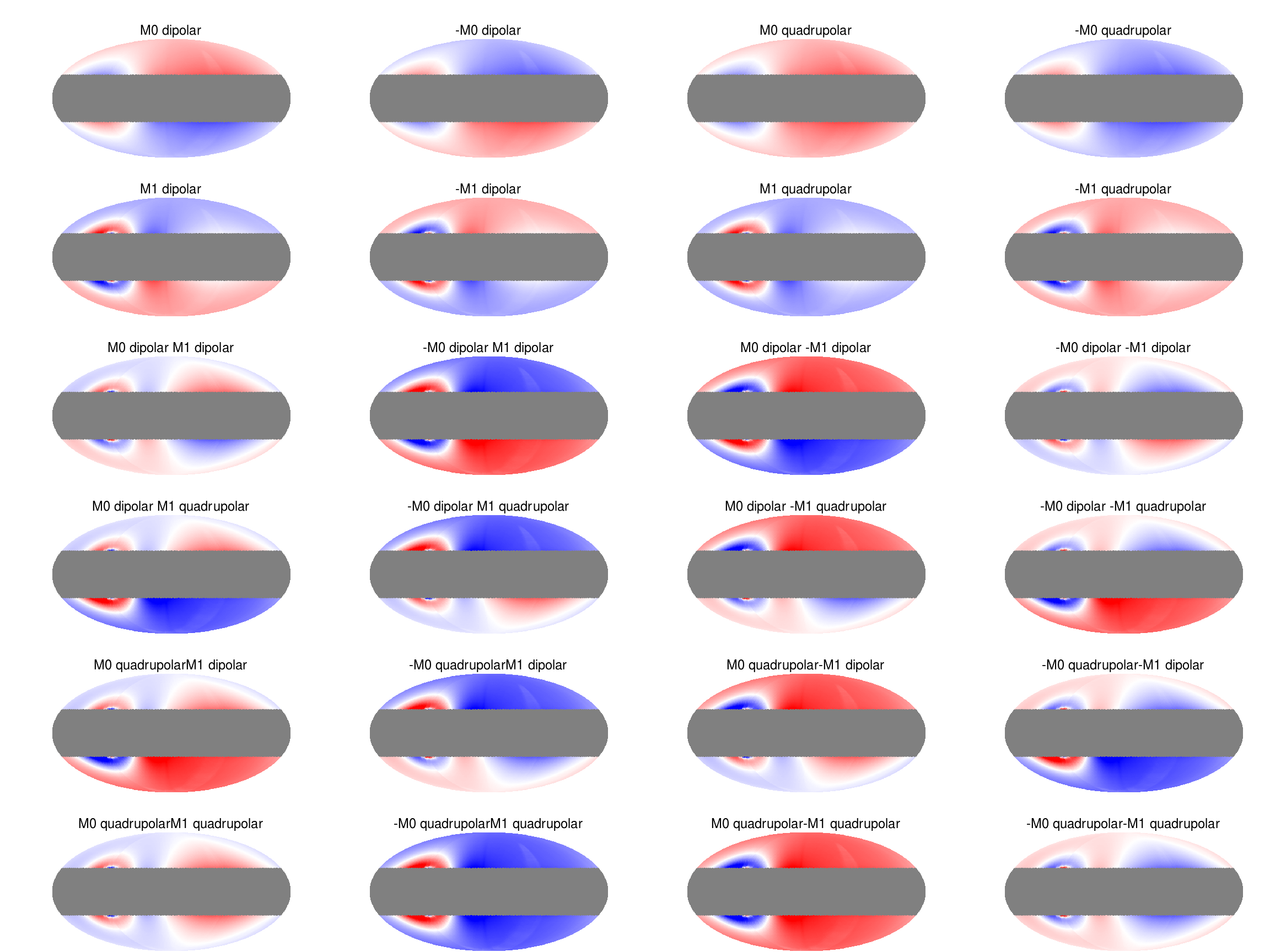} 
\caption{Simulated RM maps showing the M=0 and M=1 modes, each with $\pm$ dipolar and quadrupolar symmetry.  The two top rows show these modes individually.  The lower rows show all 16 combinations thereof. The case of -M0 dipolar added to -M1 quadrupolar, in the fourth row and rightmost column, is reproduced in Fig. \ref{fig:m0m1_model}. \label{fig:dynamomodels}}
\end{figure}

Here we explore whether a large angular scale model of the Galactic magnetic field can reproduce the main features described in section \ref{sec:RMdata}, namely:
\begin{enumerate}
\item Asymmetry across the Galactic plane.
\item A sin$(2\ell)$ RM pattern in the North.
\item A sin$(\ell+\pi)$ RM pattern in the South.
\item The amplitudes of the two patterns differ by a factor of two, with the stronger amplitude in the South.
\end{enumerate}
In this section we show that a dynamo-based model can be found which displays these general features. We have not achieved an exact match between our model and the data, and we deliberately leave this to future work. Our goal here is simply to demonstrate that dynamo models are strongly relevant to understanding the observed patterns in the RM sky.
Following the approach of \citet[sec. 3]{West_etal_2020}, we start with a combination of M0 (axisymmetric) and M1
(bisymmetric) spiral modes, each
either positive or negative in radial direction, and each either dipolar (continuous field across the midplane) or
quadrupolar, i.e. symmetric field on either side of the midplane \citep{Sokoloff_Shukurov_1990}. Combining just these two simplest spiral modes makes possible a diverse set of field configurations, some of which are asymmetric between the hemispheres, as seen in Fig. \ref{fig:dynamomodels}.  


For the $\vec{B}$ field configuration that matches  each spiral pattern we use combinations of dynamo models developed by  \citet{Henriksen:2017gp} and \citet{ Henriksen_etal_2018}, and subsequently applied to modelling the edge-on spiral galaxy NGC 4631 by \citet{Woodfinden_etal_2019}. We use the best fit case from \citet{Woodfinden_etal_2019} as a test case for this work. 
Different approaches start more or less from the same set of dynamo equations, but some use numerical techniques involving the solution of partial differential equations and some use other semi-analytic approximations, such as assuming a `zero z' approximation. The advantage of scale invariance is that it allows a quick survey of the possibilities based on algebraic equations and analytic solutions, and most importantly, it allows a coherent treatment of the disk and halo fields together. 
The dynamo models used in this work do not separate the disk field from the halo field. Rather, the two components are a result of the same scale-invariant dynamo modes. 

We solve the dynamo equations for the M=0 and M=1 cases, where M is the spiral mode, using a grid that has $n_x = 64$,  $n_y = 64$, and  $n_z = 32$ pixels, corresponding to a single hemisphere of the model magnetic field of a galaxy, and using a physical scale of 0.625~kpc/pixel (i.e., 64 pixels corresponds to 40~kpc). The model has no small-scale structure and so this relatively coarse resolution is sufficient.
The coordinate system defines the plane of the model galaxy to be parallel to the $xy$-plane, with the origin at its Galactic center. The $z$-axis is perpendicular to the plane, with $z>0$ towards the Northern Hemisphere. We scale the average strength of the output magnetic field of the m=0 mode to be 1~$\mu$G. We then scale the M = 1 mode to have the same average power, and thus the same average RM, as the M = 0 mode.

\begin{figure}
\hspace{1in} \includegraphics[width=5in]{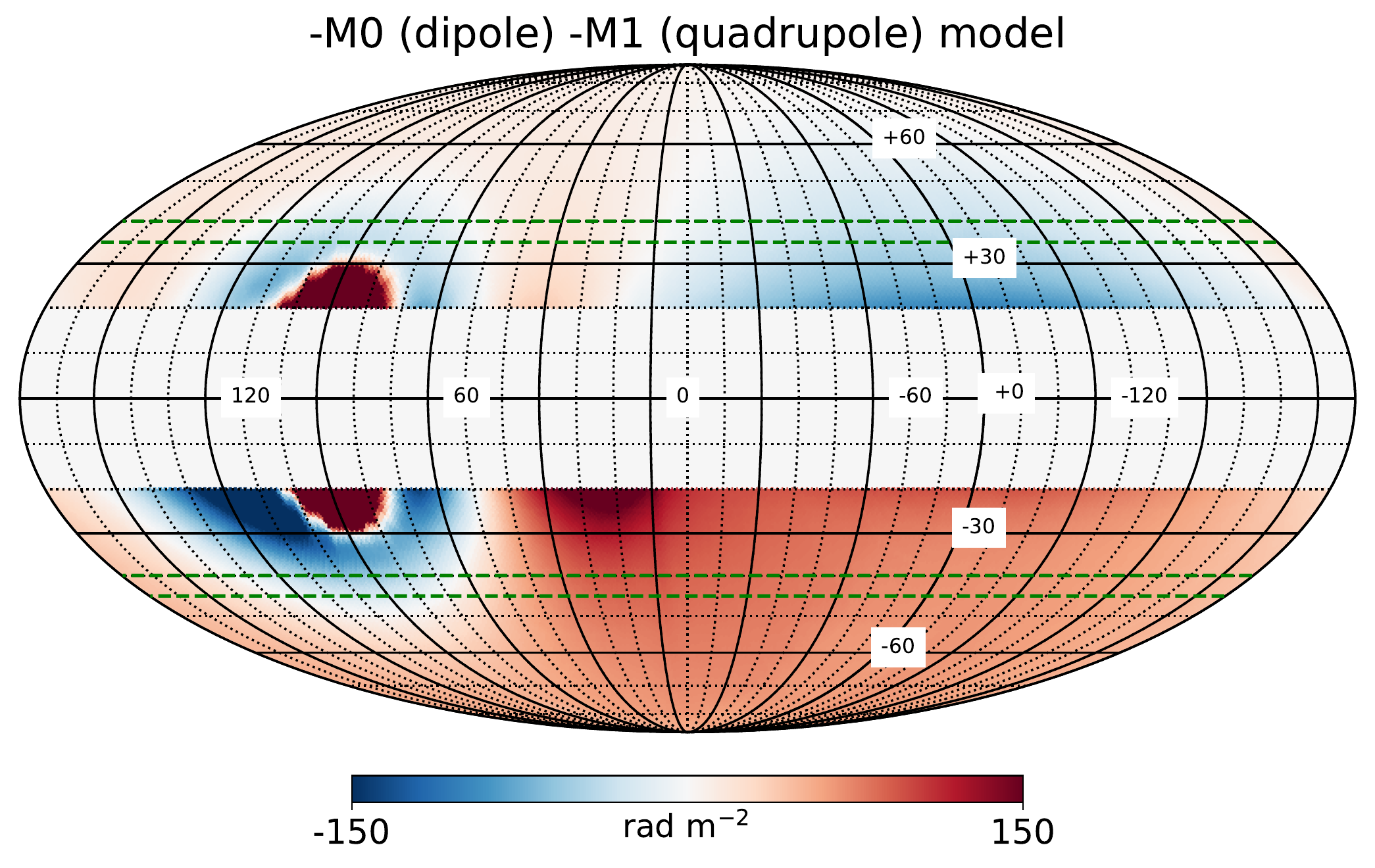}

\caption{The preferred model RM distribution resulting from a halo field
configuration including -M0 dipolar and -M1 quadrupolar components. 
The model is arbitrarily rotated in longitude. 
Latitudes $-20\arcdeg < b < +20\arcdeg$ are blanked.  Latitude ranges used to compute the azimuth dependence shown in figure \ref{fig:model_slices} are indicated by dashed green lines.
\label{fig:m0m1_model}}
\end{figure}

We find the solution for the dynamo equation for the vector potential \citep[][eq. 1]{Henriksen:2017gp} for points where $z>0$, and then assume either dipolar or quadrupolar symmetry across the disk of the galaxy 
to calculate points where $z<0$. 
We integrate the coherent field using a low-resolution Healpix projection $N_{\textrm{side}}=64$, corresponding to roughly $1^\circ$ pixels.  This angular resolution corresponds to a physical scale perpendicular to the LoS that is roughly 0.2~kpc at a distance of 10~kpc. 

We place the observer inside of this grid, at a position similar to the Sun's position in the Galaxy, i.e., $(x,y,z)=(-8,0,0)$~kpc. We then use the Hammurabi code \citep{Waelkens:2009bn} to compute the RM for an observer embedded inside this magnetic field geometry. The RM is computed by integrating volume elements along lines of sight through a grid where each element, $i$, contributes an increment of RM calculated by $RM_{i}=0.81 \ {n_{e}}\   {B_{i,\parallel}}\  \Delta r$. Here 
$n_{e}$ is the thermal
electron density of the halo, which we assume to have a constant value of 0.01~cm$^{-3}$, and $\Delta r$ is the element size (0.625 kpc). 

\begin{figure}
\hspace{1.5in}\includegraphics[width=4in]{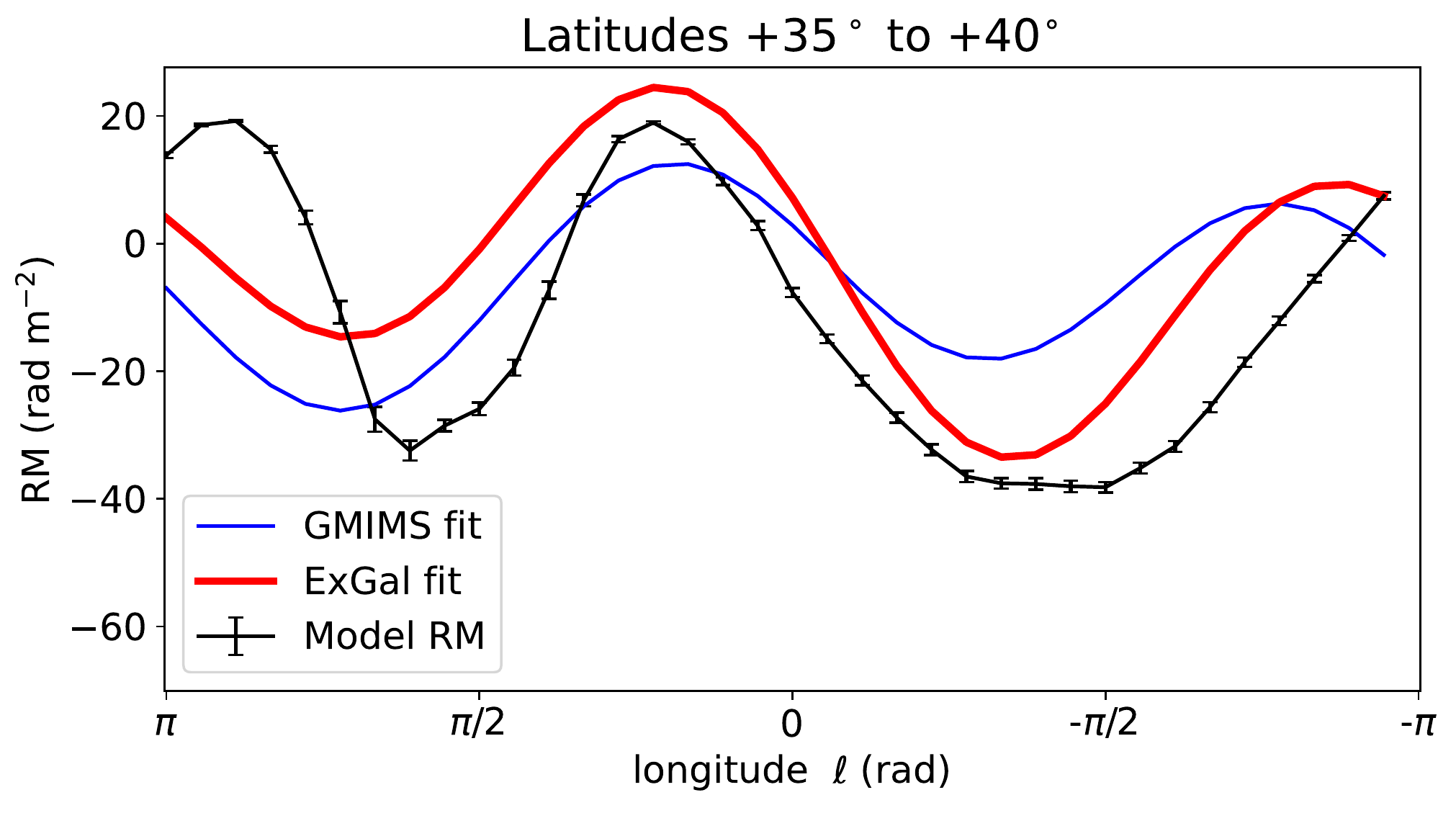}

\hspace{1.5in}\includegraphics[width=4in]{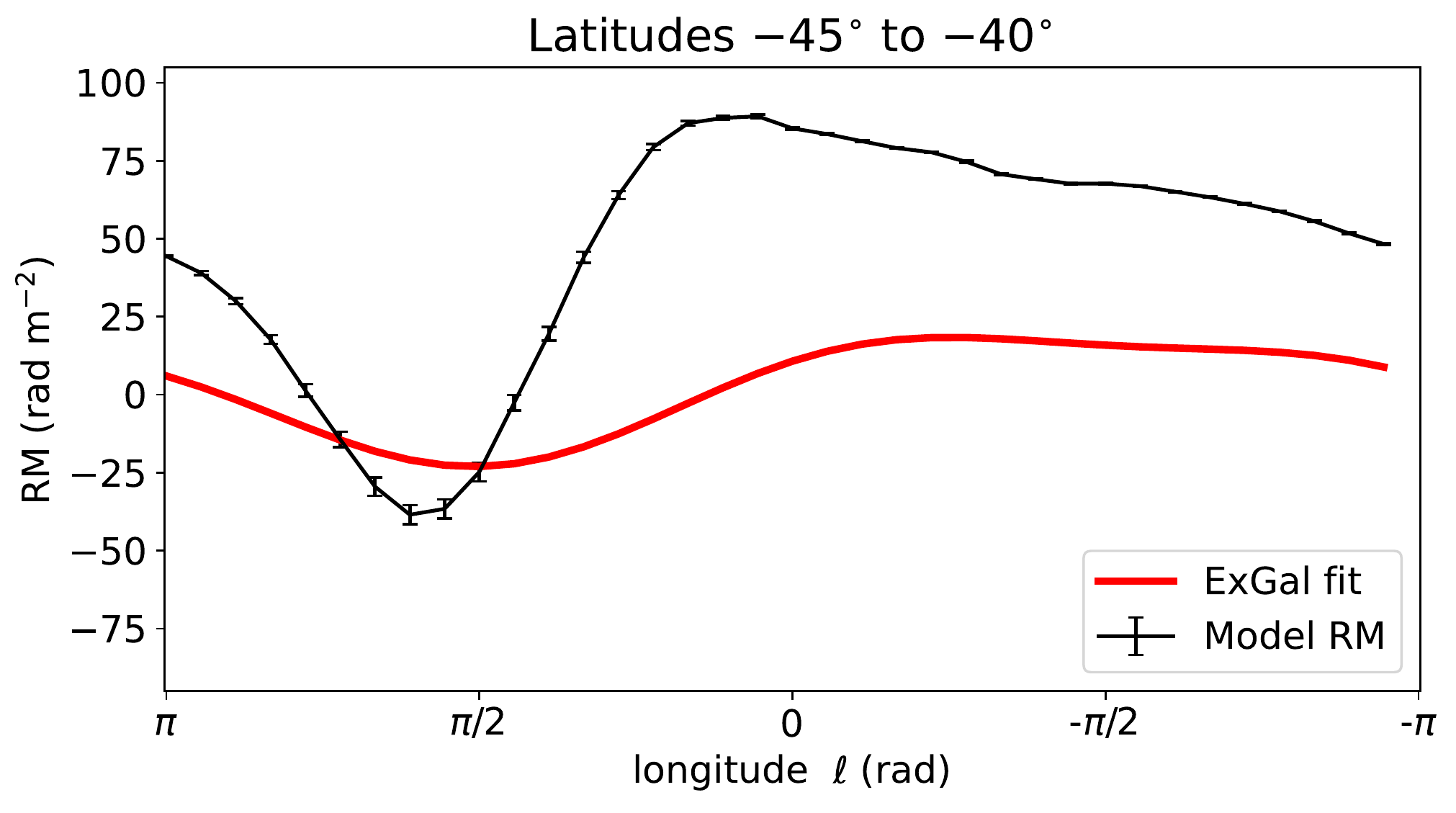}

\caption{Constant latitude binned slices through the -M0-M1 model.  The upper panel covers 
latitudes $+35\arcdeg < b < +40\arcdeg$, the red and blue curves are copied from the
upper panel of Fig. \ref{fig:example_b40}.  The lower panel is for $-45\arcdeg < b < -40\arcdeg$
corresponding to Fig. \ref{fig:ex-45}.  Error bars show the scatter of values in each longitude bin, which are necessarily small because of the low angular resolution of the model. \label{fig:model_slices} }
\end{figure}

\begin{figure}
\hspace{-.5in}\includegraphics[width=3in]{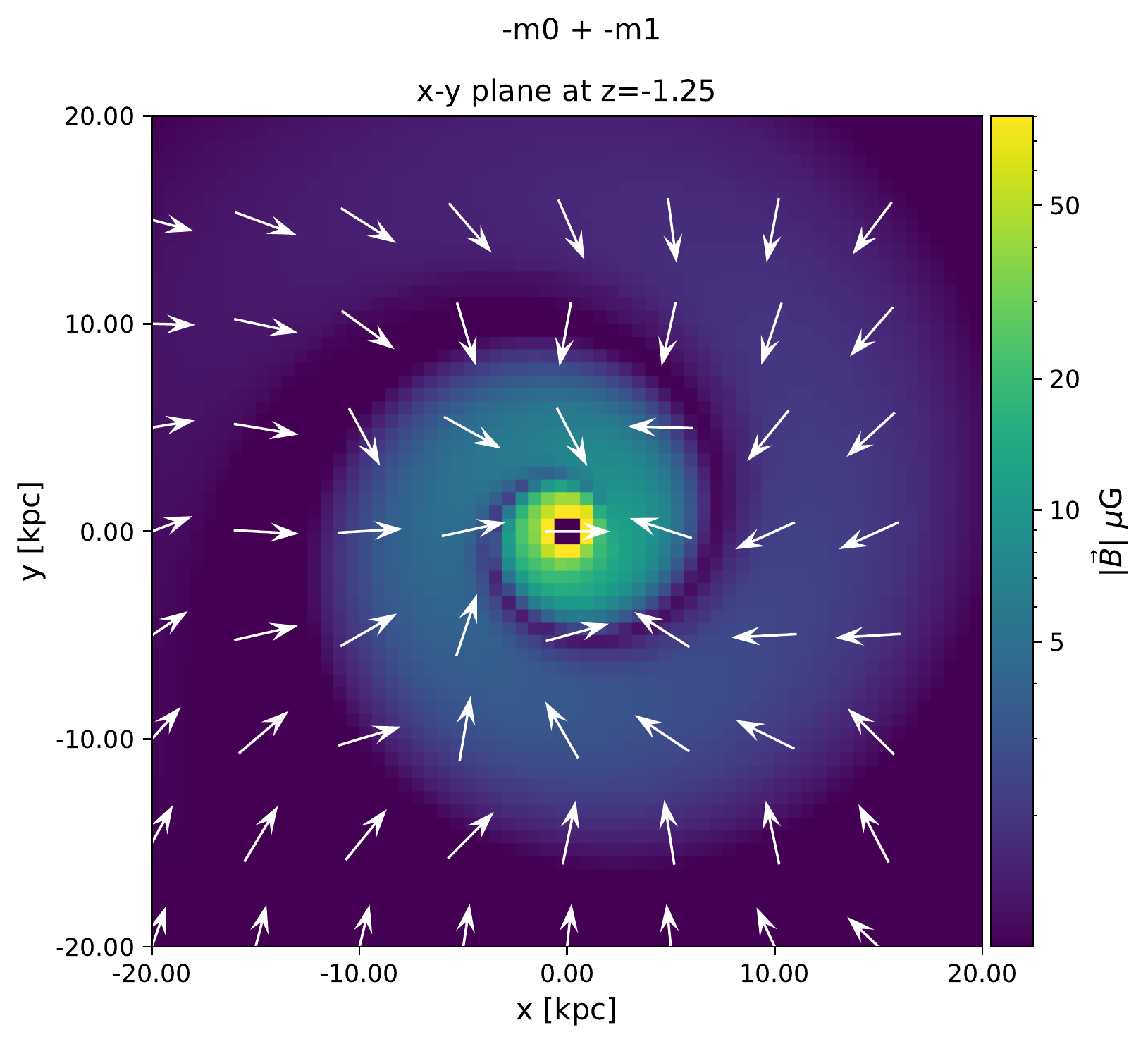}
\hspace{-.5in}\includegraphics[width=3in]{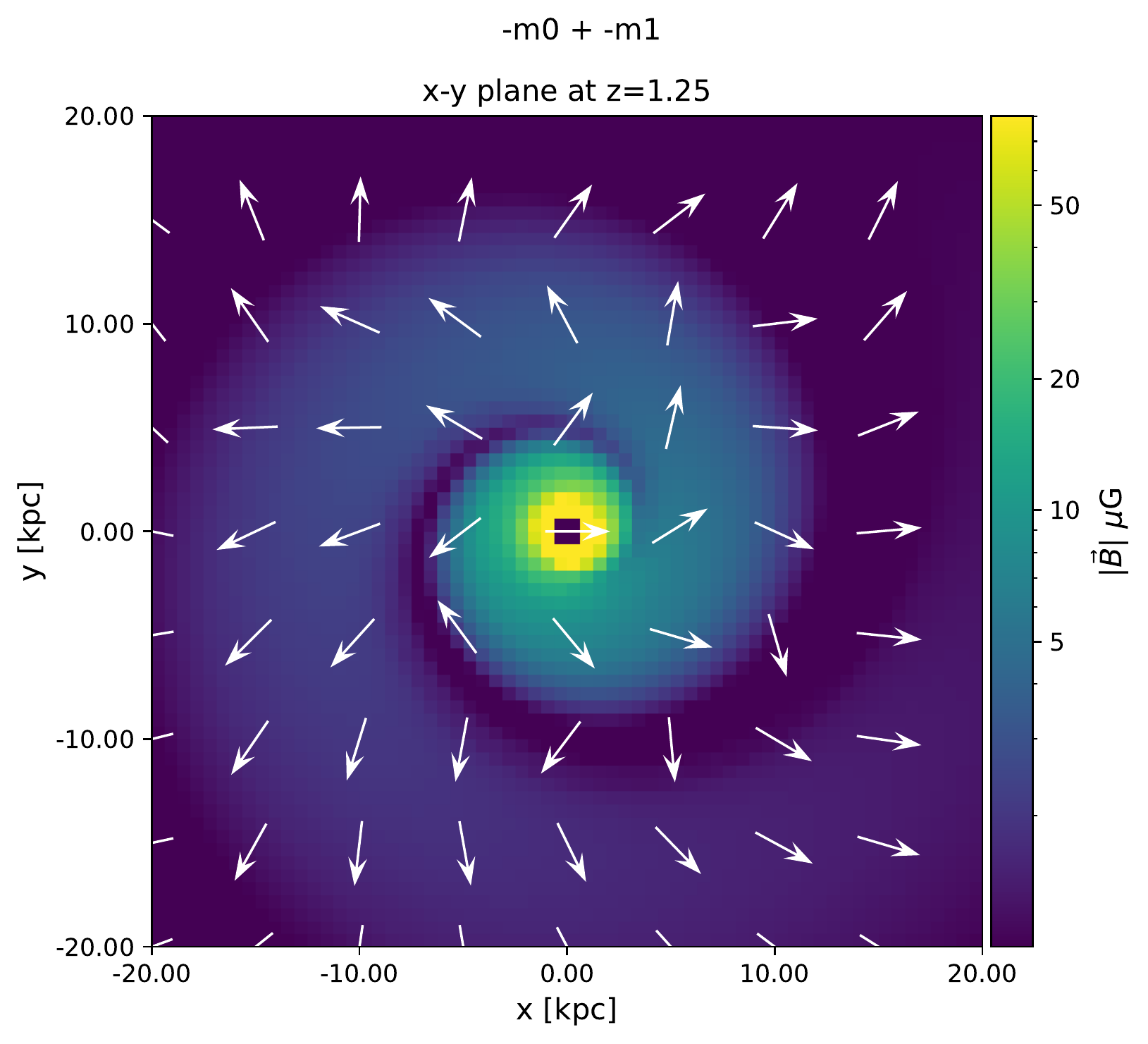}
\hspace{-.5in}\includegraphics[width=3in]{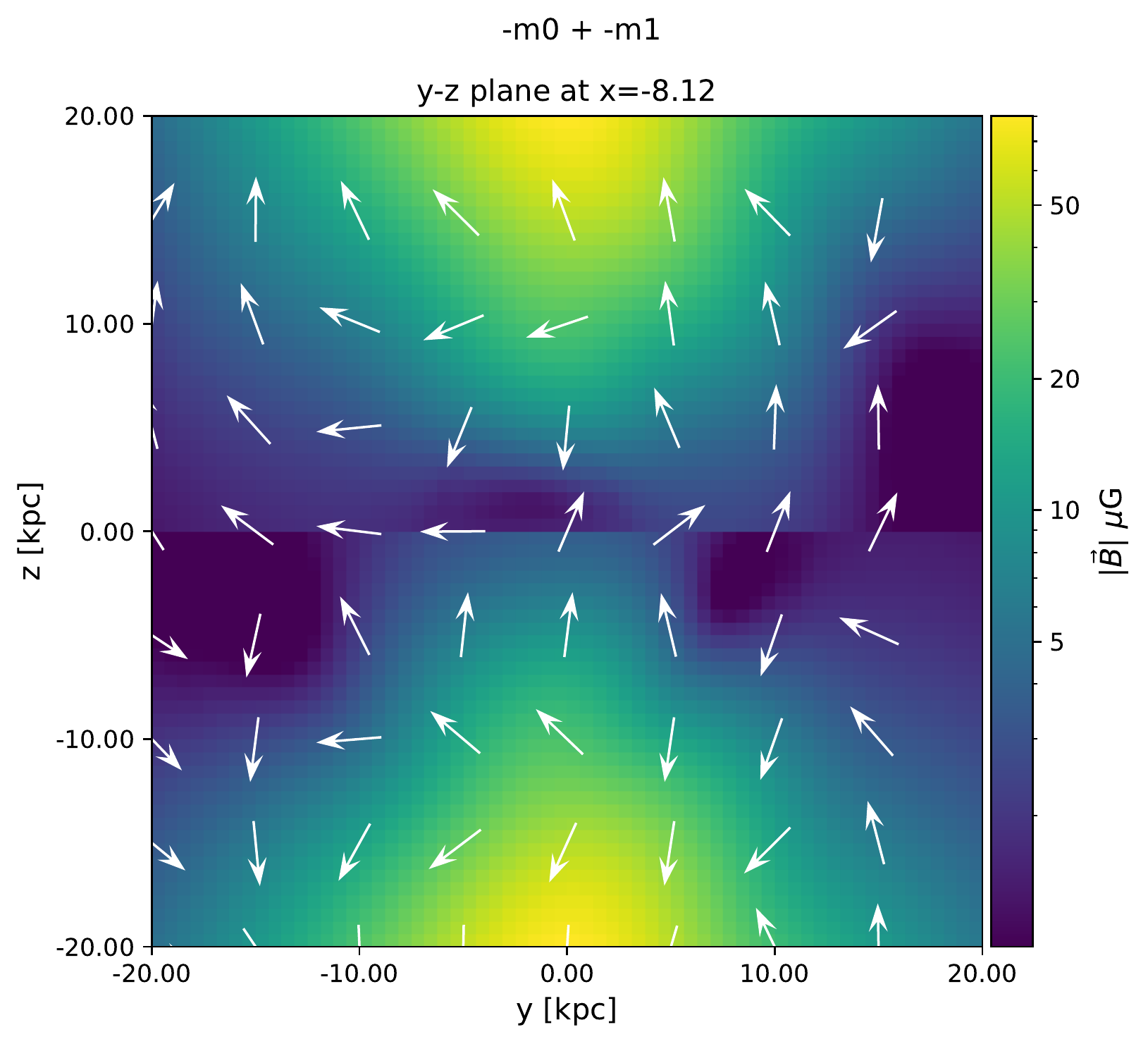}

\caption{
 Examples of the $\vec{B}$ field configuration in the -M0 (dipolar) -M1 (quadrupolar) model.  The left and middle panels
show the field direction projected on planes parallel to the midplane, displaced above and below midplane by 1.25 kpc.
The right panel shows the field on a plane perpendicular to the midplane, through a position ($x = -8.1$ kpc,
$y = 0$, $z = 0$) normal to the radial direction, roughly corresponding to the position of the observer for the computation of the RM figures \ref{fig:dynamomodels}, \ref{fig:m0m1_model}, and \ref{fig:model_slices}.   The $\vec{B}$ field model is for a 
generic spiral galaxy, it has not been fitted to the Milky Way, so the scales on the axes are representative.  
Unlike Figs. \ref{fig:dynamomodels}, \ref{fig:m0m1_model} and
\ref{fig:model_slices}, the field is not integrated along the line of sight nor projected on the sky, these panels merely illustrate sections through the model.
The background color shows the magnitude of the field ($|\vec{B}|$),
with a logarithmic scale.  The field is neither symmetric nor antisymmetric about the midplane.  
\label{fig:Bsections}}
\end{figure}

The output is a Healpix image \citep{Gorski:2005ku} of the RM across the sky, that can be compared with the extragalactic RM data. We mask latitudes below $|b|=30\arcdeg$ because we cannot adequately display them on this image, and because our primary interest is at higher latitudes.
The top two rows of Fig. \ref{fig:dynamomodels} show the M=0 and M=1 modes, each with dipolar or quadrupolar symmetry. The following rows show all 16 combinations of the M=0 plus M=1 modes, with equal amplitudes, positive or negative, in each case. In these panels, we have rotated the centre point of the longitude axis by 100$^\circ$ clockwise to more closely resemble the appearance of the pattern we observe in the real data. This rotation is equivalent to changing the viewing position within the model galaxy, i.e., the $(x,y)$ coordinates, but while maintaining the same radial distance, i.e., $r=\sqrt{x^2+y^2}=8$~kpc).  Here we can see that the M=0 or M=1 mode alone cannot reproduce a sin$(2l)$ pattern. However, the combination of $-\textbf{B}_{M=0}$ (dipolar) added to $-\textbf{B}_{M=1}$ (quadrupolar) can crudely reproduce all of the observed features, including the asymmetry across the plane. This model is shown in larger format in Fig. \ref{fig:m0m1_model}.  Slices through the model at latitudes $+35\arcdeg<b<+40\arcdeg$ and $-45\arcdeg<b<-40\arcdeg$ corresponding to Fig.s \ref{fig:example_b40} and \ref{fig:ex-45} are shown in Fig. \ref{fig:model_slices}.  The $\vec{B}$ field on three planes that make
sections through the Galaxy are shown in Fig. \ref{fig:Bsections} to illustrate the complexity of the field
in this $-M0-M1$ model. Although the model has not been adjusted to fit the data, and clearly adjustment is needed as seen by the mismatch on Fig. \ref{fig:model_slices}, the fact that the two hemispheres in this model give such different overall RM patterns motivates further work.  

\section{Discussion \label{sec:discussion}}

Comparing the RM surveys based on diffuse emission (GMIMS) and the foreground derived from extragalactic source RMs, the data in section \ref{sec:RMdata} show that the mid-latitude regions, roughly $10^{\circ} < |b| < 50^{\circ}$ in both hemispheres, are where a large-scale, coherent picture emerges.  Near the plane, $|b|<10^{\circ}$, and at both poles, $|b|>50^{\circ}$, the two surveys do not correlate, 
either because the path lengths sampled by the two techniques are different, or because of different resolutions, both in angle and on the Faraday spectrum, $\delta \phi$. However, at mid-latitudes the very good agreement between the two quite different observational methods indicates that the Milky Way magnetic field can be traced reliably with RM surveys of either type.  

If there were a one-to-one correspondence between GMIMS and extragalactic RMs, i.e. slope of one on Fig. \ref{fig:slope_n_ratio}, that would indicate that the Galactic synchrotron emission all comes from behind
the entire volume of Faraday rotating medium.  In regions where this is not the case, a comparison of the two RM maps gives information on how these two media are mixed along every line of sight.


To understand the values of the amplitude ratios and correlation slopes displayed on Fig. \ref{fig:slope_n_ratio}, we consider a 
Burn slab \citep{Burn_1966} as described in Section \ref{sec:correl}.  
Such a mixed  rotating and emitting region has a finite width, $\Delta \varphi$, in the Faraday spectrum, because emission from different points along the LoS undergoes different amounts of rotation. For a Burn slab
the detected polarisation angle is a linear function of the wavelength squared, just as it is for a background point source that undergoes rotation due to the same uniform magnetic field along the LoS. The RM derived from that slope is half the value of the corresponding point source RM, leading to a ratio of two, shown by the upper blue line on Fig. \ref{fig:slope_n_ratio}. 
For the GMIMS-HBN survey, the RMSF is broader than $\Delta \varphi$ for any realistic
slab at intermediate latitudes, so the slab is unresolved in $\varphi$,
with a single Faraday depth peak near the centre of $\Delta \varphi$, giving approximately half the value of the total RM of the slab. 

If the slab is in the foreground, and the synchrotron emission extends further
along the LoS than the Faraday rotating medium, then the ratio 
of the extragalactic RM to the diffuse RM can be less than two.  The ratio is one in the extreme where the
slab becomes simply a foreground screen of rotating plasma with all the emission coming from
beyond the screen.  

The two points on Fig. \ref{fig:slope_n_ratio} 
above $b=+40\arcdeg$ that give
slopes of 0.71$\pm$0.14 and 0.56$\pm$0.16 (see the lower right panel of Fig. \ref{fig:example_b40}), are revealing;  a value less than one in this ratio implies a field reversal along the
LoS, beyond the diffuse emission.  
If there are one or more sign changes in the magnetic field component along the LoS then the interpretation of the RM ratios can get much more complicated. In that case the diffuse RM could be arbitrarily high, either positive or negative, and the extragalactic RM could be zero, or vice versa, giving $C_2$ ratios of zero or $\pm \infty$, or anything in between.  Modelling the Faraday spectrum with better resolution would be needed to interpret the RMs in that case \citep[e.g.][Appendix C]{Bracco_etal_2022,Erceg_etal_2022}. This suggests that at positive latitudes we may be looking through a field reversal that is several hundred pc above the midplane.  The absence of correlation between the GMIMS and extragalactic survey RM at the highest latitudes could be explained if the diffuse emission comes from both behind and in front of the reversal.

The simplest and most striking result of the analysis of the RMs
in section \ref{sec:RMdata} is that the typical values of the RM in the Galactic
Southern Hemisphere are bigger than in the North.
On Table \ref{tab:longitude_fits} and Fig. \ref{fig:constants_vs_lat} 
the $C_1$ term has values
of 50 to 60 rad m$^{-2}$ for $-5\arcdeg > b > -30\arcdeg$, dropping smoothly to 
$\sim 30$ rad m$^{-2}$ at $b= -40\arcdeg$.  This is not clear in the GMIMS data because of the
sparse coverage of southern declinations, but for the extragalactic data
on the lower right hand panel of Fig. \ref{fig:constants_vs_lat} the blue curve at the
southern latitudes is a factor of two above the red curve, and
a factor of four above the blue curve at the corresponding Northern latitudes. 
The different amplitudes of the RMs in the two hemispheres is expected in the -M0-M1 model of section \ref{sec:models} (Figs. \ref{fig:m0m1_model} and \ref{fig:model_slices}).  

In the South, the
correspondence of the main negative RM area with the insides of Loop II and Loop IIIs
is striking in the right panels of Figs. \ref{fig:exgal_ylm} and \ref{fig:gmims_ylm},
in the range $60\arcdeg < \ell < 120\arcdeg$.  Positive RMs dominate the whole longitude range
$180\arcdeg < \ell < 360\arcdeg$, but in the Orion-Eridanus region they are the strongest,
mostly at southern latitudes but reaching into the North in the
longitude range $240\arcdeg<\ell<
260\arcdeg$.  These two structures by themselves may cause the very large values of RM in
the South that are not seen in the North.  Thus the Northern Hemisphere may be a
window to the halo field ($|z|>0.3$ kpc) 
whereas in the South the RMs are dominated by  structures in the nearby disk.
So both the effect of synchrotron loops with distances less than $\sim 1$ kpc (Sec. \ref{sec:loops}) and the global models including the halo field ($|z|>0.5$ kpc) (Sec. \ref{sec:models}) 
may be needed to 
understand the difference between the RM patterns in the Northern and Southern Hemispheres.

This synthesis of the two frameworks, on the one hand, nearby structures on large angular scales, and, on the other, a global 
halo-field pattern, is supported by the ratios of values of
$C_2$ and the correlation slopes shown on Fig. \ref{fig:slope_n_ratio}.
All the Southern Hemisphere slopes are consistent with
a value of two, suggesting that the synchrotron
emission and the Faraday rotation 
are mixed in a single      
emitting-rotating medium.  The lower values in the Northern Hemisphere, between one
and two but mostly closer to one, indicate that there the diffuse synchrotron emission is 
mostly coming from beyond the medium that causes the Faraday rotation,
thus at distances greater than $\sim$ 1 kpc.  Distance estimates to more interstellar
dust clouds \citep{Lallement_etal_2018,Pelgrims_etal_2020} and observations of the three-dimensional structure in the $\vec{B}$ field around these clouds  \citep{Tahani_etal_2019,Tahani_etal_2022a,Tahani_etal_2022b} are rapidly improving our understanding of the field in the nearby disk.  As larger surveys become available, e.g. POSSUM \citep{Gaensler_etal_2010,Anderson_etal_2021} and SPICE-RACS (Thomson et al. in preparation), these large-scale and small-scale RM studies can be unified.


\section{Conclusions} \label{sec:conclusions}
Comparing the RM map of the Milky Way derived from large samples of extragalactic source RMs with the first-moment map derived from the GMIMS Faraday depth spectra of the diffuse synchrotron emission
shows that these two data sets often give very different results.  Where they agree is on the large angular
scales, at intermediate Galactic latitudes.  Surprisingly, the patterns that show up in both
surveys are neither symmetric nor antisymmetric between the Northern and Southern Galactic
Hemispheres.  In the North, there is a strong $\sin{(2 \ell)}$ pattern in the RMs between 
$b=+10$\arcdeg and $b=+50$\arcdeg.  The pattern is closely, but not precisely aligned with
the Galactic center, i.e. 20\arcdeg$>\phi_2>$5\arcdeg, see Fig. \ref{fig:phases}.  In contrast,
the Southern Hemisphere shows a $\sin{(\ell + \pi)}$ pattern, with the $\pi$ phase offset
indicating that RMs in the first and second Galactic quadrants are mostly negative, while in the third and
fourth quadrants they are mostly positive.  Again, the pattern is aligned closely with longitude zero.  Pulsar RMs
match these patterns when we consider only pulsars well above and below the thin disk ($|z|>\ \sim$0.6 kpc).

The question raised by comparison of GMIMS and
extragalactic RMs is: why are the Northern and Southern Hemispheres so different?  In sections \ref{sec:loops} and \ref{sec:models} we take a first step toward a coherent picture of the global magnetic field configuration in the disk and halo of the Milky Way.

Large angular scale, nearby structures like the radio continuum loops show
morphological correspondence with RM features.  Expanding the RMs in 
low-order spherical harmonics makes this correspondence clearer, as shown in Figs \ref{fig:exgal_ylm}
and \ref{fig:gmims_ylm}.  Loops II, III, and IIIs trace the boundaries of large areas of negative
RMs, and Loop I, i.e. the North Polar Spur, has high positive RM.  The boundary of the Orion-Eridanus
superbubble is traced by positive RMs.  To the extent that distances are known, these loops and
bubbles are within one kpc of the sun \citep[][and references therein]{West_etal_2021}, with the possible exception of parts of Loop I \citep{Lallement_2022}.

If the mid-latitude RM patterns cannot be explained entirely by these nearby structures, then
they may help to determine the pattern of the $\vec{B}$ field in the lower Galactic halo ($|z| > \sim$ 0.5 kpc). 
If the vertical ($z$) component of the $\vec{B}$ field is continuous at $z=0$, as in
a dipole (M0) configuration, then it is hard to reconcile the asymmetry between the hemispheres
in their RM patterns.  Adding a quadrupole (M1) field, that has antisymmetry between the hemispheres, can introduce a discontinuity in combined $B_z$ at $z=0$, so it helps to match the data, as in the -M0-M1 model discussed in section \ref{sec:models}.   

At low latitudes ($|b|<$5\arcdeg) the RM pattern is more complicated
\citep[reviewed by][]{Han_2017}.  The $\vec{B}$ field in the disk is primarily
spiral or azimuthal, but with at least one field reversal
just a few hundred pc inside the solar circle
\citep[e.g.][]{Simard-Normandin_1980,Sofue_Fujimoto_1983, Rand_Kulkarni_1989, Weisberg_etal_2004, Xu_Han_2019}.  
An azimuthal field naturally gives a $\sin{(\ell)}$ pattern, and
a nearby reversal introduces a $\sin{(2\ell)}$ and higher order terms \citep{VanEck_etal_2011} that may be connected with the inter-arm halo field described by \citet{Mao_etal_2012}.

An attractive conjecture is that the field reversal seen in the disk just $\sim$0.15 to 0.3 kpc
inside the solar circle at $z=0$ might move to larger Galactic
radius at positive $z$, so that it passes directly above the solar circle at $z \sim$0.3 kpc.  
\citet{Ordog_etal_2017} have shown that the Sagittarius Arm reversal ($\ell \sim $52\arcdeg\xspace to 72\arcdeg)
is not cylindrical \citep[see also][]{Ma_etal_2020}. The latitude of the reversal boundary is a linear function of longitude, with slope about 0.5,
implying that its height above the plane, $z$, increases linearly with Galactic radius.  A similar 
radial slope of the nearby field reversal would be a natural explanation for the $\sin{(2\ell)}$ pattern in the RMs at mid-latitudes in the Northern Galactic Hemisphere.  

Nearby edge-on spiral galaxies show ``X-shaped'' $\vec{B}$ fields in their halos, \citep{Ferriere_Terral_2014,Krause_etal_2020}, reviewed by \citet[][sec. 4.13]{Beck_2015}.  A particularly dramatic example is NGC 4631 \citep{Mora-Partiarroyo_etal_2019}, that shows field reversals in its northern halo.  The pattern
has been modeled with dynamo components similar to those illustrated in Fig. \ref{fig:dynamomodels} \citep{Woodfinden_etal_2019}.  This is what the local field reversal would look like if it extends radially at high z.  
The smooth shift in the longitude of zero-phase as
a function of latitude (Fig. \ref{fig:phases}) could indicate a
variation of the pitch angle of the reversal with height above the plane, $z$.

The ratio between the Galactic and extragalactic RMs suggests that, in the Northern Hemisphere, the synchrotron emission is mostly beyond the Faraday rotating medium, whereas the two are spatially mixed in the South (Sect. \ref{sec:discussion}).  Moving beyond a single homogeneous slab or screen geometry leads to more complex models for the
juxtaposition of the rotating and emitting regions.  An example is the model of
\citet{Basu_etal_2019}, that includes the effects of random fields. The much finer resolution, $\delta \varphi$, in the Faraday spectra
observed by LOFAR at $\lambda \sim 1$ m gives sufficient precision to motivate
such detailed interpretation \citep{Erceg_etal_2022, Bracco_etal_2022, Sobey_etal_2019}.  The broad
RMSF of the GMIMS-HBN observations does
not justify a similar fine-scale analysis of the Faraday spectrum in these data.

 When the GMIMS High Band North (DRAO) survey is supplemented by a low band survey
in the Northern Hemisphere, 
 then models of the propagation of the polarized radiation through the magneto-ionic medium can be improved. Better sampling of the extragalactic  RM will be provided by POSSUM \citep{Gaensler_etal_2010,Anderson_etal_2021}, a much larger radio polarization survey than all previous observations put together. Similarly, the parameters of the dynamo models of the magnetic field structures described briefly in Sect. \ref{sec:models} can be adjusted to better fit the RM data from all three kinds of polarized emission: pulsars, extragalactic sources, and the diffuse synchrotron radiation.

\section{Acknowledgements}

The authors are grateful to Bryan Gaensler, Wolfgang Reich and an anonymous referee for helpful comments on the manuscript.

A.B. acknowledges support from the European Union’s Horizon 2020 research and innovation program through the Marie Skłodowska-Curie Grant agreement No. 843008 and from the European Research Council through the Advanced Grant MIST (FP7/2017-2022, No. 742719).
J-L. H. is supported by the National Natural Science Foundation of China (Grant Numbers 11988101, 11833009).
A.S.H. is supported by a Natural Sciences and Engineering Research Council (NSERC) Discovery Grant.
A.O. is supported by the Dunlap Institute for Astronomy and Astrophysics at the University of Toronto.

\software{
This research made use of SciPy \citep{SciPy2020}, NumPy \citep{van2011numpy}, Matplotlib, a Python library for publication quality graphics \citep{Hunter_2007}, and Astropy,\footnote{\url{http://www.astropy.org}} a community-developed core Python package for Astronomy \citep{astropy:2013}. }


\begin{thebibliography}{}
\expandafter\ifx\csname natexlab\endcsname\relax\def\natexlab#1{#1}\fi
\providecommand{\url}[1]{\href{#1}{#1}}
\providecommand{\dodoi}[1]{doi:~\href{http://doi.org/#1}{\nolinkurl{#1}}}
\providecommand{\doeprint}[1]{\href{http://ascl.net/#1}{\nolinkurl{http://ascl.net/#1}}}
\providecommand{\doarXiv}[1]{\href{https://arxiv.org/abs/#1}{\nolinkurl{https://arxiv.org/abs/#1}}}

\bibitem[{{Anderson} {et~al.}(2021){Anderson}, {Heald}, {Eilek}, {Lenc},
  {Gaensler}, {Rudnick}, {Van Eck}, {O'Sullivan}, {Stil}, {Chippendale},
  {Riseley}, {Carretti}, {West}, {Farnes}, {Harvey-Smith}, {McClure-Griffiths},
  {Bock}, {Bunton}, {Koribalski}, {Tremblay}, {Voronkov}, \&
  {Warhurst}}]{Anderson_etal_2021}
{Anderson}, C.~S., {Heald}, G.~H., {Eilek}, J.~A., {et~al.} 2021, \pasa, 38,
  e020, \dodoi{10.1017/pasa.2021.4}

\bibitem[{{Andrae} {et~al.}(2010){Andrae}, {Schulze-Hartung}, \&
  {Melchior}}]{Andrae_etal_2010}
{Andrae}, R., {Schulze-Hartung}, T., \& {Melchior}, P. 2010, arXiv e-prints,
  arXiv:1012.3754.
\newblock \doarXiv{1012.3754}

\bibitem[{{Astropy Collaboration} {et~al.}(2013){Astropy Collaboration},
  {Robitaille}, {Tollerud}, {Greenfield}, {Droettboom}, {Bray}, {Aldcroft},
  {Davis}, {Ginsburg}, {Price-Whelan}, {Kerzendorf}, {Conley}, {Crighton},
  {Barbary}, {Muna}, {Ferguson}, {Grollier}, {Parikh}, {Nair}, {Unther},
  {Deil}, {Woillez}, {Conseil}, {Kramer}, {Turner}, {Singer}, {Fox}, {Weaver},
  {Zabalza}, {Edwards}, {Azalee Bostroem}, {Burke}, {Casey}, {Crawford},
  {Dencheva}, {Ely}, {Jenness}, {Labrie}, {Lim}, {Pierfederici}, {Pontzen},
  {Ptak}, {Refsdal}, {Servillat}, \& {Streicher}}]{astropy:2013}
{Astropy Collaboration}, {Robitaille}, T.~P., {Tollerud}, E.~J., {et~al.} 2013,
  \aap, 558, A33, \dodoi{10.1051/0004-6361/201322068}

\bibitem[{{Barlow}(1989)}]{Barlow_1989}
{Barlow}, R.~J. 1989, {Statistics. A guide to the use of statistical methods in
  the physical sciences} (Wiley, New York)

\bibitem[{{Basu} {et~al.}(2019){Basu}, {Fletcher}, {Mao}, {Burkhart}, {Beck},
  \& {Schnitzeler}}]{Basu_etal_2019}
{Basu}, A., {Fletcher}, A., {Mao}, S.~A., {et~al.} 2019, Galaxies, 7, 89,
  \dodoi{10.3390/galaxies7040089}

\bibitem[{{Beck}(2015)}]{Beck_2015}
{Beck}, R. 2015, \aapr, 24, 4, \dodoi{10.1007/s00159-015-0084-4}

\bibitem[{{Berkhuijsen}(1971)}]{Berkhuijsen_1971}
{Berkhuijsen}, E.~M. 1971, \aap, 14, 359

\bibitem[{{Berkhuijsen}(1973)}]{Berkhuijsen1973}
---. 1973, \aap, 24, 143

\bibitem[{{Bracco} {et~al.}(2022){Bracco}, {Ntormousi}, {Jeli{\'c}},
  {Padovani}, {{\v{S}}iljeg}, {Erceg}, {Turi{\'c}}, {Ceraj}, \&
  {{\v{S}}nidari{\'c}}}]{Bracco_etal_2022}
{Bracco}, A., {Ntormousi}, E., {Jeli{\'c}}, V., {et~al.} 2022, arXiv e-prints,
  arXiv:2204.02774.
\newblock \doarXiv{2204.02774}

\bibitem[{{Brentjens} \& {de Bruyn}(2005)}]{Brentjens_deBruyn_2005}
{Brentjens}, M.~A., \& {de Bruyn}, A.~G. 2005, \aap, 441, 1217,
  \dodoi{10.1051/0004-6361:20052990}

\bibitem[{{Brown} {et~al.}(2007){Brown}, {Haverkorn}, {Gaensler}, {Taylor},
  {Bizunok}, {McClure-Griffiths}, {Dickey}, \& {Green}}]{Brown_etal_2007}
{Brown}, J.~C., {Haverkorn}, M., {Gaensler}, B.~M., {et~al.} 2007, \apj, 663,
  258, \dodoi{10.1086/518499}

\bibitem[{{Burn}(1966)}]{Burn_1966}
{Burn}, B.~J. 1966, \mnras, 133, 67, \dodoi{10.1093/mnras/133.1.67}

\bibitem[{{Cordes} \& {Lazio}(2002)}]{Cordes_Lazio_2002}
{Cordes}, J.~M., \& {Lazio}, T.~J.~W. 2002, arXiv e-prints, astro.
\newblock \doarXiv{astro-ph/0207156}

\bibitem[{{Dennis} \& {Land}(2008)}]{Dennis_Land_2008}
{Dennis}, M.~R., \& {Land}, K. 2008, \mnras, 383, 424,
  \dodoi{10.1111/j.1365-2966.2007.12484.x}

\bibitem[{{Dickey} {et~al.}(2019){Dickey}, {Landecker}, {Thomson}, {Wolleben},
  {Sun}, {Carretti}, {Douglas}, {Fletcher}, {Gaensler}, {Gray}, {Haverkorn},
  {Hill}, {Mao}, \& {McClure-Griffiths}}]{Dickey_etal_2019}
{Dickey}, J.~M., {Landecker}, T.~L., {Thomson}, A. J.~M., {et~al.} 2019, \apj,
  871, 106, \dodoi{10.3847/1538-4357/aaf85f}

\bibitem[{{Drake} \& {Wright}(2020)}]{Drake_2020}
{Drake}, K.~P., \& {Wright}, G.~B. 2020, Journal of Computational Physics, 416,
  109544, \dodoi{10.1016/j.jcp.2020.109544}

\bibitem[{{Erceg} {et~al.}(2022){Erceg}, {Jeli{\'c}}, {Haverkorn}, {Bracco},
  {Shimwell}, {Tasse}, {Dickey}, {Ceraj}, {Drabent}, {Hardcastle}, \&
  {Turi{\'c}}}]{Erceg_etal_2022}
{Erceg}, A., {Jeli{\'c}}, V., {Haverkorn}, M., {et~al.} 2022, \aap, 663, A7,
  \dodoi{10.1051/0004-6361/202142244}

\bibitem[{{Ferri{\`e}re}(2016)}]{Ferriere_2016}
{Ferri{\`e}re}, K. 2016, in Journal of Physics Conference Series, Vol. 767,
  Journal of Physics Conference Series, 012006

\bibitem[{{Ferri{\`e}re} \& {Terral}(2014)}]{Ferriere_Terral_2014}
{Ferri{\`e}re}, K., \& {Terral}, P. 2014, \aap, 561, A100,
  \dodoi{10.1051/0004-6361/201322966}

\bibitem[{{Ferri{\`e}re} {et~al.}(2021){Ferri{\`e}re}, {West}, \&
  {Jaffe}}]{Ferriere_etal_2021}
{Ferri{\`e}re}, K., {West}, J.~L., \& {Jaffe}, T.~R. 2021, \mnras, 507, 4968,
  \dodoi{10.1093/mnras/stab1641}

\bibitem[{{Finkbeiner}(2003)}]{Finkbeiner_2003}
{Finkbeiner}, D.~P. 2003, \apjs, 146, 407, \dodoi{10.1086/374411}

\bibitem[{{Gaensler} {et~al.}(2010){Gaensler}, {Landecker}, {Taylor}, \&
  {POSSUM Collaboration}}]{Gaensler_etal_2010}
{Gaensler}, B.~M., {Landecker}, T.~L., {Taylor}, A.~R., \& {POSSUM
  Collaboration}. 2010, in American Astronomical Society Meeting Abstracts,
  Vol. 215, American Astronomical Society Meeting Abstracts \#215, 470.13

\bibitem[{{Gaensler} {et~al.}(2008){Gaensler}, {Madsen}, {Chatterjee}, \&
  {Mao}}]{Gaensler_etal_2008}
{Gaensler}, B.~M., {Madsen}, G.~J., {Chatterjee}, S., \& {Mao}, S.~A. 2008,
  \pasa, 25, 184, \dodoi{10.1071/AS08004}

\bibitem[{{Gardner} {et~al.}(1969){Gardner}, {Morris}, \&
  {Whiteoak}}]{Gardner_etal_1969}
{Gardner}, F.~F., {Morris}, D., \& {Whiteoak}, J.~B. 1969, Australian Journal
  of Physics, 22, 79, \dodoi{10.1071/PH690079}

\bibitem[{{G{\'o}rski} {et~al.}(2005){G{\'o}rski}, {Hivon}, {Banday},
  {Wandelt}, {Hansen}, {Reinecke}, \& {Bartelmann}}]{Gorski:2005ku}
{G{\'o}rski}, K.~M., {Hivon}, E., {Banday}, A.~J., {et~al.} 2005, \apj, 622,
  759, \dodoi{10.1086/427976}

\bibitem[{{Han}(2001)}]{Han_2001}
{Han}, J.~L. 2001, \apss, 278, 181, \dodoi{10.1023/A:1013102711400}

\bibitem[{{Han}(2017)}]{Han_2017}
---. 2017, \araa, 55, 111, \dodoi{10.1146/annurev-astro-091916-055221}

\bibitem[{{Han} {et~al.}(1997){Han}, {Manchester}, {Berkhuijsen}, \&
  {Beck}}]{Han_etal_1997}
{Han}, J.~L., {Manchester}, R.~N., {Berkhuijsen}, E.~M., \& {Beck}, R. 1997,
  \aap, 322, 98

\bibitem[{{Han} {et~al.}(2006){Han}, {Manchester}, {Lyne}, {Qiao}, \& {van
  Straten}}]{Han_etal_2006}
{Han}, J.~L., {Manchester}, R.~N., {Lyne}, A.~G., {Qiao}, G.~J., \& {van
  Straten}, W. 2006, \apj, 642, 868, \dodoi{10.1086/501444}

\bibitem[{{Han} {et~al.}(1999){Han}, {Manchester}, \& {Qiao}}]{Han_etal_1999}
{Han}, J.~L., {Manchester}, R.~N., \& {Qiao}, G.~J. 1999, \mnras, 306, 371,
  \dodoi{10.1046/j.1365-8711.1999.02544.x}

\bibitem[{{Han} {et~al.}(2018){Han}, {Manchester}, {van Straten}, \&
  {Demorest}}]{Han_etal_2018}
{Han}, J.~L., {Manchester}, R.~N., {van Straten}, W., \& {Demorest}, P. 2018,
  \apjs, 234, 11, \dodoi{10.3847/1538-4365/aa9c45}

\bibitem[{{Han} \& {Qiao}(1994)}]{Han_Qiao_1994}
{Han}, J.~L., \& {Qiao}, G.~J. 1994, \aap, 288, 759

\bibitem[{{Harvey-Smith} {et~al.}(2011){Harvey-Smith}, {Madsen}, \&
  {Gaensler}}]{Harvey-Smith_etal_2011}
{Harvey-Smith}, L., {Madsen}, G.~J., \& {Gaensler}, B.~M. 2011, \apj, 736, 83,
  \dodoi{10.1088/0004-637X/736/2/83}

\bibitem[{{Haslam} {et~al.}(1982){Haslam}, {Salter}, {Stoffel}, \&
  {Wilson}}]{Haslam_1982}
{Haslam}, C.~G.~T., {Salter}, C.~J., {Stoffel}, H., \& {Wilson}, W.~E. 1982,
  \aaps, 47, 1

\bibitem[{{Haverkorn}(2015)}]{Haverkorn_2015}
{Haverkorn}, M. 2015, in Astrophysics and Space Science Library, Vol. 407,
  Magnetic Fields in Diffuse Media, ed. A.~{Lazarian}, E.~M. {de Gouveia Dal
  Pino}, \& C.~{Melioli}, 483

\bibitem[{{Haverkorn} {et~al.}(2008){Haverkorn}, {Brown}, {Gaensler}, \&
  {McClure-Griffiths}}]{Haverkorn_etal_2008}
{Haverkorn}, M., {Brown}, J.~C., {Gaensler}, B.~M., \& {McClure-Griffiths},
  N.~M. 2008, \apj, 680, 362, \dodoi{10.1086/587165}

\bibitem[{{Henriksen}(2017)}]{Henriksen:2017gp}
{Henriksen}, R.~N. 2017, \mnras, 469, 4806, \dodoi{10.1093/mnras/stx1169}

\bibitem[{{Henriksen} \& {Irwin}(2021)}]{Henriksen_Irwin_2021}
{Henriksen}, R.~N., \& {Irwin}, J. 2021, \apj, 920, 133,
  \dodoi{10.3847/1538-4357/ac173f}

\bibitem[{{Henriksen} {et~al.}(2018){Henriksen}, {Woodfinden}, \&
  {Irwin}}]{Henriksen_etal_2018}
{Henriksen}, R.~N., {Woodfinden}, A., \& {Irwin}, J.~A. 2018, \mnras, 476, 635,
  \dodoi{10.1093/mnras/sty256}

\bibitem[{{Hunter}(2007)}]{Hunter_2007}
{Hunter}, J.~D. 2007, Computing in Science and Engineering, 9, 90,
  \dodoi{10.1109/MCSE.2007.55}

\bibitem[{{Hutschenreuter} \& {En{\ss}lin}(2020)}]{Hutschenreuter_Ensslin_2020}
{Hutschenreuter}, S., \& {En{\ss}lin}, T.~A. 2020, \aap, 633, A150,
  \dodoi{10.1051/0004-6361/201935479}

\bibitem[{{Hutschenreuter} {et~al.}(2022){Hutschenreuter}, {Anderson}, {Betti},
  {Bower}, {Brown}, {Br{\"u}ggen}, {Carretti}, {Clarke}, {Clegg}, {Costa},
  {Croft}, {Eck}, {Gaensler}, {de Gasperin}, {Haverkorn}, {Heald}, {Hull},
  {Inoue}, {Johnston-Hollitt}, {Kaczmarek}, {Law}, {Ma}, {MacMahon}, {Mao},
  {Riseley}, {Roy}, {Shanahan}, {Shimwell}, {Stil}, {Sobey}, {O'Sullivan},
  {Tasse}, {Vacca}, {Vernstrom}, {Williams}, {Wright}, \&
  {En{\ss}lin}}]{Hutschenreuter_etal_2021}
{Hutschenreuter}, S., {Anderson}, C.~S., {Betti}, S., {et~al.} 2022, \aap, 657,
  A43, \dodoi{10.1051/0004-6361/202140486}

\bibitem[{{Indrani} \& {Deshpande}(1999)}]{Indrani_Deshpande_1999}
{Indrani}, C., \& {Deshpande}, A.~A. 1999, \na, 4, 33,
  \dodoi{10.1016/S1384-1076(98)00038-4}

\bibitem[{{Jaffe}(2019)}]{Jaffe_2019}
{Jaffe}, T.~R. 2019, Galaxies, 7, 52, \dodoi{10.3390/galaxies7020052}

\bibitem[{{Jansson} \& {Farrar}(2012)}]{Jansson_Farrar_2012}
{Jansson}, R., \& {Farrar}, G.~R. 2012, \apj, 757, 14,
  \dodoi{10.1088/0004-637X/757/1/14}

\bibitem[{{Joubaud} {et~al.}(2019){Joubaud}, {Grenier}, {Ballet}, \&
  {Soler}}]{Joubaud_etal_2019}
{Joubaud}, T., {Grenier}, I.~A., {Ballet}, J., \& {Soler}, J.~D. 2019, \aap,
  631, A52, \dodoi{10.1051/0004-6361/201936239}

\bibitem[{{Krause} {et~al.}(2020){Krause}, {Irwin}, {Schmidt}, {Stein},
  {Miskolczi}, {Carolina Mora-Partiarroyo}, {Wiegert}, {Beck}, {Stil}, {Heald},
  {Li}, {Damas-Segovia}, {Vargas}, {Rand}, {West}, {Walterbos}, {Dettmar},
  {English}, \& {Woodfinden}}]{Krause_etal_2020}
{Krause}, M., {Irwin}, J., {Schmidt}, P., {et~al.} 2020, \aap, 639, A112,
  \dodoi{10.1051/0004-6361/202037780}

\bibitem[{{Lallement}(2022)}]{Lallement_2022}
{Lallement}, R. 2022, arXiv e-prints, arXiv:2203.01312.
\newblock \doarXiv{2203.01312}

\bibitem[{{Lallement} {et~al.}(2018){Lallement}, {Capitanio}, {Ruiz-Dern},
  {Danielski}, {Babusiaux}, {Vergely}, {Elyajouri}, {Arenou}, \&
  {Leclerc}}]{Lallement_etal_2018}
{Lallement}, R., {Capitanio}, L., {Ruiz-Dern}, L., {et~al.} 2018, \aap, 616,
  A132, \dodoi{10.1051/0004-6361/201832832}

\bibitem[{{Landecker} {et~al.}(2010){Landecker}, {Reich}, {Reid}, {Reich},
  {Wolleben}, {Kothes}, {Uyan{\i}ker}, {Gray}, {Del Rizzo}, {F{\"u}rst},
  {Taylor}, \& {Wielebinski}}]{Landecker_etal_2010}
{Landecker}, T.~L., {Reich}, W., {Reid}, R.~I., {et~al.} 2010, \aap, 520, A80,
  \dodoi{10.1051/0004-6361/200913921}

\bibitem[{{Lenc} {et~al.}(2016){Lenc}, {Gaensler}, {Sun}, {Sadler}, {Willis},
  {Barry}, {Beardsley}, {Bell}, {Bernardi}, {Bowman}, {Briggs}, {Callingham},
  {Cappallo}, {Carroll}, {Corey}, {de Oliveira-Costa}, {Deshpande}, {Dillon},
  {Dwarkanath}, {Emrich}, {Ewall-Wice}, {Feng}, {For}, {Goeke}, {Greenhill},
  {Hancock}, {Hazelton}, {Hewitt}, {Hindson}, {Hurley-Walker},
  {Johnston-Hollitt}, {Jacobs}, {Kapi{\'n}ska}, {Kaplan}, {Kasper}, {Kim},
  {Kratzenberg}, {Line}, {Loeb}, {Lonsdale}, {Lynch}, {McKinley}, {McWhirter},
  {Mitchell}, {Morales}, {Morgan}, {Morgan}, {Murphy}, {Neben}, {Oberoi},
  {Offringa}, {Ord}, {Paul}, {Pindor}, {Pober}, {Prabu}, {Procopio}, {Riding},
  {Rogers}, {Roshi}, {Udaya Shankar}, {Sethi}, {Srivani}, {Staveley-Smith},
  {Subrahmanyan}, {Sullivan}, {Tegmark}, {Thyagarajan}, {Tingay}, {Trott},
  {Waterson}, {Wayth}, {Webster}, {Whitney}, {Williams}, {Williams}, {Wu},
  {Wyithe}, \& {Zheng}}]{Lenc_etal_2016}
{Lenc}, E., {Gaensler}, B.~M., {Sun}, X.~H., {et~al.} 2016, \apj, 830, 38,
  \dodoi{10.3847/0004-637X/830/1/38}

\bibitem[{{Ma} {et~al.}(2020){Ma}, {Mao}, {Ordog}, \& {Brown}}]{Ma_etal_2020}
{Ma}, Y.~K., {Mao}, S.~A., {Ordog}, A., \& {Brown}, J.~C. 2020, \mnras, 497,
  3097, \dodoi{10.1093/mnras/staa2105}

\bibitem[{{Manchester} {et~al.}(2005){Manchester}, {Hobbs}, {Teoh}, \&
  {Hobbs}}]{Manchester_etal_2005}
{Manchester}, R.~N., {Hobbs}, G.~B., {Teoh}, A., \& {Hobbs}, M. 2005, \aj, 129,
  1993, \dodoi{10.1086/428488}

\bibitem[{{Mao} {et~al.}(2012){Mao}, {McClure-Griffiths}, {Gaensler}, {Brown},
  {van Eck}, {Haverkorn}, {Kronberg}, {Stil}, {Shukurov}, \&
  {Taylor}}]{Mao_etal_2012}
{Mao}, S.~A., {McClure-Griffiths}, N.~M., {Gaensler}, B.~M., {et~al.} 2012,
  \apj, 755, 21, \dodoi{10.1088/0004-637X/755/1/21}

\bibitem[{{McClure-Griffiths} {et~al.}(2000){McClure-Griffiths}, {Dickey},
  {Gaensler}, {Green}, {Haynes}, \& {Wieringa}}]{McClure-Griffiths_etal_2000}
{McClure-Griffiths}, N.~M., {Dickey}, J.~M., {Gaensler}, B.~M., {et~al.} 2000,
  \aj, 119, 2828, \dodoi{10.1086/301413}

\bibitem[{McKinven(2021)}]{McKinven_2021}
McKinven, R. 2021, PhD thesis, University of Toronto

\bibitem[{{Mora-Partiarroyo} {et~al.}(2019){Mora-Partiarroyo}, {Krause},
  {Basu}, {Beck}, {Wiegert}, {Irwin}, {Henriksen}, {Stein}, {Vargas}, {Heesen},
  {Walterbos}, {Rand}, {Heald}, {Li}, {Kamieneski}, \&
  {English}}]{Mora-Partiarroyo_etal_2019}
{Mora-Partiarroyo}, S.~C., {Krause}, M., {Basu}, A., {et~al.} 2019, \aap, 632,
  A11, \dodoi{10.1051/0004-6361/201935961}

\bibitem[{{Ocker} {et~al.}(2020){Ocker}, {Cordes}, \&
  {Chatterjee}}]{Ocker_etal_2020}
{Ocker}, S.~K., {Cordes}, J.~M., \& {Chatterjee}, S. 2020, \apj, 897, 124,
  \dodoi{10.3847/1538-4357/ab98f9}

\bibitem[{{Oppermann} {et~al.}(2012){Oppermann}, {Junklewitz}, {Robbers},
  {Bell}, {En{\ss}lin}, {Bonafede}, {Braun}, {Brown}, {Clarke}, {Feain},
  {Gaensler}, {Hammond}, {Harvey-Smith}, {Heald}, {Johnston-Hollitt}, {Klein},
  {Kronberg}, {Mao}, {McClure-Griffiths}, {O'Sullivan}, {Pratley}, {Robishaw},
  {Roy}, {Schnitzeler}, {Sotomayor-Beltran}, {Stevens}, {Stil}, {Sunstrum},
  {Tanna}, {Taylor}, \& {Van Eck}}]{Oppermann_etal_2012}
{Oppermann}, N., {Junklewitz}, H., {Robbers}, G., {et~al.} 2012, \aap, 542,
  A93, \dodoi{10.1051/0004-6361/201118526}

\bibitem[{{Oppermann} {et~al.}(2015){Oppermann}, {Junklewitz}, {Greiner},
  {En{\ss}lin}, {Akahori}, {Carretti}, {Gaensler}, {Goobar}, {Harvey-Smith},
  {Johnston-Hollitt}, {Pratley}, {Schnitzeler}, {Stil}, \&
  {Vacca}}]{Oppermann_etal_2015}
{Oppermann}, N., {Junklewitz}, H., {Greiner}, M., {et~al.} 2015, \aap, 575,
  A118, \dodoi{10.1051/0004-6361/201423995}

\bibitem[{Ordog(2020)}]{Ordog_2020}
Ordog, A. 2020, PhD thesis, University of Calgary, \dodoi{10.11575/PRISM/38243}

\bibitem[{{Ordog} {et~al.}(2019){Ordog}, {Booth}, {Van Eck}, {Brown}, \&
  {Landecker}}]{Ordog_etal_2019}
{Ordog}, A., {Booth}, R., {Van Eck}, C., {Brown}, J.-A., \& {Landecker}, T.
  2019, Galaxies, 7, 43, \dodoi{10.3390/galaxies7020043}

\bibitem[{{Ordog} {et~al.}(2017){Ordog}, {Brown}, {Kothes}, \&
  {Landecker}}]{Ordog_etal_2017}
{Ordog}, A., {Brown}, J.~C., {Kothes}, R., \& {Landecker}, T.~L. 2017, \aap,
  603, A15, \dodoi{10.1051/0004-6361/201730740}

\bibitem[{{Pakmor} {et~al.}(2018){Pakmor}, {Guillet}, {Pfrommer}, {G{\'o}mez},
  {Grand}, {Marinacci}, {Simpson}, \& {Springel}}]{Pakmor_etal_2018}
{Pakmor}, R., {Guillet}, T., {Pfrommer}, C., {et~al.} 2018, \mnras, 481, 4410,
  \dodoi{10.1093/mnras/sty2601}

\bibitem[{{Panopoulou} {et~al.}(2021){Panopoulou}, {Dickinson}, {Readhead},
  {Pearson}, \& {Peel}}]{Panopoulou_2021}
{Panopoulou}, G.~V., {Dickinson}, C., {Readhead}, A.~C.~S., {Pearson}, T.~J.,
  \& {Peel}, M.~W. 2021, \apj, 922, 210, \dodoi{10.3847/1538-4357/ac273f}

\bibitem[{{Pelgrims} {et~al.}(2020){Pelgrims}, {Ferri{\`e}re}, {Boulanger},
  {Lallement}, \& {Montier}}]{Pelgrims_etal_2020}
{Pelgrims}, V., {Ferri{\`e}re}, K., {Boulanger}, F., {Lallement}, R., \&
  {Montier}, L. 2020, \aap, 636, A17, \dodoi{10.1051/0004-6361/201937157}

\bibitem[{{Purcell} {et~al.}(2015){Purcell}, {Gaensler}, {Sun}, {Carretti},
  {Bernardi}, {Haverkorn}, {Kesteven}, {Poppi}, {Schnitzeler}, \&
  {Staveley-Smith}}]{Purcell_2015}
{Purcell}, C.~R., {Gaensler}, B.~M., {Sun}, X.~H., {et~al.} 2015, \apj, 804,
  22, \dodoi{10.1088/0004-637X/804/1/22}

\bibitem[{{Rand} \& {Kulkarni}(1989)}]{Rand_Kulkarni_1989}
{Rand}, R.~J., \& {Kulkarni}, S.~R. 1989, \apj, 343, 760,
  \dodoi{10.1086/167747}

\bibitem[{{Remazeilles} {et~al.}(2015){Remazeilles}, {Dickinson}, {Banday},
  {Bigot-Sazy}, \& {Ghosh}}]{Remazeilles_2014}
{Remazeilles}, M., {Dickinson}, C., {Banday}, A.~J., {Bigot-Sazy}, M.~A., \&
  {Ghosh}, T. 2015, \mnras, 451, 4311, \dodoi{10.1093/mnras/stv1274}

\bibitem[{{Riseley} {et~al.}(2020){Riseley}, {Galvin}, {Sobey}, {Vernstrom},
  {White}, {Zhang}, {Gaensler}, {Heald}, {Anderson}, {Franzen}, {Hancock},
  {Hurley-Walker}, {Lenc}, \& {Van Eck}}]{Riseley_2020}
{Riseley}, C.~J., {Galvin}, T.~J., {Sobey}, C., {et~al.} 2020, \pasa, 37, e029,
  \dodoi{10.1017/pasa.2020.20}

\bibitem[{{Simard-Normandin} \& {Kronberg}(1980)}]{Simard-Normandin_1980}
{Simard-Normandin}, M., \& {Kronberg}, P.~P. 1980, \apj, 242, 74,
  \dodoi{10.1086/158445}

\bibitem[{{Sobey} {et~al.}(2019){Sobey}, {Bilous}, {Grie{\ss}meier}, {Hessels},
  {Karastergiou}, {Keane}, {Kondratiev}, {Kramer}, {Michilli}, {Noutsos},
  {Pilia}, {Polzin}, {Stappers}, {Tan}, {van Leeuwen}, {Verbiest},
  {Weltevrede}, {Heald}, {Alves}, {Carretti}, {En{\ss}lin}, {Haverkorn},
  {Iacobelli}, {Reich}, \& {Van Eck}}]{Sobey_etal_2019}
{Sobey}, C., {Bilous}, A.~V., {Grie{\ss}meier}, J.~M., {et~al.} 2019, \mnras,
  484, 3646, \dodoi{10.1093/mnras/stz214}

\bibitem[{{Sofue} \& {Fujimoto}(1983)}]{Sofue_Fujimoto_1983}
{Sofue}, Y., \& {Fujimoto}, M. 1983, \apj, 265, 722, \dodoi{10.1086/160718}

\bibitem[{{Sokoloff} \& {Shukurov}(1990)}]{Sokoloff_Shukurov_1990}
{Sokoloff}, D., \& {Shukurov}, A. 1990, \nat, 347, 51, \dodoi{10.1038/347051a0}

\bibitem[{{Sokoloff} {et~al.}(1998){Sokoloff}, {Bykov}, {Shukurov},
  {Berkhuijsen}, {Beck}, \& {Poezd}}]{Sokoloff_etal_1998}
{Sokoloff}, D.~D., {Bykov}, A.~A., {Shukurov}, A., {et~al.} 1998, \mnras, 299,
  189, \dodoi{10.1046/j.1365-8711.1998.01782.x}

\bibitem[{{Spoelstra}(1984)}]{Spoelstra_1984}
{Spoelstra}, T.~A.~T. 1984, \aap, 135, 238

\bibitem[{{Stil} \& {Taylor}(2007)}]{Stil_Taylor_2007}
{Stil}, J.~M., \& {Taylor}, A.~R. 2007, \apjl, 663, L21, \dodoi{10.1086/519791}

\bibitem[{{Stil} {et~al.}(2011){Stil}, {Taylor}, \& {Sunstrum}}]{Stil_2011}
{Stil}, J.~M., {Taylor}, A.~R., \& {Sunstrum}, C. 2011, \apj, 726, 4,
  \dodoi{10.1088/0004-637X/726/1/4}

\bibitem[{{Sun} \& {Reich}(2010)}]{Sun_Reich_2010}
{Sun}, X.-H., \& {Reich}, W. 2010, Research in Astronomy and Astrophysics, 10,
  1287, \dodoi{10.1088/1674-4527/10/12/009}

\bibitem[{{Sun} {et~al.}(2008){Sun}, {Reich}, {Waelkens}, \&
  {En{\ss}lin}}]{Sun_etal_2008}
{Sun}, X.~H., {Reich}, W., {Waelkens}, A., \& {En{\ss}lin}, T.~A. 2008, \aap,
  477, 573, \dodoi{10.1051/0004-6361:20078671}

\bibitem[{{Sun} {et~al.}(2015){Sun}, {Landecker}, {Gaensler}, {Carretti},
  {Reich}, {Leahy}, {McClure-Griffiths}, {Crocker}, {Wolleben}, {Haverkorn},
  {Douglas}, \& {Gray}}]{Sun_etal_2015}
{Sun}, X.~H., {Landecker}, T.~L., {Gaensler}, B.~M., {et~al.} 2015, \apj, 811,
  40, \dodoi{10.1088/0004-637X/811/1/40}

\bibitem[{{Tahani} {et~al.}(2019){Tahani}, {Plume}, {Brown}, {Soler}, \&
  {Kainulainen}}]{Tahani_etal_2019}
{Tahani}, M., {Plume}, R., {Brown}, J.~C., {Soler}, J.~D., \& {Kainulainen}, J.
  2019, \aap, 632, A68, \dodoi{10.1051/0004-6361/201936280}

\bibitem[{{Tahani} {et~al.}(2022{\natexlab{a}}){Tahani}, {Lupypciw}, {Glover},
  {Plume}, {West}, {Kothes}, {Inutsuka}, {Lee}, {Robishaw}, {Knee}, {Brown},
  {Doi}, {Grenier}, \& {Haverkorn}}]{Tahani_etal_2022a}
{Tahani}, M., {Lupypciw}, W., {Glover}, J., {et~al.} 2022{\natexlab{a}}, \aap,
  660, A97, \dodoi{10.1051/0004-6361/202141170}

\bibitem[{{Tahani} {et~al.}(2022{\natexlab{b}}){Tahani}, {Glover}, {Lupypciw},
  {West}, {Kothes}, {Plume}, {Inutsuka}, {Lee}, {Grenier}, {Knee}, {Brown},
  {Doi}, {Robishaw}, \& {Haverkorn}}]{Tahani_etal_2022b}
{Tahani}, M., {Glover}, J., {Lupypciw}, W., {et~al.} 2022{\natexlab{b}}, \aap,
  660, L7, \dodoi{10.1051/0004-6361/202243322}

\bibitem[{{Taylor} {et~al.}(2009){Taylor}, {Stil}, \&
  {Sunstrum}}]{Taylor_etal_2009}
{Taylor}, A.~R., {Stil}, J.~M., \& {Sunstrum}, C. 2009, \apj, 702, 1230,
  \dodoi{10.1088/0004-637X/702/2/1230}

\bibitem[{{Thomson} {et~al.}(2018){Thomson}, {McClure-Griffiths}, {Federrath},
  {Dickey}, {Carretti}, {Gaensler}, {Haverkorn}, {Kesteven}, \&
  {Staveley-Smith}}]{Thomson_etal_2018}
{Thomson}, A. J.~M., {McClure-Griffiths}, N.~M., {Federrath}, C., {et~al.}
  2018, \mnras, 479, 5620, \dodoi{10.1093/mnras/sty1865}

\bibitem[{{Thomson} {et~al.}(2019){Thomson}, {Landecker}, {Dickey},
  {McClure-Griffiths}, {Wolleben}, {Carretti}, {Fletcher}, {Federrath}, {Hill},
  {Mao}, {Gaensler}, {Haverkorn}, {Clark}, {Van Eck}, \&
  {West}}]{Thomson_etal_2019}
{Thomson}, A. J.~M., {Landecker}, T.~L., {Dickey}, J.~M., {et~al.} 2019,
  \mnras, 487, 4751, \dodoi{10.1093/mnras/stz1438}

\bibitem[{{Thomson} {et~al.}(2021){Thomson}, {Landecker}, {McClure-Griffiths},
  {Dickey}, {Campbell}, {Carretti}, {Clark}, {Federrath}, {Gaensler}, {Han},
  {Haverkorn}, {Hill}, {Mao}, {Ordog}, {Pratley}, {Reich}, {Van Eck}, {West},
  \& {Wolleben}}]{Thomson_etal_2021}
{Thomson}, A. J.~M., {Landecker}, T.~L., {McClure-Griffiths}, N.~M., {et~al.}
  2021, \mnras, 507, 3495, \dodoi{10.1093/mnras/stab1805}

\bibitem[{{Tribble}(1991)}]{Tribble_1991}
{Tribble}, P.~C. 1991, \mnras, 250, 726, \dodoi{10.1093/mnras/250.4.726}

\bibitem[{{Vallee} \& {Bignell}(1983)}]{Vallee_Bignell_1983}
{Vallee}, J.~P., \& {Bignell}, R.~C. 1983, \apj, 272, 131,
  \dodoi{10.1086/161269}

\bibitem[{{van der Walt} {et~al.}(2011){van der Walt}, {Colbert}, \&
  {Varoquaux}}]{van2011numpy}
{van der Walt}, S., {Colbert}, S.~C., \& {Varoquaux}, G. 2011, Computing in
  Science and Engineering, 13, 22, \dodoi{10.1109/MCSE.2011.37}

\bibitem[{{Van Eck} {et~al.}(2011){Van Eck}, {Brown}, {Stil}, {Rae}, {Mao},
  {Gaensler}, {Shukurov}, {Taylor}, {Haverkorn}, {Kronberg}, \&
  {McClure-Griffiths}}]{VanEck_etal_2011}
{Van Eck}, C.~L., {Brown}, J.~C., {Stil}, J.~M., {et~al.} 2011, \apj, 728, 97,
  \dodoi{10.1088/0004-637X/728/2/97}

\bibitem[{{Van Eck} {et~al.}(2019){Van Eck}, {Haverkorn}, {Alves}, {Beck},
  {Best}, {Carretti}, {Chy{\.z}y}, {En{\ss}lin}, {Farnes}, {Ferri{\`e}re},
  {Heald}, {Iacobelli}, {Jeli{\'c}}, {Reich}, {R{\"o}ttgering}, \&
  {Schnitzeler}}]{van_Eck_etal_2019}
{Van Eck}, C.~L., {Haverkorn}, M., {Alves}, M.~I.~R., {et~al.} 2019, \aap, 623,
  A71, \dodoi{10.1051/0004-6361/201834777}

\bibitem[{{Vidal} {et~al.}(2015){Vidal}, {Dickinson}, {Davies}, \&
  {Leahy}}]{Vidal_2015}
{Vidal}, M., {Dickinson}, C., {Davies}, R.~D., \& {Leahy}, J.~P. 2015, \mnras,
  452, 656, \dodoi{10.1093/mnras/stv1328}

\bibitem[{{Virtanen} {et~al.}(2020{\natexlab{a}}){Virtanen}, {Gommers},
  {Oliphant}, {Haberland}, {Reddy}, {Cournapeau}, {Burovski}, {Peterson},
  {Weckesser}, {Bright}, {van der Walt}, {Brett}, {Wilson}, {Millman},
  {Mayorov}, {Nelson}, {Jones}, {Kern}, {Larson}, {Carey}, {Polat}, {Feng},
  {Moore}, {VanderPlas}, {Laxalde}, {Perktold}, {Cimrman}, {Henriksen},
  {Quintero}, {Harris}, {Archibald}, {Ribeiro}, {Pedregosa}, {van Mulbregt}, \&
  {SciPy 1. 0 Contributors}}]{Jones_etal_2001}
{Virtanen}, P., {Gommers}, R., {Oliphant}, T.~E., {et~al.} 2020{\natexlab{a}},
  Nature Methods, 17, 261, \dodoi{10.1038/s41592-019-0686-2}

\bibitem[{{Virtanen} {et~al.}(2020{\natexlab{b}}){Virtanen}, {Gommers},
  {Oliphant}, {Haberland}, {Reddy}, {Cournapeau}, {Burovski}, {Peterson},
  {Weckesser}, {Bright}, {van der Walt}, {Brett}, {Wilson}, {Millman},
  {Mayorov}, {Nelson}, {Jones}, {Kern}, {Larson}, {Carey}, {Polat}, {Feng},
  {Moore}, {VanderPlas}, {Laxalde}, {Perktold}, {Cimrman}, {Henriksen},
  {Quintero}, {Harris}, {Archibald}, {Ribeiro}, {Pedregosa}, {van Mulbregt}, \&
  {SciPy 1. 0 Contributors}}]{SciPy2020}
---. 2020{\natexlab{b}}, Nature Methods, 17, 261,
  \dodoi{10.1038/s41592-019-0686-2}

\bibitem[{{Waelkens} {et~al.}(2009){Waelkens}, {Jaffe}, {Reinecke}, {Kitaura},
  \& {En{\ss}lin}}]{Waelkens:2009bn}
{Waelkens}, A., {Jaffe}, T., {Reinecke}, M., {Kitaura}, F.~S., \& {En{\ss}lin},
  T.~A. 2009, \aap, 495, 697, \dodoi{10.1051/0004-6361:200810564}

\bibitem[{{Weisberg} {et~al.}(2004){Weisberg}, {Cordes}, {Kuan}, {Devine},
  {Green}, \& {Backer}}]{Weisberg_etal_2004}
{Weisberg}, J.~M., {Cordes}, J.~M., {Kuan}, B., {et~al.} 2004, \apjs, 150, 317,
  \dodoi{10.1086/379802}

\bibitem[{{West} {et~al.}(2020){West}, {Henriksen}, {Ferri{\`e}re},
  {Woodfinden}, {Jaffe}, {Gaensler}, \& {Irwin}}]{West_etal_2020}
{West}, J.~L., {Henriksen}, R.~N., {Ferri{\`e}re}, K., {et~al.} 2020, \mnras,
  499, 3673, \dodoi{10.1093/mnras/staa3068}

\bibitem[{{West} {et~al.}(2021){West}, {Landecker}, {Gaensler}, {Jaffe}, \&
  {Hill}}]{West_etal_2021}
{West}, J.~L., {Landecker}, T.~L., {Gaensler}, B.~M., {Jaffe}, T., \& {Hill},
  A.~S. 2021, \apj, 923, 58, \dodoi{10.3847/1538-4357/ac2ba2}

\bibitem[{{Wolleben} {et~al.}(2010){Wolleben}, {Landecker}, {Hovey}, {Messing},
  {Davison}, {House}, {Somaratne}, \& {Tashev}}]{Wolleben_etal_2010}
{Wolleben}, M., {Landecker}, T.~L., {Hovey}, G.~J., {et~al.} 2010, \aj, 139,
  1681, \dodoi{10.1088/0004-6256/139/4/1681}

\bibitem[{{Wolleben} {et~al.}(2021){Wolleben}, {Landecker}, {Douglas}, {Gray},
  {Ordog}, {Dickey}, {Hill}, {Carretti}, {Brown}, {Gaensler}, {Han},
  {Haverkorn}, {Kothes}, {Leahy}, {McClure-Griffiths}, {McConnell}, {Reich},
  {Taylor}, {Thomson}, \& {West}}]{Wolleben_etal_2021}
{Wolleben}, M., {Landecker}, T.~L., {Douglas}, K.~A., {et~al.} 2021, \aj, 162,
  35, \dodoi{10.3847/1538-3881/abf7c1}

\bibitem[{{Woodfinden} {et~al.}(2019){Woodfinden}, {Henriksen}, {Irwin}, \&
  {Mora-Partiarroyo}}]{Woodfinden_etal_2019}
{Woodfinden}, A., {Henriksen}, R.~N., {Irwin}, J., \& {Mora-Partiarroyo}, S.~C.
  2019, \mnras, 487, 1498, \dodoi{10.1093/mnras/stz1366}

\bibitem[{{Xu} \& {Han}(2014)}]{Xu_Han_2014}
{Xu}, J., \& {Han}, J.-L. 2014, Research in Astronomy and Astrophysics, 14,
  942, \dodoi{10.1088/1674-4527/14/8/005}

\bibitem[{{Xu} \& {Han}(2019)}]{Xu_Han_2019}
{Xu}, J., \& {Han}, J.~L. 2019, \mnras, 486, 4275,
  \dodoi{10.1093/mnras/stz1060}

\bibitem[{{Yao} {et~al.}(2017){Yao}, {Manchester}, \& {Wang}}]{Yao_etal_2017}
{Yao}, J.~M., {Manchester}, R.~N., \& {Wang}, N. 2017, \apj, 835, 29,
  \dodoi{10.3847/1538-4357/835/1/29}

\end{thebibliography}

\appendix

\section{The effect of binning the RM values before correlation
\label{sec:appendix_bin_sizes}}

\begin{figure}
\hspace{1in}\includegraphics[width=5in]{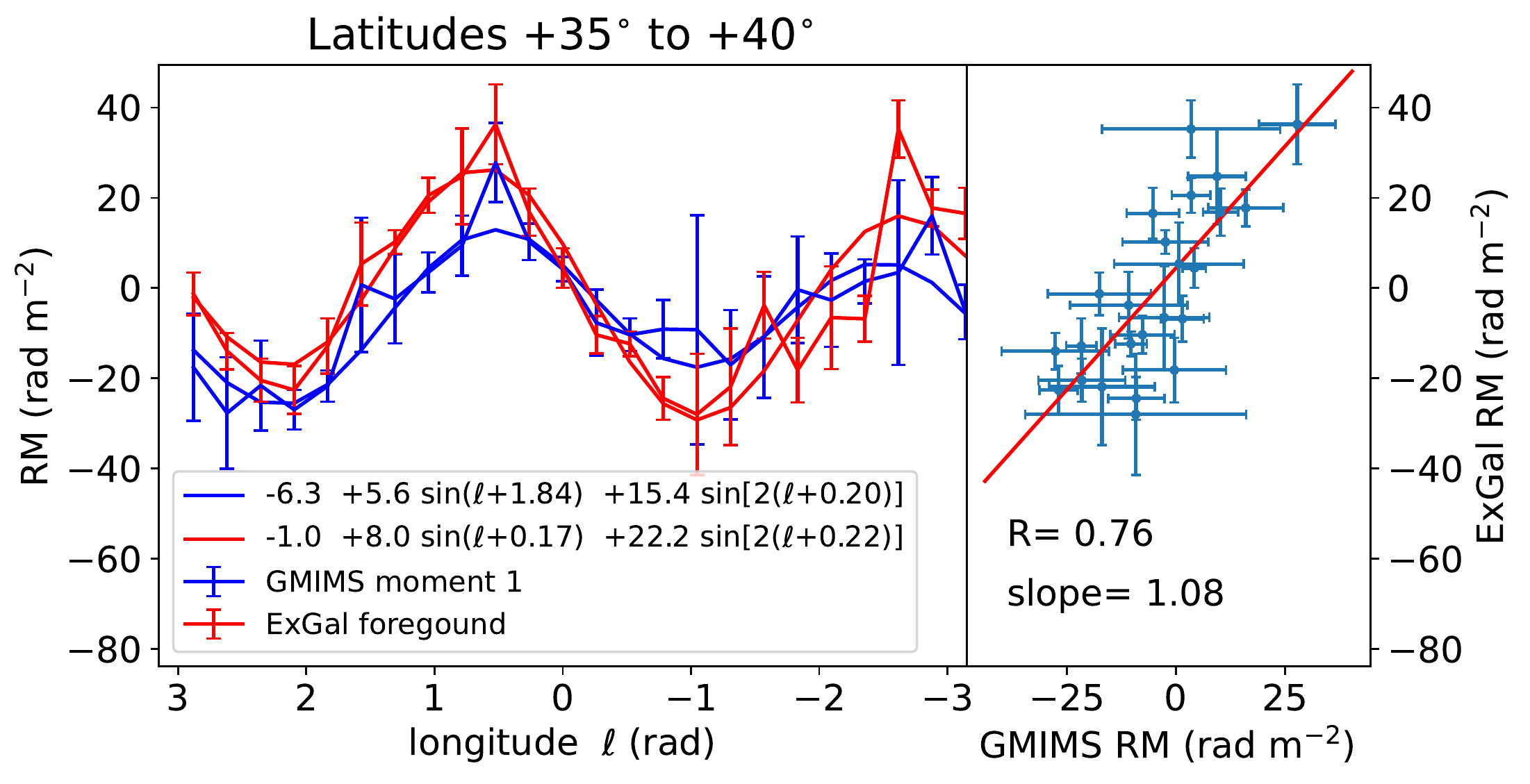}

\hspace{1in}\includegraphics[width=5in]{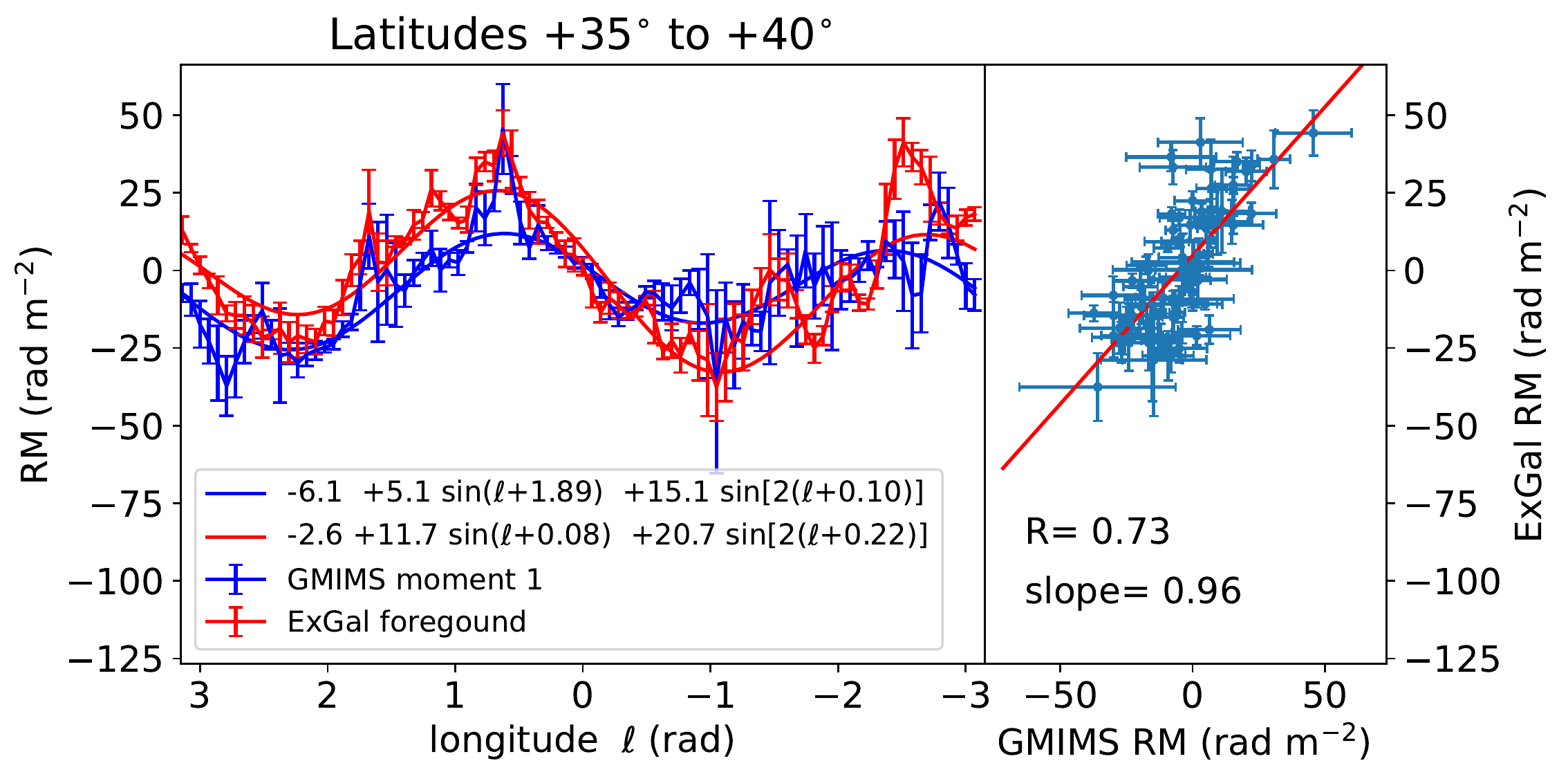}
\caption{Illustration of the effect of changing the bin width in longitude
from 15\arcdeg\xspace (24 bins, top panel) to 4\arcdeg\xspace (90 bins, lower panel).
Although there is more structure in the data on scales of a few degrees with 
the smaller bins, the results from fitting sine curves are changed only slightly.
\label{fig:24_90bins_b35}}
\end{figure}

\begin{figure}
\hspace{1.3in}\includegraphics[width=4in]{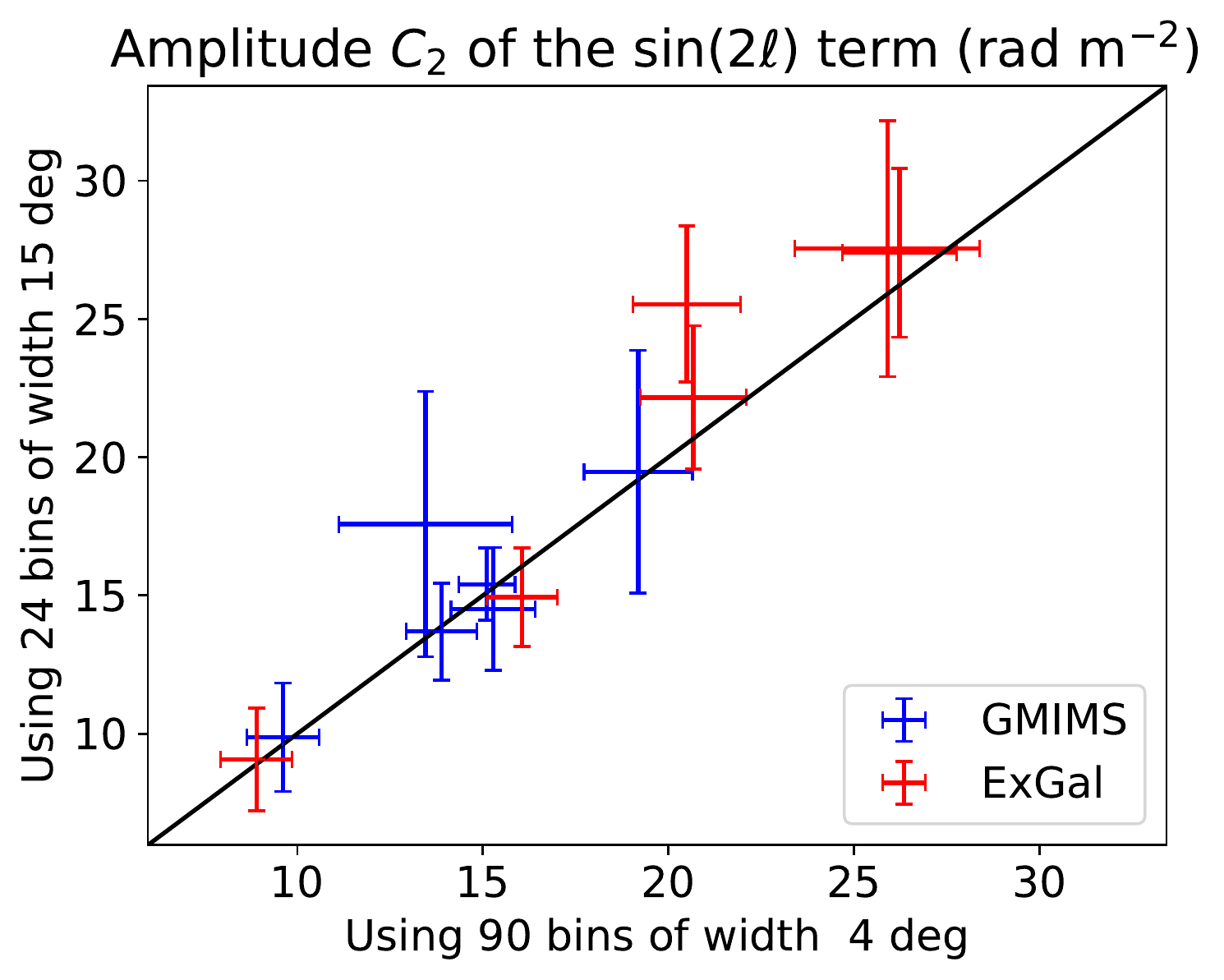}
\caption{Comparing the fitted values of coefficient $C_2$ (Eq. \ref{eq:5paramfit})
at different latitudes for different bin widths, 4\arcdeg\xspace and 15\arcdeg\xspace,
as on Fig. \ref{fig:24_90bins_b35}.  Only latitudes giving values of $C_2 > 5 \sigma$
are shown.  For all but one point the results agree within error bars of $\pm 1 \sigma$.
From this and similar comparisons with many different bin widths, we conclude that
the bin size has minimal effect on the least squares fitting in section \ref{sec:RMdata}.
\label{fig:C_2_comparison_24vs90}}
\end{figure}

Grouping the values of RM from the extragalactic and GMIMS surveys into 
bins 5\arcdeg\xspace to 10\arcdeg\xspace on a side and then computing the median of the
values in each bin greatly reduces the scatter in the points on Fig.s
\ref{fig:example_b40} and \ref{fig:b40_noNPS}.  The bin size has very little
effect on the results of fitting for the constants in Eq. \ref{eq:5paramfit}.
To confirm the robustness of the results on Table \ref{tab:longitude_fits},
we repeat the analysis of section \ref{sec:RMdata} using all integer factors 
of 360 between 24 and 90 for the number of longitude bins for each 
5\arcdeg\xspace latitude band.  Two examples are shown in Fig. \ref{fig:24_90bins_b35}, with 24 and 90 bins respectively, for latitudes +35\arcdeg$<b<$+40\arcdeg,
corresponding to the upper panel of Fig. \ref{fig:example_b40}.  The upper
panel of Fig. \ref{fig:24_90bins_b35} has 24 bins of width 15\arcdeg,
the lower panel has 90 bins of width 4\arcdeg.  The fitted functions have
quite similar values for the amplitudes and phases of both the $\sin{(\ell)}$
and $\sin{(2\ell)}$ terms.  Considering the full latitude range 
20\arcdeg$<b<$50\arcdeg\xspace where the fitted value of $C_2$ is 
greater than five times the error gives Fig.
\ref{fig:C_2_comparison_24vs90}.
The fitted values of $C_2$, i.e. the amplitude of the $\sin{(2\ell)}$ term,
are plotted for each latitude using longitude
bins of 4\arcdeg\xspace vs. 15\arcdeg.
The results are quite consistent, with only one of the 12 points showing more
than one-sigma difference.

\begin{figure}
\hspace{1.3in}\includegraphics[width=4in]{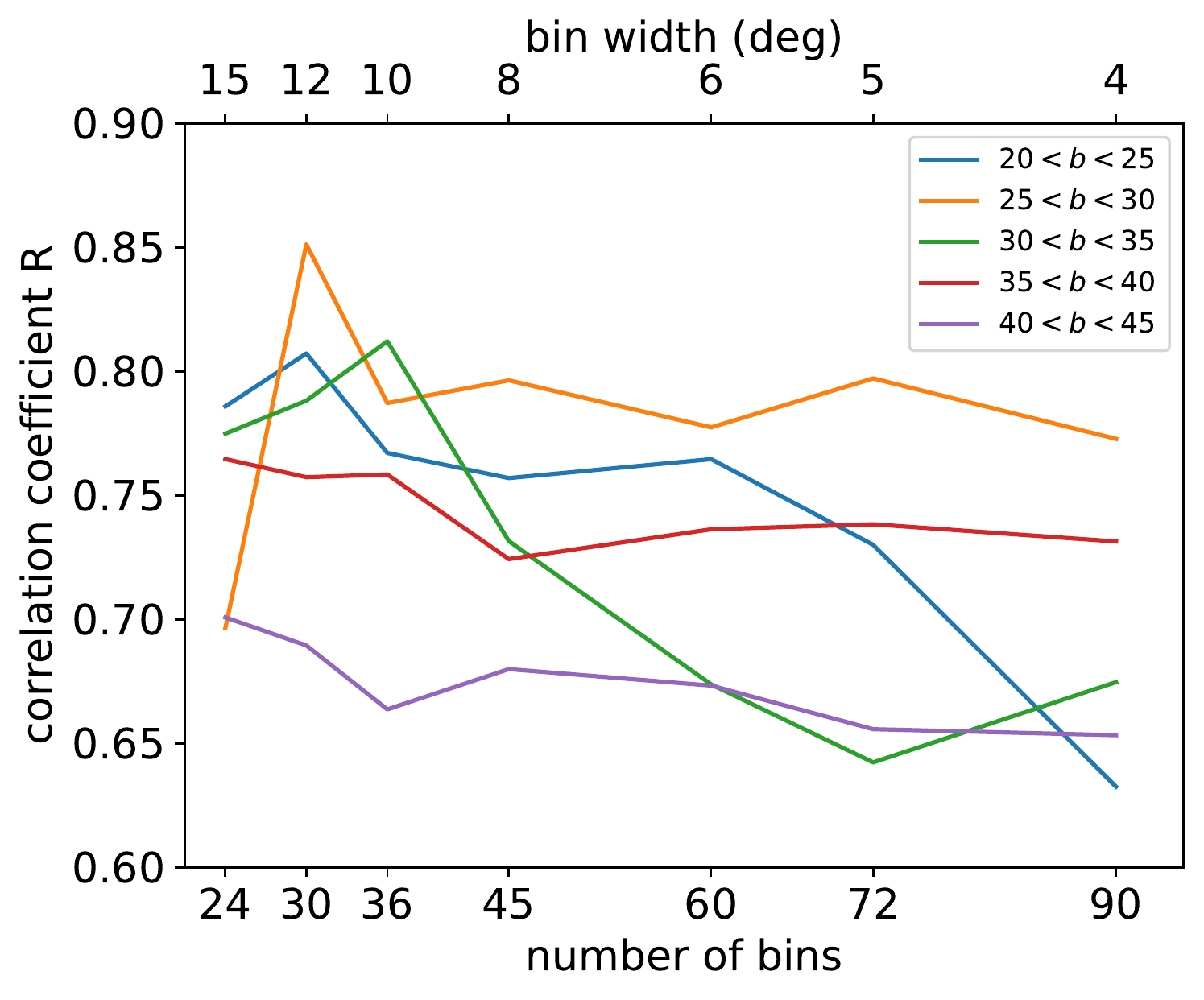}
\caption{The correlation between RMs from the two surveys as a function of the
bin widths.  There is a small but significant reduction in the correlation coefficient
as the bins get smaller.
\label{fig:correl_vs_nbins}}
\end{figure}

Decreasing the size of the bins 
has a moderate effect on the correlation between the extragalactic and 
GMIMS values of RM.  This is shown on Fig. \ref{fig:correl_vs_nbins}.
Each trace shows the result of computing correlations like those shown on
Fig. \ref{fig:scatter}, for different bin widths from 15\arcdeg\xspace to
4\arcdeg.  For most latitudes the correlation coefficient, $R$, decreases
as the bins get narrower, from roughly 0.75 for bins $\sim$10\arcdeg\xspace wide
or more, down to about 0.70 for bin widths of 4\arcdeg\xspace to 5\arcdeg.
Although the effect is small, the fact that narrower bins show less correlation
suggests that the large angle patterns are well correlated between the two
surveys, while the structure in RMs at angles less than about 5\arcdeg\xspace
is less well correlated.  This result might be expected if the small
angular-scale variations come from structures that cover narrow intervals
along the LoS.  The two surveys weight the polarized emission 
from different distances differently, so small features have different
effects in the two.

\begin{figure}
\hspace{.2in}\includegraphics[width=6.4in]{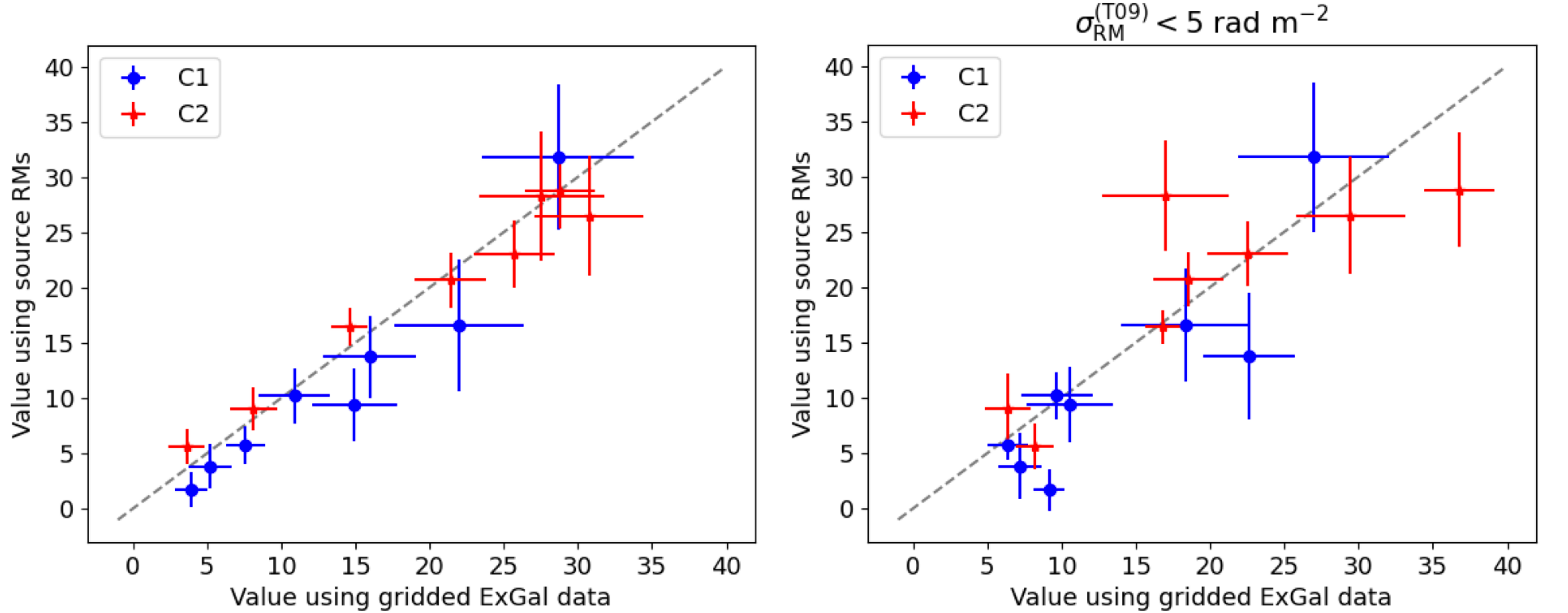}
\caption{Comparing the fitted values of coefficient $C_1$ and $C_2$ (Eq. \ref{eq:5paramfit})
at northern latitudes ($15\arcdeg < b < 55\arcdeg$) for 10 degree bins using the gridded extragalactic RMs vs. binning the extragalactic sources directly from \citet{Taylor_etal_2009}.  The left panel uses all 
sources with equal weighting.
The right panel uses only the 
sources with the highest precision in RM, i.e. those
with measurement error $\sigma_{RM} < 5$ rad m$^{-2}$.
\label{fig:check_nogrid}}
\end{figure}

In addition to checking the effect of the bin widths, it is interesting to check whether the smoothing and interpolation
of the extragalactic RMs to make the continuous RM function of
\citet{Hutschenreuter_etal_2021} has a significant effect on the
fitting results.  For this we use the catalog of \citet{Taylor_etal_2009} and bin the RMs of the sources
directly, then compute the median of the RMs in each bin.
The result is shown on the left panel of Fig. \ref{fig:check_nogrid}.
The values of the coefficients $C_1$ and $C_2$ are consistent
within the error bars, except for those points that have 
poorly determined values of the constants (less than 5 $\sigma$)
from the least squares fitting (Eq. \ref{eq:5paramfit}). The right panel of Fig. \ref{fig:check_nogrid} shows a similar result for a smaller sample of extragalactic sources, only those with nominal high precision ($\sigma_{RM}<$ 5 rad m$^{-2}$) in the RM measurements by \citet{Taylor_etal_2009}.  Reducing the size of the sample has the effect of degrading the correlation with the binned result, in spite of the higher precision of the individual measurements.

\section{Chi-square, Goodness of Fit, and the RM Distribution Function \label{sec:appendix_chisquared}}

The quality of the least squares fits of the median RMs in each longitude bin is considered here, using the method of reduced chi-square
($\chi^2_{\nu}$), following the technique described in \citet[][section 8.3]{Barlow_1989}.  Defining $\chi^2$ as:
\begin{equation} \label{eq:chisq}
\chi^2 \ = \ \sum_{i=1}^n \frac{[y_i \ - \ f(\ell_i)]^2}{\sigma_i^2}
\end{equation}
where $n$ is the number of data points (in this case $n$=36 except at latitudes where there are no samples in one or more bins because they are
below the declination limit of the GMIMS survey) and the $y_i$ are the medians of the binned values of the RM at each longitude, $\ell_i$.  The rms
errors of the measured values are $\sigma_i$, and the fitted function $f(\ell_i)$ has the form of Eq. \ref{eq:5paramfit}, with the
values of the five parameters, $C_0,\ C_1,\ C_2,\ \phi_1$, and $\phi_2$ as described in Sec. \ref{sec:2longitude}.
Dividing by the number of degrees of freedom, $\nu = n - k$, where $k=5$ is the number of parameters in the model.
gives the reduced chi-square statistic:
\begin{equation}\label{eq:reduced_chisq}
\chi^2_{\nu} \ = \ \frac{\chi^2}{\nu}
\end{equation}
If the fitted function, $f$, completely describes the underlying variation between the data points, $y_i$, (but not the  errors) then we expect $\chi^2_{\nu} \sim 1$.
Under ideal conditions \citep{Andrae_etal_2010} the excess of $\chi^2_{\nu}$ above one can be interpreted as a probability that the
fitted function fully explains the underlying variation among the measured values, as a statistical test of null hypothesis.
Even though these conditions do not apply here, it is useful to consider the quality of the fitted functions for each latitude, with parameters given on 
Table \ref{tab:longitude_fits}, by computing $\chi^2_{\nu}$ and comparing it to a value that would give 10\% probability, $P = 0.1$, that the data
are fully described by the model, and that the residuals are purely due to the errors, $\sigma_i$, in the data points.  

As \citet{Barlow_1989} points out, for large $\nu$ ($\nu > \sim 30$) the quantity $\sqrt{2 \chi^2}$ is distributed roughly as a Gaussian with mean
value $\sqrt{2 \nu - 1}$ and standard deviation equal to one.  With this approximation the probability $P$ gives $\chi^2_{\nu}$ using the inverse of the error function, $erf^{-1}$
({\tt erfinv} in {\sc scipy.special}), as:
\begin{equation}\label{eq:erf}
\chi^2_{\nu} \ \simeq \ \frac{\left[ \sqrt{2 \nu - 1} \ + \ \sqrt{2} \cdot erf^{-1}(1 - 2 P) \right]^2}{2 \nu} \ \ \ \ \ \ \ \ \ (\textrm{for}\ \nu > \sim 30)
\end{equation}
This gives $\chi^2_{\nu} < 1.33$ for $P < 0.10$ with $\nu = 31$ (the more precise {\sc scipy} routine {\tt chdtri} gives 1.34).  The model must be linear in the fitted
parameters to allow interpretation of $P$ as a probability, but Eq. \ref{eq:5paramfit} is not linear in the phases, $\phi_1$ and $\phi_2$.
So the conclusion that $P=$0.1 or 10\% probability of the residuals exceeding this value by chance, corresponding to the 1.33 threshold in $\chi^2_{\nu}$, is only for comparison purposes.

\begin{table}
\caption{Reduced Chi-Squared Results \label{tab:chisq}}
\begin{center}
\begin{tabular}{|l|cc|cc|}
\hline
& \multicolumn{2}{c|}{GMIMS}  
& \multicolumn{2}{c|}{ExGal} \\
Latitude Range & $\chi^2_{\nu}$ & $\Psi_{\nu}^2$ & $\chi^2_{\nu}$ & $\Psi_{\nu}^2$ \\ \hline
-60\arcdeg$ < b \leq$-55\arcdeg & 1.2$\pm$0.4 & 3.5$\pm$0.3 & 2.5$\pm$0.3 & 6.5$\pm$0.2 \\
-55\arcdeg$ < b \leq$-50\arcdeg & 1.8$\pm$0.4 & 3.8$\pm$0.3 & 2.2$\pm$0.3 & 12.4$\pm$0.2 \\
-50\arcdeg$ < b \leq$-45\arcdeg & 1.7$\pm$0.4 & 4.6$\pm$0.3 & 1.3$\pm$0.3 & 7.1$\pm$0.2 \\
-45\arcdeg$ < b \leq$-40\arcdeg & 0.5$\pm$0.4 & 1.8$\pm$0.3 & 2.1$\pm$0.3 & 7.2$\pm$0.2 \\
-40\arcdeg$ < b \leq$-35\arcdeg & 0.8$\pm$0.4 & 1.5$\pm$0.3 & 2.6$\pm$0.3 & 11.4$\pm$0.2 \\
-35\arcdeg$ < b \leq$-30\arcdeg & 0.8$\pm$0.3 & 1.9$\pm$0.3 & 2.9$\pm$0.3 & 15.1$\pm$0.2 \\
-30\arcdeg$ < b \leq$-25\arcdeg & 0.7$\pm$0.3 & 2.3$\pm$0.3 & 2.4$\pm$0.3 & 13.1$\pm$0.2 \\
-25\arcdeg$ < b \leq$-20\arcdeg & 0.6$\pm$0.3 & 1.7$\pm$0.3 & 1.8$\pm$0.3 & 9.1$\pm$0.2 \\
-20\arcdeg$ < b \leq$-15\arcdeg & 0.8$\pm$0.3 & 1.8$\pm$0.3 & 2.8$\pm$0.3 & 8.4$\pm$0.2 \\
-15\arcdeg$ < b \leq$-10\arcdeg & 0.5$\pm$0.3 & 0.8$\pm$0.3 & 3.3$\pm$0.3 & 6.6$\pm$0.2 \\
-10\arcdeg$ < b \leq$-5\arcdeg & 0.6$\pm$0.3 & 1.0$\pm$0.3 & 1.3$\pm$0.3 & 4.8$\pm$0.2 \\
-5\arcdeg$ < b \leq$ 0\arcdeg & 0.8$\pm$0.3 & 2.4$\pm$0.3 & 0.7$\pm$0.3 & 2.8$\pm$0.2 \\
 0\arcdeg$ < b \leq$ 5\arcdeg & 0.5$\pm$0.3 & 1.2$\pm$0.3 & 0.9$\pm$0.3 & 2.6$\pm$0.2 \\
 5\arcdeg$ < b \leq$10\arcdeg & 0.4$\pm$0.3 & 1.6$\pm$0.3 & 1.7$\pm$0.3 & 4.7$\pm$0.2 \\
10\arcdeg$ < b \leq$15\arcdeg & 1.1$\pm$0.3 & 3.2$\pm$0.3 & 1.7$\pm$0.3 & 4.1$\pm$0.2 \\
15\arcdeg$ < b \leq$20\arcdeg & 0.5$\pm$0.3 & 2.7$\pm$0.3 & 1.8$\pm$0.3 & 4.9$\pm$0.2 \\
20\arcdeg$ < b \leq$25\arcdeg & 0.7$\pm$0.3 & 1.9$\pm$0.3 & 1.8$\pm$0.3 & 4.9$\pm$0.2 \\
25\arcdeg$ < b \leq$30\arcdeg & 1.2$\pm$0.3 & 2.9$\pm$0.3 & 1.7$\pm$0.3 & 8.4$\pm$0.2 \\
30\arcdeg$ < b \leq$35\arcdeg & 0.9$\pm$0.3 & 4.5$\pm$0.3 & 2.6$\pm$0.3 & 9.1$\pm$0.2 \\
35\arcdeg$ < b \leq$40\arcdeg & 0.8$\pm$0.3 & 6.4$\pm$0.2 & 2.9$\pm$0.3 & 10.3$\pm$0.2 \\
40\arcdeg$ < b \leq$45\arcdeg & 1.3$\pm$0.3 & 5.9$\pm$0.2 & 1.5$\pm$0.3 & 9.0$\pm$0.2 \\
45\arcdeg$ < b \leq$50\arcdeg & 1.2$\pm$0.3 & 5.0$\pm$0.2 & 1.6$\pm$0.3 & 4.3$\pm$0.2 \\
50\arcdeg$ < b \leq$55\arcdeg & 0.9$\pm$0.3 & 2.4$\pm$0.2 & 1.8$\pm$0.3 & 4.7$\pm$0.2 \\
55\arcdeg$ < b \leq$60\arcdeg & 0.9$\pm$0.3 & 1.7$\pm$0.2 & 1.6$\pm$0.3 & 4.0$\pm$0.2 \\
\hline
\end{tabular}
\end{center}
\end{table}

Table \ref{tab:chisq} gives values for $\chi^2_{\nu}$ for all latitude fits, for both the GMIMS and extragalactic
data, corresponding to Table \ref{tab:longitude_fits}.  
Errors on the $\chi^2_{\nu}$ values on Table \ref{tab:chisq} are
simply $\sqrt{2/\nu}$ assuming that the variance of $\chi^2$ is $2 \cdot \nu$. 
Of particular interest is the range of Northern latitudes where the $\sin{2\ell}$ term dominates the fitted
function, for both the GMIMS and extragalactic data, i.e. $+20 < b < +50$.  The GMIMS
data are consistent with $\chi^2_{\nu} = 1.0$ for this range.
The extragalactic values of $\chi^2_{\nu}$ are greater than one, typically by more than their errors.  This suggests that the data could
be better fitted with a function with more parameters, e.g. more terms in a Fourier series.  This is not surprising since 
the $\sin{\ell}$ and $\sin{2\ell}$ functions cannot model the variations of RM on angular scales smaller than about $\frac{\pi}{2}$ radians,
but there is a great deal of structure in the RM distribution on all angular scales \citep[e.g.][]{Haverkorn_etal_2008}.  
For comparison, Table \ref{tab:chisq} shows $\chi^2_{\nu}$ for a different model, given by simply removing the mean value of the RMs
at each latitude, i.e.
\begin{equation} \label{eq:chisq_mean_only}
\Psi^2_{\nu} \ = \ \sum_{i=1}^n \frac{[y_i \ - \ \bar{y}]^2}{\sigma_i^2}
\end{equation} 
where $\bar{y}$ is the mean of the $y_i$ values, and for this case $\nu = 35$.
In all cases, $\Psi^2_{\nu} > \chi^2_{\nu}$, typically by a factor of three to five, which shows that fitting functions with the form
of Eq. \ref{eq:5paramfit} is well justified, although further fitting on smaller angular scales would also be justified for the extragalactic data in particular.  For the GMIMS data there is also structure on angular
scales smaller than one radian, as can be seen by the contrast between the upper and lower panels for Fig. 
\ref{fig:gmims_ylm}, so adding more terms to Equation \ref{eq:5paramfit} would certainly decrease 
the values of $\chi^2_{\nu}$.  The fact that these values are already less than one on Table \ref{tab:chisq}
suggests that the error bars on the data points, i.e. the errors on the median RM values in each bin, have been overestimated.  

\begin{figure}
\hspace{1.1in} \includegraphics[width=5in]{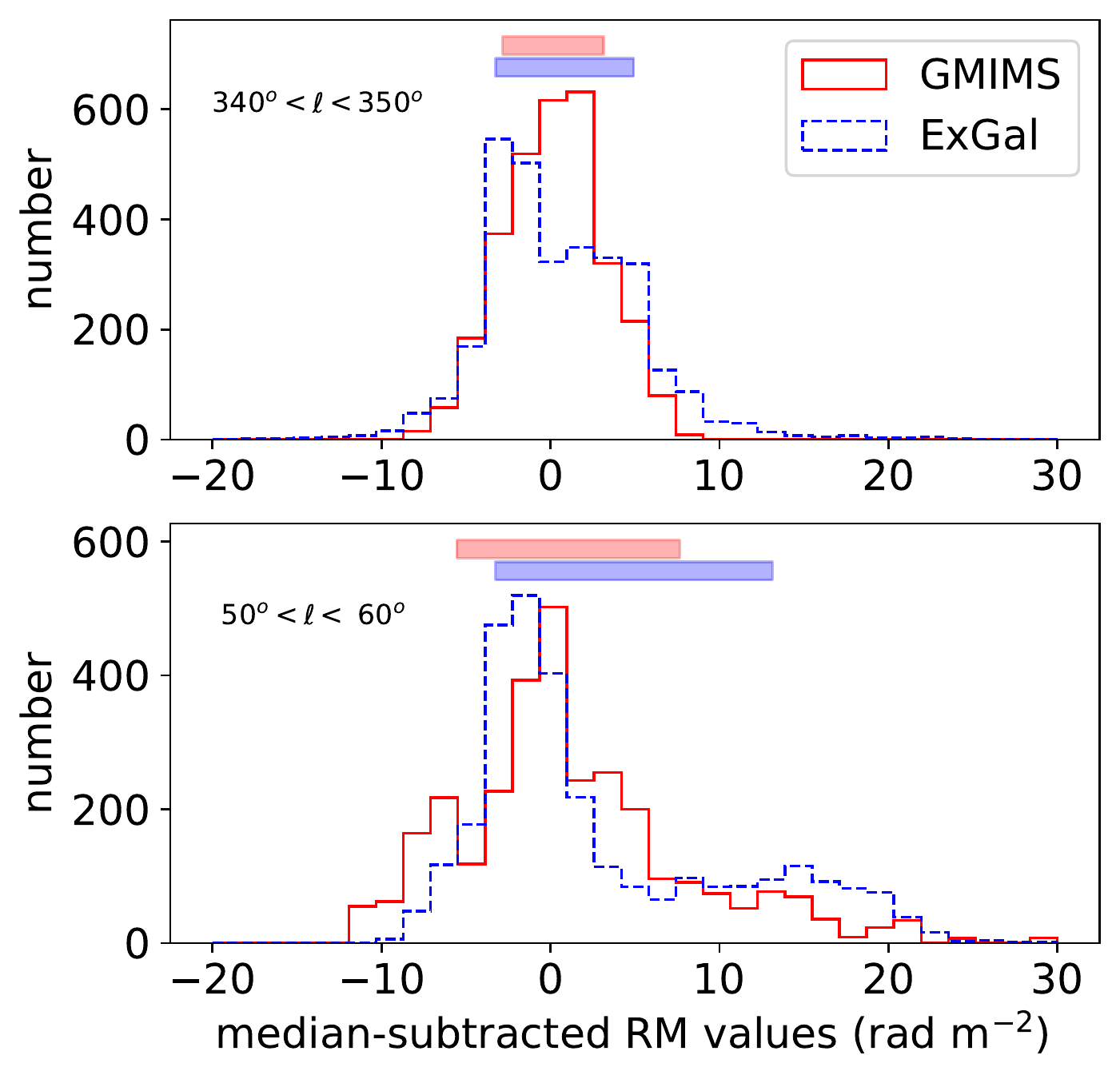}

\caption{Two examples of the distributions of values in longitude bins, both for the range 35\arcdeg$< b <$40\arcdeg, with the GMIMS histogram in red and the extragalactic in blue.  The upper panel shows the distribution of samples in the 
longitude bin from 340\arcdeg $< \ell <$350\arcdeg, the lower panel shows the same for the bin from 50\arcdeg$< \ell <$
60\arcdeg that includes part of the North Polar Spur.  In both cases the medians have been subtracted from the values, so that all the histograms have zero median.  The rectangles at
the top show the range from the 16th to the 84th percentiles, that we use to determine $\pm 1 \sigma$ for the error bars on Fig. 
\ref{fig:example_b40} and for the determination of $\chi^2_{\nu}$ on Table \ref{tab:chisq}.
\label{fig:two_nonGaussian} }
\end{figure}

A fundamental source of misinterpretation in the application of the $\chi^2$ statistic is the possibility that the errors of the fitted data points,
$\sigma_i$, do not reflect variances of Gaussian distributions.
Studying the distributions of the points in the bins shows that sometimes they are roughly normally distributed and sometimes not.  Figure 
\ref{fig:two_nonGaussian} shows two typical cases, both taken from longitude bins for the latitude range 35\arcdeg$< b <$40\arcdeg.
The upper panel shows distributions that are approximately normal, but the lower panel shows a case for which both
samples show a long positive-going tail.  In some other bins one or both distributions are bimodal.  A proper statistical analysis of 
goodness of fit to a model should be done using Monte-Carlo methods based on the probability distributions themselves, as in the 
Markov-Chain-Monte-Carlo analysis of \citet{Thomson_etal_2019}.

The rich texture of the RM patterns on the sky causes the non-Gaussian RM distributions in the bins.  It also contributes to the relatively high values of
the $\chi^2_{\nu}$ for the extragalactic data on Table \ref{tab:chisq}.  Modelling the RM patterns of both surveys on angular
scales smaller than $\sim 1$ radian is beyond the scope of this paper.  As surveys with a greater areal density of RMs become available the effective angular resolution
of the RM maps will improve, allowing more detailed modelling of structures in the Galactic magnetic field on a wide range of scales.

\end{document}